\documentclass[]{aa1}		 % for the 2-column author version

\usepackage{natbib}
\usepackage{lscape}
\usepackage{graphicx}
\usepackage{longtable}
\usepackage{supertabular}
\usepackage{tabularx}
\usepackage{txfonts}
\usepackage{extarrows}
\usepackage{color}
\usepackage[table]{xcolor}
\usepackage{hyperref}
\bibpunct{(}{)}{;}{a}{}{,}
\usepackage{booktabs}

\definecolor{Red}{rgb}{1, 0, 0}
\definecolor{Green}{rgb}{0, 1, 0}
\definecolor{Blue}{rgb}{0, 0, 1}
\definecolor{Black}{rgb}{0, 0, 0}
\definecolor{White}{rgb}{1, 1, 1}
\definecolor{Grey}{rgb}{0.5, 0.5, 0.5}
\definecolor{Yellow}{rgb}{1, 1, 0}
\definecolor{Magenta}{rgb}{1, 0, 1}
\definecolor{Cyan}{rgb}{0, 1, 1}
\definecolor{Orange}{rgb}{0.8, 0.3, 0}
\definecolor{DarkGreen}{rgb}{0.2, 0.7, 0.1}
\definecolor{Pink}{rgb}{1, 0.4, 0.7}
\definecolor{TabHeaderColor}{rgb}{0.7, 0.20,  0.14} 	% bright pompeian red: 170/255, 50/255, 35/255

\begin{document}

\title{The baryon budget of galaxies across the first billion years}

\subtitle{Theoretical predictions for gas phases, depletion times, stellar return fractions}

\titlerunning{The early baryon cycle}

\authorrunning{U.~Maio \& C.~P\'eroux}

\author{
	Umberto Maio\inst{1, 2}\thanks{Corresponding author: \href{mailto:umberto.maio@inaf.it}{\texttt{umberto.maio@inaf.it}}}
	\and 
	C\'eline P\'eroux\inst{3,4}
}

\institute{
    	INAF-Italian National Institute of Astrophysics, Observatory of Trieste, via G. Tiepolo 11, 34143 Trieste, Italy % \email{umberto.maio@inaf.it}
	\and
    	Institute for Fundamental Physics of the Universe, via Beirut 2, 34151 Trieste, Italy
	\and
    	European Southern Observatory, Karl-Schwarzschild-Stra{\ss}e 2, 85748 Garching bei Muenchen, Germania
	\and
    	Aix Marseille Universit\'e, CNRS, LAM (Laboratoire d’Astrophysique de Marseille) UMR 7326, 13388, Marseille, France
}

\date{Received 9 October 2025 / Accepted 15 February 2026}

%\abstract{}{}{}{}{}
%  5 {} tokens are mandatory: Context - Aims - Methods - Results - Conclusions
\abstract
{ {\it Context.}
ALMA and JWST observations of galaxies in the first billion years of the Universe provide key constraints on the baryon cycle during the epoch of reionisation. A complete census of the baryonic phases in early galaxies is essential to understand the efficiency and timescales of star formation.
}
{ {\it Aims.}
In this work, we study cosmic matter at redshift $ z > 5 $ to investigate the different phases in which early gas and stars reside, their corresponding mass budgets, the resulting depletion times, and the expected stellar return fraction as a function of stellar age.}
{ {\it Methods.}
We used the {\sc ColdSIM} hydrodynamical time-dependent non-equilibrium chemistry simulations to perform a detailed analysis of the cold, warm, hot, and stellar phases for both bound structures (galaxies and their circumgalactic medium) and the diffuse intergalactic medium. We further investigated the cold HI and H$_2$ components, explicitly computed in our simulations, and examined their relations with the host mass, star formation, metallicity, and depletion timescales.
We also provide observational insights and discuss the implications for stellar mass functions, Population~III star formation, and changes in the initial mass function.
}
{ {\it Results.}
Cosmic gas prior to reionisation is mostly cold, while at later epochs the warm phase becomes dominant as a consequence of enhanced star formation activity and increasing UV reionising radiation.
Stellar return fractions at these times are $\sim 0.15$--0.20, a factor of two lower than the values usually adopted.
Mass functions and mass density parameters in bound objects increase with cosmic time, closely tracing the overall structure formation process.
Cold, warm, and hot gas masses as well as HI and H$_2$ components show increasing trends with mass and star formation rate, while HI and H$_2$ depletion times decrease down to 0.01--0.1~Gyr with a weak dependence on metallicity.
The resulting star formation efficiency remains at the level of a few percent independently of $z$ and gas-to-star fractions decline with mass, influenced by local feedback and environment.
Our findings are consistent with ALMA, VLA and IRAM surveys at later epochs, including ALFALFA, xCOLDGASS, GASS, xGASS, EDGE-CALIFA, PHIBBS, and ASPECS.
}
{ {\it Conclusions.}
Gas phases are quantitatively related to the underlying stellar populations and can be used to infer unknown quantities. In the appendix, we provide fit functions describing the trends of the stellar return fraction, the main sequence, phase mass relations, gas-to-star fractions, and depletion timescales.
}


\maketitle

% =============================================================================

%**********************************************************************

\section{Introduction} \label{sect:introduction}

%**********************************************************************

\noindent
Baryonic structure formation relies on a scenario in which cosmic gas falls into the galaxy potential wells from large scales, cools down, and condenses into stars.
Formed stars release energy and mass into the surrounding medium via star formation feedback that, together with active galactic nucleus (AGN) activity, leads baryon outflows.
This determines a continuous mass transfer among different baryon phases and scales, during which matter flows in and out of galaxies: the so-called baryon cycle \cite[e.g.][]{Tumlinson2017, PH2020}.
This latter essentially mediates the interplay between cosmic mass that is gravitationally bound in galaxies and the intergalactic medium (IGM).
Primordial times witness the very first episodes of the cosmic baryon cycle, which is likely in place by the first half-gigayear of the Universe.
\\
Recent JWST and ALMA detections \cite[][]{Chemerynska2024, Carniani2024, Schouws2025arxiv, Schouws2025OIII, Helton2025col, Naidu2025, Conselice2025, Fujimoto2025, Gandolfi2025} suggest that the first galaxies have already formed by redshift $ z\simeq 14 $ and present relatively complex metallicity ($Z$) patterns with a star-forming main sequence established over timescales of 10--100~Myr.
The onset of such early star formation episodes requires significant amounts of cold molecular (H$_2$) mass. Indeed, neutral hydrogen (HI) and helium are not efficient coolants below about $10^4\,\rm K$, but HI and residual free electrons trigger the formation of large quantities of H$_2$ both in pristine population~III (PopIII) and in polluted population~II-I (PopII-I) regimes \cite[][]{Saslaw1967, HM1979, Abel1997, GP1998, Maio2010, Wise2012, Huyan2025, Vanzella2026}.
\\
Cold or cooling gas is commonly observed in galaxies through HI 21~cm line or molecular transitions.
Warm gas is usually seen in star-forming sites via photoionisation emission in H$_\alpha$ or high-ionisation metal species and expected in optically thick regions of MHD-driven accretion discs.
With high-energy instruments, such as eRosita or XRISM, hot (2--10~keV) gas can be found in optically selected objects at $z \simeq 0$--2, X-ray emitting galaxy clusters, and AGN coronas \cite[][]{Jiang2019, Comparat2023}.
Resolved active-galaxy studies in the local Universe have detected kiloparsec-scale molecular outflows with strong contributions from the H$_2$ phase at temperatures of $ T \gtrsim 10^3\,\rm K$ \cite[][]{Cicone2014, Cicone2021, Lamperti2022, Zanchettin2025, Koribalski2025, Langan2026}, while MUSE observations indicate the presence of $\sim$10$^2\,\rm K$ molecular gas in galaxies that are both accreting and expelling material \cite[][]{Weng2022}.
Despite the evidence that baryons can be found in a broad variety of phases and objects, quantitative estimates are challenging.
\\
The relations between cold neutral gas (detected in absorption) and warm ionised gas (probed in emission) are currently being explored.
Recent comparative studies involving controlled samples at $z\simeq 2$--4 have found a typical variance in the probed gas abundances of $0.2$-0.3~dex with dependencies on the line diagnostic used \cite[][]{Schady2024}.
This approach opens the possibility of combining both absorption and emission techniques to better investigate high-$z$ faint galaxies \cite[][]{Kulkarni2022}.
Nevertheless, theoretical studies focusing on high-$z$ gas phases are still lacking and the contribution of cold, warm, and hot gas to the whole cosmic mass budget is to be determined.
\\
It is known from both observational results and theoretical calculations that at $z < 6$, after reionisation completes, the cosmic mass density of cold neutral gas flattens around $10^{-3}$ of the cosmic critical density \cite[][]{Wolfe1995, Rao2000, StorrieLombardi2000,  Peroux2001, Nagamine2004, Tacconi2020, Heintz2022}.
The condensed baryon budget is dominated by the HI component, while corresponding H$_2$ values are estimated to be much lower \cite[between a factor of a few and some dex;][]{Walter2020, Decarli2020, Maio2022, Boogaard2023, Messias2024}.
Previous studies have shown that warm or hot gas can account for a large fraction of baryons in low-redshift structures
\cite[e.g.][]{ Dev2024, Deepak2025}, but at high $z$ the situation is extremely uncertain and their share, distribution, and evolution is debated \cite[][]{Peroux2024}.
\\
The interplay among different gas phases, the conversion of cold gas into stars and the release of stellar mass into warm and hot gas depend on a few physical quantities, i.e. gas depletion times ($t_{\rm depl}$), stellar ages ($t_{\rm age}$), and the stellar return fraction ($R$).
A clear assessment of these variables is crucial for our understanding of the early baryon cycle, as well as for addressing the connection between gas and stars.
Low-$z$ and cosmic-noon studies have suggested that typical depletion times might be of the order of some gigayears \cite[][]{Tacconi2013, Tacconi2018, Castignani2025, Kaur2025}, while commonly adopted values for $R$ are obtained by averaging over a chosen initial stellar mass function (IMF) and imposing instantaneous recycling of massive stars \cite[][]{Maeder1992, MD2014}.
That means that stellar ages and time-dependent yields are not taken into account.
Thus, resulting $R$ values, typically in the range of $R\simeq 0.3$--0.4 for \cite{Salpeter1955} and \cite{Kroupa2001} or \cite{Chabrier2003} IMFs, can be considered valid only on timescales of 5--10~Gyr.
Due to the much shorter cosmic times at $z \gtrsim 5$, those findings are unreliable for such an epoch.
Hence, depletion times and stellar return fractions should be revised and reconciled with high-$z$ timescales.
This is also relevant for estimates of the star formation rate (SFR) in the early Universe.
\\
Throughout this work, we shed light on these issues by using ad hoc hydrodynamical simulations of primordial structure formation taken by the {\sc ColdSIM} project \cite[e.g.][]{Maio2022, Casavecchia2024, Parente2026}.
We adopt a $\Lambda$ cold dark matter ($\Lambda$CDM) model with present-day cosmological-constant, matter, and baryon density parameters of $\Omega_{\rm 0,\Lambda} = 0.726 $, $\Omega_{\rm 0,m} = 0.274 $, and $\Omega_{\rm 0,b} =  0.0458 $, respectively.
The expansion parameter normalised to 100~km/s/Mpc is $h = 0.702$, while the mass variance within 8 Mpc/$h$ radius is $\sigma_8 = 0.816$ and the power spectrum index is $n = 0.968$
\cite[][]{Planck2020}.
The comoving cosmological critical density is $ \rho_{\rm 0,crit } = 277.4 \, h^2 \rm M_\odot\, kpc^{-3} = 136.7 \, \rm M_\odot\, kpc^{-3} $ and the value of the metallicity in the solar neighbourhood is $Z_\odot = 0.02$.
All logarithms are in base 10.
\\
The text is organised as follows.
In Sect.~\ref{sect:simulations}, we describe the numerical simulations used and the analysis performed.
In Sect.~\ref{sect:results} we show our main results and in Sect.~\ref{sect:discussion} we discuss them.
We then summarise our findings and conclude in Sect.~\ref{sect:conclusions}.
Fitting relations and supplementary material are collected in the appendix.

%************************************************************************

\section{Method} \label{sect:simulations}

%************************************************************************

We employed the detailed and accurate {\sc ColdSIM} numerical simulations of primordial cosmic structure formation.
The implementation includes gravitational-force calculations and smoothed-particle hydrodynamics inherited from the {\sc P-Gadget-3} code \cite[][]{Springel2005} and additional extensions including:
non-equilibrium atomic and molecular chemistry (for e$^-$, H, H$^+$, H$^-$, He, He$^+$, He$^{++}$, H$_2$, H$_2^+$, D, D$+$, HD, and HeH$^+$) explicitly followed in cold, warm, and hot gas phases;
dust grain catalysis;
photoelectric and cosmic-ray heating;
HI and H$_2$ density-dependent self-shielding;
gas cooling from resonant and sub-millimetre fine-structure line emission;
star formation and stellar evolution of PopII-I and PopIII stars according to a Salpeter IMF;
remnant masses and metal spreading for He, C, N, O, Ne, Mg, Si, S, Ca, and Fe, from type-II and type-Ia supernovae (SNe) and asymptotic-giant-branch (AGB) events according to metal-dependent yields and lifetimes;
thermal feedback; and galactic winds (kinetic feedback) at 350~km/s
\cite[][]{Maio2022, Maio2023, Casavecchia2024, Casavecchia2025, Parente2026}.
PopIII and PopII-I regimes are differentiated in terms of a critical metallicity $Z_{\rm crit} = 10^{-4} \,\rm Z_\odot$ and differ for the stellar metal yields adopted. Yields for zero-metallicity stars are used in the PopIII case ($Z < Z_{\rm crit}$) and yields for metal-enriched stars in the PopII-I case ($Z \ge Z_{\rm crit}$).
We remind the reader that $Z_{\rm crit} $ values are quite uncertain \cite[][]{Bromm2001zcrit, Omukai2005, SS2006, Chiaki2014, Chon2024}, but, due to the rapid metal enrichment by early SNe, different assumptions do not lead to significantly different results \cite[][]{Maio2010}.
\\
We consider three cosmic volumes initialised at redshift $z=99$ according to the $\Lambda$CDM paradigm.
Our reference (Ref) set-up has a comoving box side of $ L = 10\,{\rm Mpc}/h$ and an initial number of gas and dark-matter particles of $ N_{\rm part} = 2\times 512^3$.
The high-resolution (HR) run has the same $L$, and $  N_{\rm part} = 2\times 1000^3$, while the large-box (LB) run is characterised by $L = 50\, {\rm Mpc}/h$ and $ N_{\rm part} = 2\times 1000^3$.
We performed our analysis on the Ref run and checked the effects of resolution or volume sampling by comparisons with the HR and LB simulations in Appendix~\ref{appendixA} and \ref{appendixSelection}.
The resulting gas ($m_{\rm gas}$) and dark-matter ($m_{\rm dm}$) mass resolutions are listed in Tab.~\ref{tab:sims}.
Cosmic structures were identified by searching for clustered particles via a friend-of-friend technique with a linking length of $\sim$0.2 the mean inter-particle separation.
A substructure finder algorithm checked for minimal values of the local gravitational potential and gave us basic properties of gravitationally bound structures \cite[][]{Dolag2009}, such as the position, virial radius ($R_{\rm vir}$), virial mass ($M_{\rm vir}$), stellar mass ($M_{\rm star}$), SFR, and $Z$.
To avoid numerical artefacts that could have led to spurious results, we considered bound structures to be well behaved only if they were sampled with at least 300 particles and their $R_{\rm vir} > 0$.
\begin{table}  %  sims
\centering
\caption{\small Simulation set-up for the HR, Ref, and LB runs.}
\label{tab:sims}
\begin{tabular}{l  c  c  c  c }
	\toprule
	\toprule
	\textcolor{TabHeaderColor}{\bf Name} & \textcolor{TabHeaderColor}{$\bf L \rm [Mpc]$}	& \textcolor{TabHeaderColor}{$\bf N_{\rm \bf part}$}	& \textcolor{TabHeaderColor}{$\bf m_{\rm \bf gas} [{\rm M_\odot}/{\it h}]$} 	&  \textcolor{TabHeaderColor}{$\bf m_{\rm \bf dm} [{\rm M_\odot}/{\it h}]$} \\
	\midrule
	HR		& 14		& 	$2\times1000^3$		& 	$1.27\times 10^4 $	 	& 	$ 6.34\times 10^4$ \\
	Ref		& 14		& 	$2\times512^3$		& 	$ 9.47\times 10^4 $	& 	$4.72\times 10^5$ \\
	LB		& 71		& 	$2\times1000^3$		& 	$1.58\times 10^6 $		& 	$ 7.93\times 10^6$ \\
	\bottomrule
\end{tabular}
\end{table}
\begin{figure*}[]
 \centering
 \includegraphics[width=0.33\textwidth]{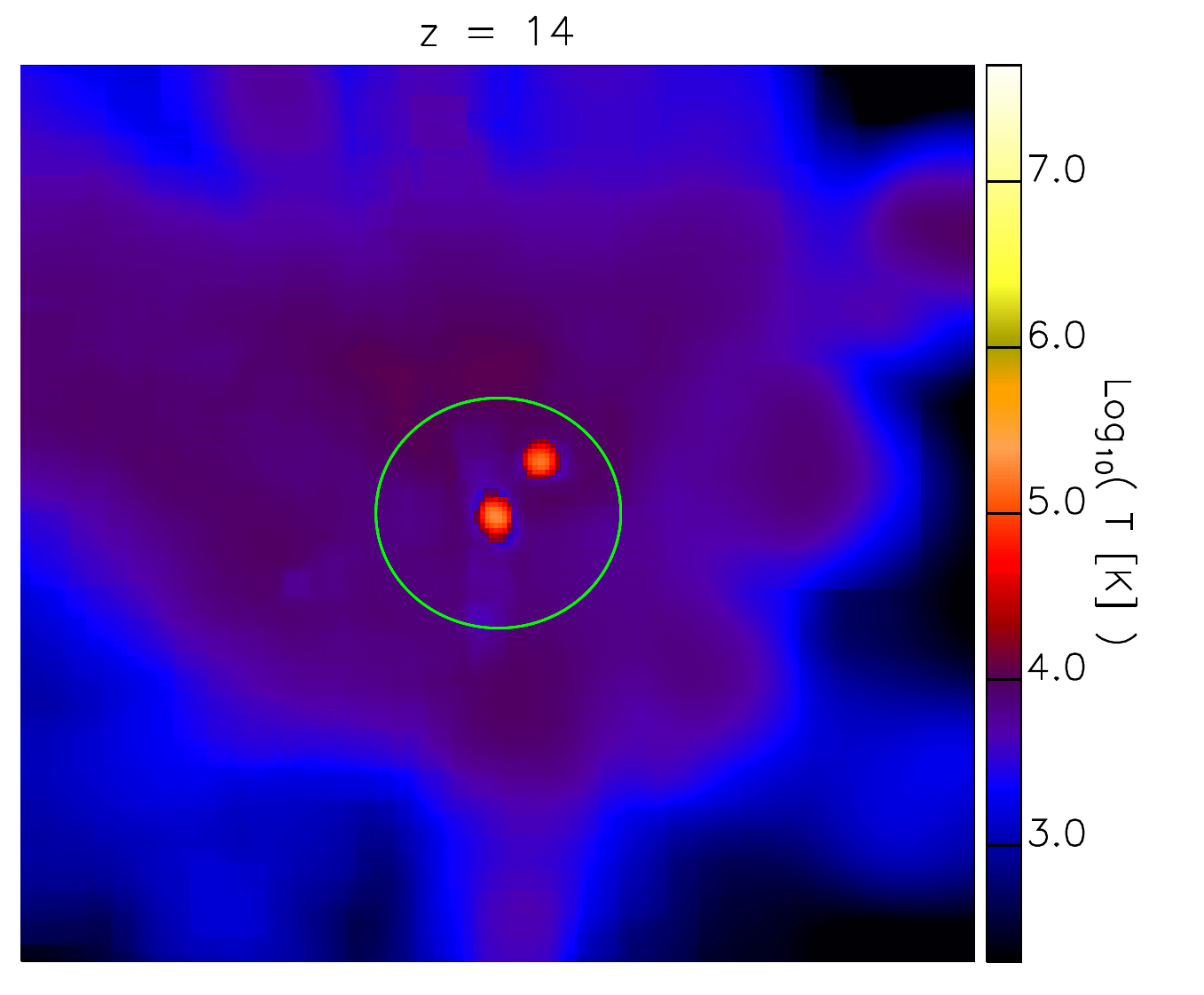} \hspace{-1.4cm}
 \includegraphics[width=0.33\textwidth]{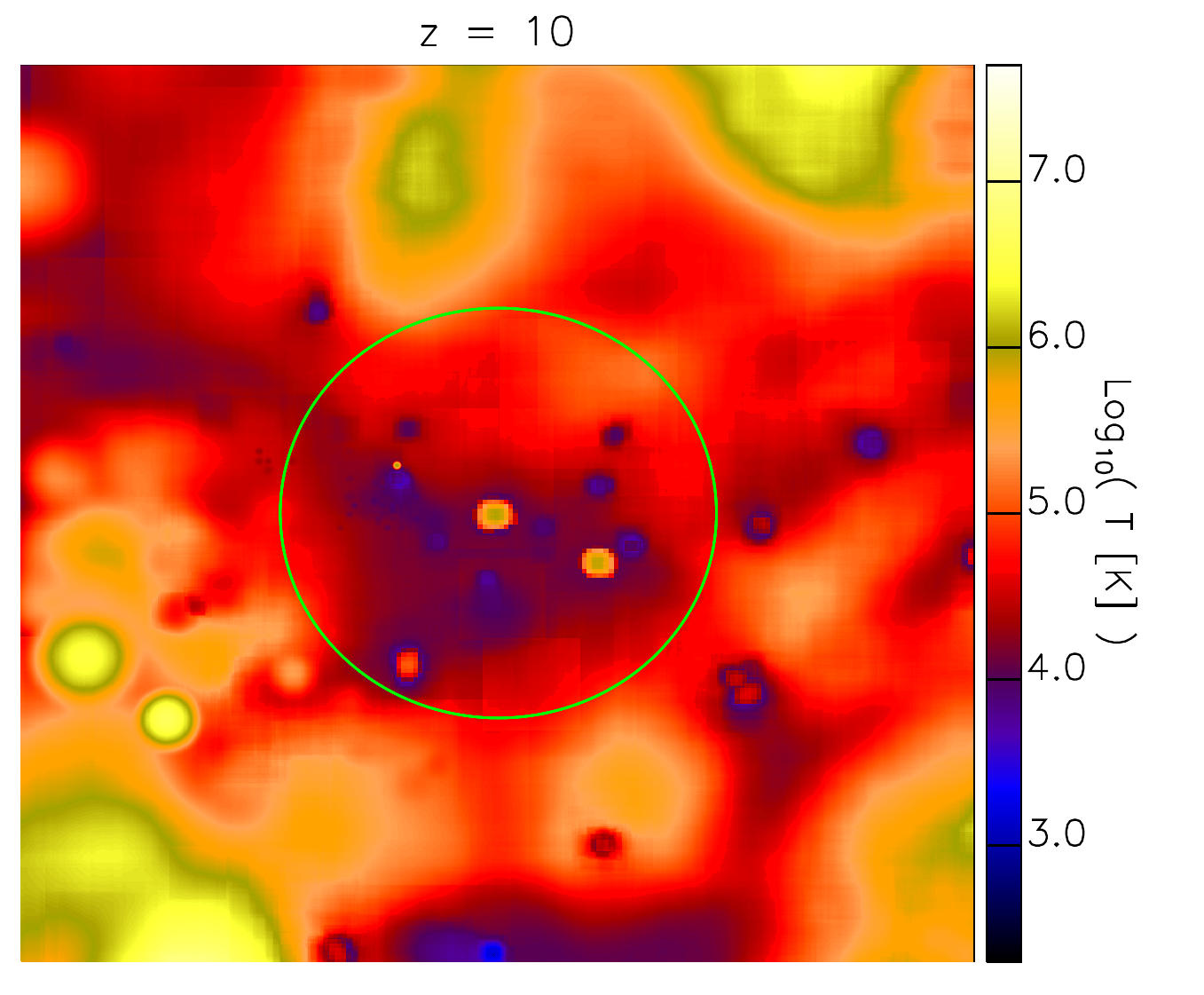} \hspace{-1.4cm}
 \includegraphics[width=0.33\textwidth]{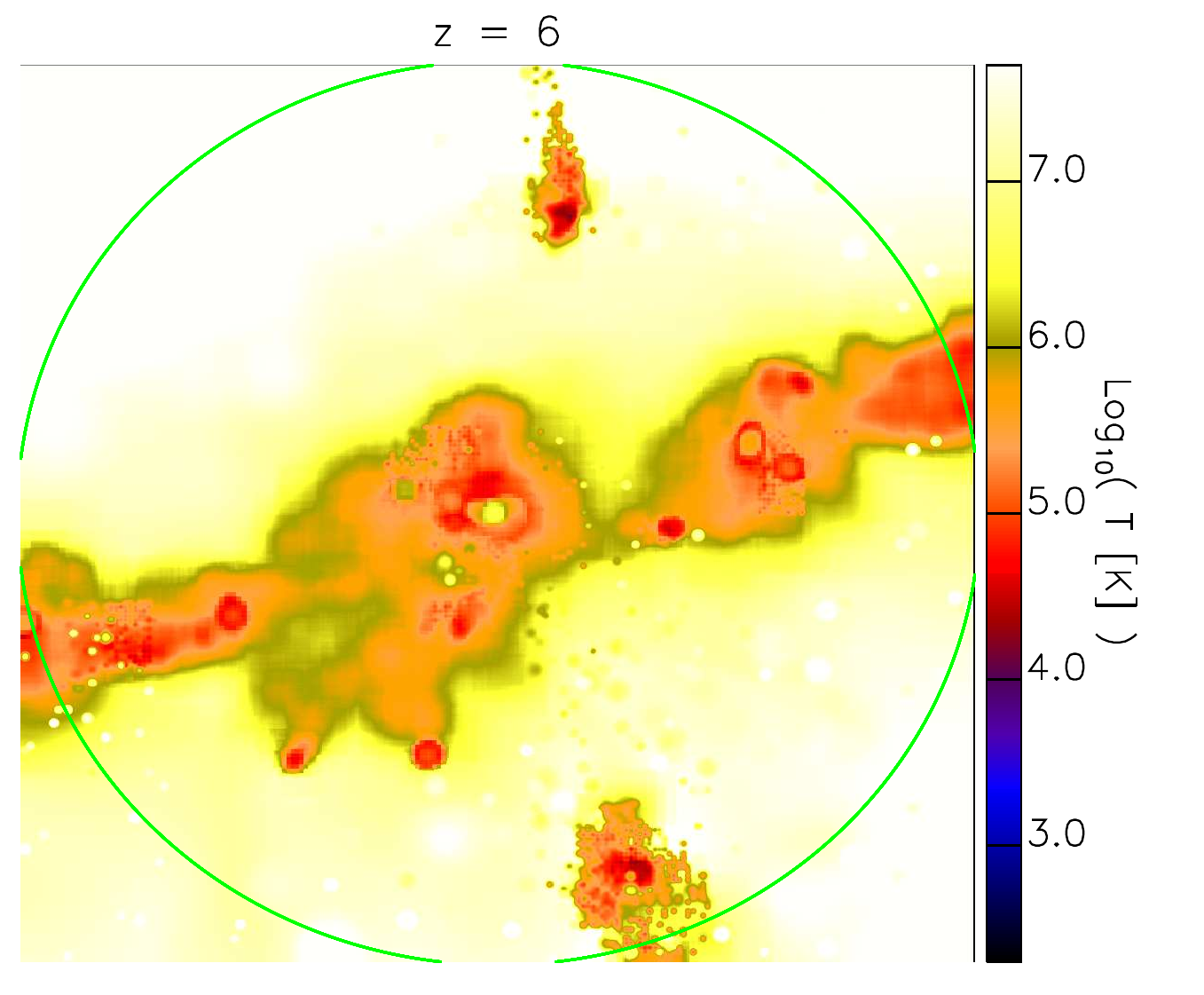} 
 \caption{\small Mass-weighted temperature maps at $ z \simeq 14 $, 10 and 6 smoothed on a grid of 1024 pixels a side for the gas in the largest halo that has a physical virial radius (circles) of $\sim2.4$, 5.9, and 18.7~kpc and a total mass of $ 1.4 \times 10^{9} $, $ 7.3 \times 10^{9} $, and $ 6.9 \times 10^{10}  \,\rm M_\odot $, respectively, from left to right. }
 \label{fig:maps}
\end{figure*}
A pictorial view of such cosmic objects is given in Fig.~\ref{fig:maps}, where the projected mass-weighted temperature is shown for the most massive halo at $z\simeq 6$ traced back to $z\simeq 10$ and 14	 when its total mass is $ 6.9 \times 10^{10} $, $ 7.3 \times 10^{9} $, and $ 1.4 \times 10^{9}  \,\rm M_\odot $, and its physical virial radius is 18.7, 5.9, and 2.4~kpc, respectively. For all the simulated objects, we additionally evaluated the masses in the cold, warm, and hot gas phase defined as the ones with gas temperatures of $ T < 10^4\,\rm K$, $ 10^4 \le T / {\rm K} \le 10^7 $, and $ T > 10^7\,\rm K$ (Tab.~\ref{tab:phases}).
\begin{table} %  phases
\centering
\caption{\small Gas phase definitions.}
\label{tab:phases}
\begin{tabular}{c  c  c  c }
	\toprule
	\toprule
	\textcolor{TabHeaderColor}{\bf Gas phase} 	& \textcolor{TabHeaderColor}{\bf cold}		&	\textcolor{TabHeaderColor}{\bf warm}			& \textcolor{TabHeaderColor}{\bf hot}\\ 
	\midrule
	Temperature	& $T<10^4\rm K$ 	& $10^4 \le T/{\rm K} \le 10^7$	& $ T > 10^7\rm K$ \\
	\bottomrule
\end{tabular}
\end{table}
From the abundances derived by our chemistry network, we explicitly computed the HI and H$_2$ masses, too.
Our modelling produces a main sequence characterised by a rather flat specific-SFR (sSFR) trend with $M_{\rm star}$, in agreement with available observations (Appendix~\ref{appendixMStdepl}).
The simulated sequence at $z>5$ can be fitted by
\begin{align}
 { \rm Log \left(sSFR/Gyr^{-1} \right) } & = 1.871 - 0.187 \, {\rm Log} \left(M_{\rm star}/ {\rm M_\odot}\right)
\label{eq:MSfit}
\end{align}
and analogous relations hold at different $z$ (Appendix~\ref{appendixMStdepl}).\\ 
The resulting SFR density (SFRD) is in line with recent observational data, as shown in our previous works \cite[e.g. Fig.~2 in][]{Casavecchia2024}, where resolution impacts for the Ref, LB, and HR runs are discussed, as well.

%*****************************************************************************

\section{Results} \label{sect:results}

%*****************************************************************************

\noindent
In the next sections, we present our main results about the cosmic baryon census and the links among different mass phases.
We first present global trends (Sect.~\ref{sect:global}), then we study the relations between baryon phases and galaxy properties (Sect.~\ref{sect:relations}). Finally, we address the picture for the baryon cycle (Sect.~\ref{sect:cycle}).

\subsection{Global trends} \label{sect:global}

\noindent
Mass density parameters and mass functions are shown below.

% \vspace{-0.25cm}
\subsubsection{Mass density parameters} \label{sect:omegas}
\begin{figure}
\centering
 \includegraphics[width=0.5\textwidth]{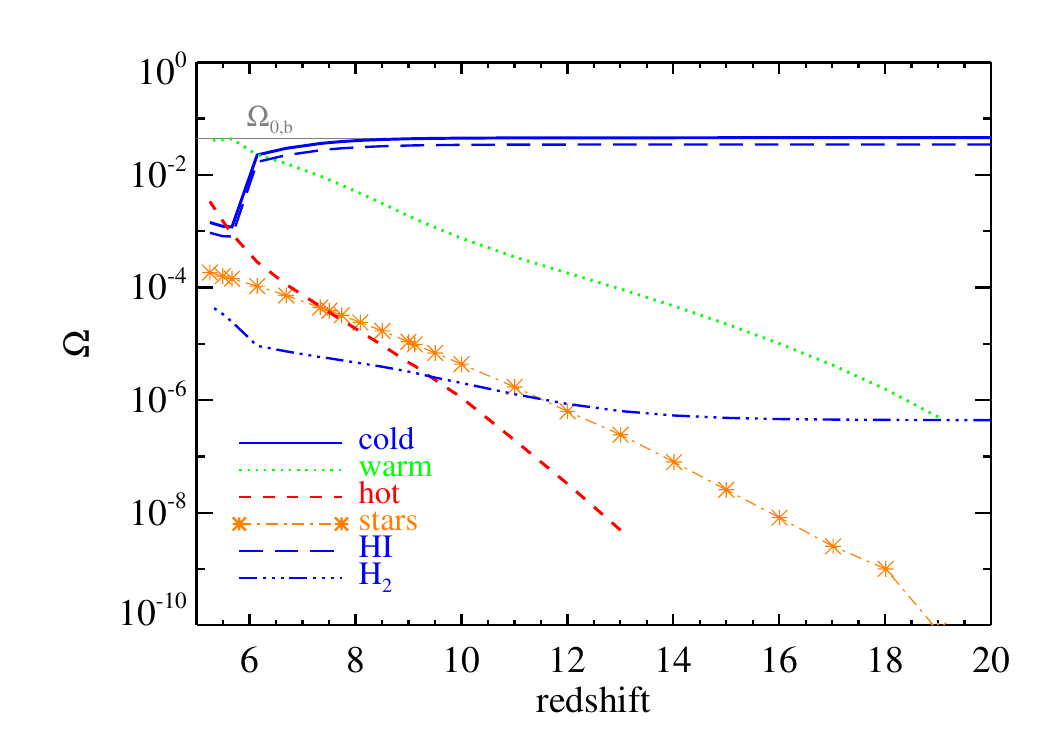}
 \caption{\small Baryon density parameters as function of redshift for cold (solid line), warm (dotted line), hot (short-dashed line), stellar (dot-dashed line with asterisks), HI (long-dashed line), and H$_2$ (dot-dot-dot-dashed line) phases. The horizontal grey line marks the value of $\Omega_{\rm 0,b}$. }
 \label{fig:omegas}
\end{figure}
\noindent
Baryon mass density parameters in the simulated cosmic volume were computed as:
$ \Omega_{\rm phase} = \rho_{\rm phase} / \rho_{\rm 0,crit} $, with $ \rho_{\rm phase} $ comoving mass density of the given baryonic phase considered (cold, warm, hot, stars, HI, H$_2$).
\\
Results for $ \Omega_{\rm cold} $, $ \Omega_{\rm warm} $, $ \Omega_{\rm hot} $, $ \Omega_{\rm stars} $, $ \Omega_{\rm HI} $, and $ \Omega_{\rm H_2} $ are shown in Fig.~\ref{fig:omegas}.
From the trends it is clear that at $z \gtrsim 6$ the cold phase dominates the cosmic mass budget, with a mass density parameter that closely traces $ \Omega_{0,b} $.
At $z \lesssim 6$, reionisation gets completed and $ \Omega_{\rm cold} $ drops by almost two orders of magnitudes, stabilising around $10^{-3}$ \cite[possibly linked to surviving neutral islands;][]{MalloyLidz2015, Becker2015, Bosman2018, Kulkarni2019, Nasir2020, Gnedin2025, Becker2024, Spina2024, Zhu2024, Davies2026}, while the warm component catches up and becomes the most important mass reservoir at later times.
At early epochs, $ \Omega_{\rm warm} $ is driven by star formation feedback and its contribution is as low as $ \sim 10^{-6} $ at $z\sim 20$, about 5~dex smaller than $\Omega_{\rm cold} $.
Ongoing star formation heats up cold material and triggers the conversion of increasingly large amounts of cold gas into warm gas between $z\sim 20$ and 6.
This is reflected by the behaviour of $ \Omega_{\rm stars} $, which accounts for a fraction of a percent of the mass involved in star-forming events (as traced by the warm phase).
\begin{table} %  IMF
\centering
\caption{\small Impact of IMF variation on stellar masses.}
\label{tab:properties}
\begin{tabular}{l l}
\toprule
\toprule
\textcolor{TabHeaderColor}{\bf IMF variation} & \textcolor{TabHeaderColor}{\bf Stellar-mass change (factor)} \\
\midrule
 from Salpeter to Chabrier 		& reduce by 0.243 dex  (1.75) \\
 from Chabrier to Kroupa 		& reduce by 0.057 dex  (1.14) \\
 from Salpeter to Kroupa 		& reduce by 0.3 dex (2.00) \\
 from diet Salpeter to Salpeter 	& increase by 0.15 dex  (1.41) \\
 from diet Salpeter to Chabrier	& reduce by 0.093 dex (1.24) \\
\bottomrule
\vspace{0.5cm}
\end{tabular}
\end{table}
We note that the stellar budget was derived from a Salpeter IMF and that possible IMF variations \cite[][]{Kroupa2001, Chabrier2003, Bell2001, Bell2003, Gallazzi2008, Herrmann2016, Blackwell2026} impact $ \Omega_{\rm stars} $ by up to 0.3~dex (a factor of 2), as summarised in Tab.~\ref{tab:properties}.
\\
Hot gas mass generated during cosmic structure assembly is subdominant, since it traces larger and rarer objects.
The cold phase hosts both HI and H$_2$, two crucial species for the conversion of gas into stars.
While HI typically accounts for 3/4 the cold mass, H$_2$ molecules are almost ubiquitous.
Their production gets enhanced during the early gas collapse at $z\sim 15$--$20$ and increases by roughly two orders of magnitudes between $z\sim 15 $ and 5.

%********************************************************

\subsubsection{Mass functions} \label{sect:phasemassfunctions}

\begin{figure}
 \centering
 \includegraphics[width=0.5\textwidth] {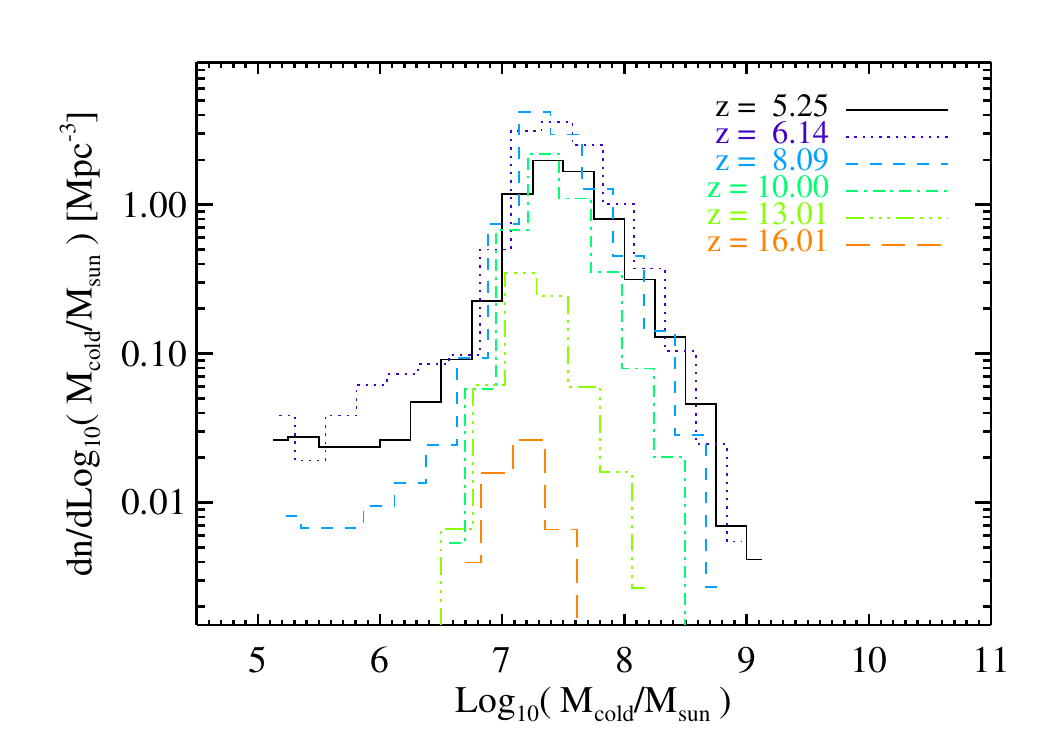}\\
 \vspace{-0.5cm}
 \includegraphics[width=0.5\textwidth] {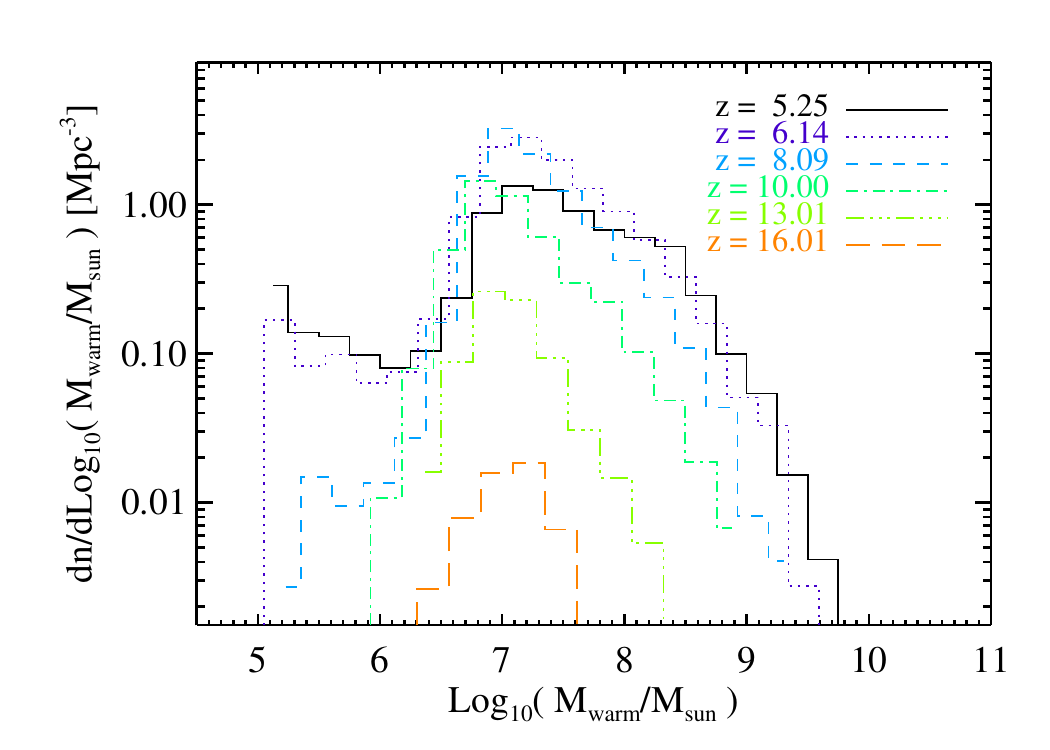}\\
 \vspace{-0.5cm}
 \includegraphics[width=0.5\textwidth] {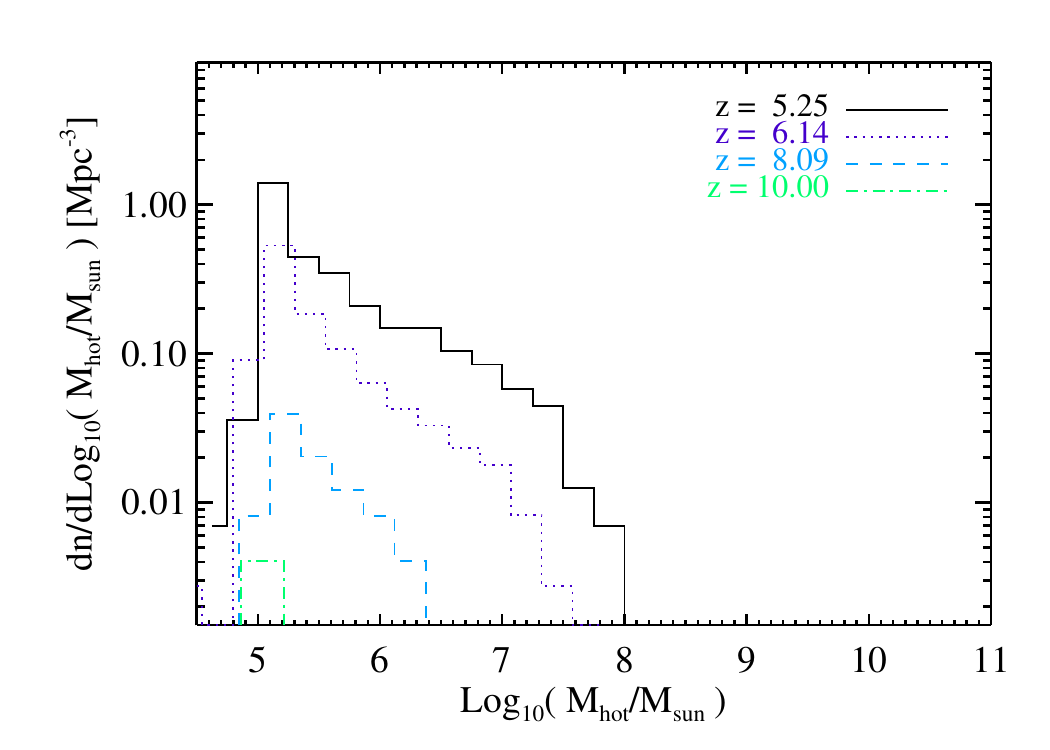}
 \caption{\small Cold (top), warm (centre), and hot (bottom) gas mass functions at $z = $5.25, 6.14, 8.09, 10.00, 13.01, and 16.01. }
 \label{fig:phasemassfunctions}
\end{figure}
\begin{figure}
 \centering
 \includegraphics[width=0.5\textwidth] {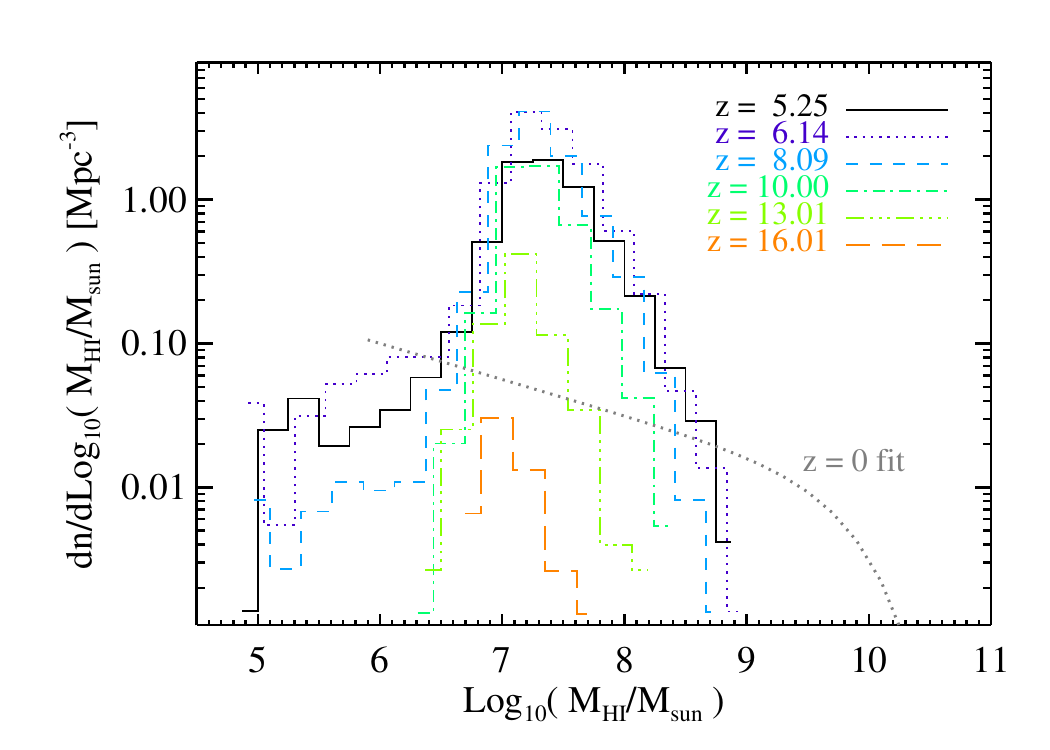}\\
 \vspace{-0.5cm}
 \includegraphics[width=0.5\textwidth] {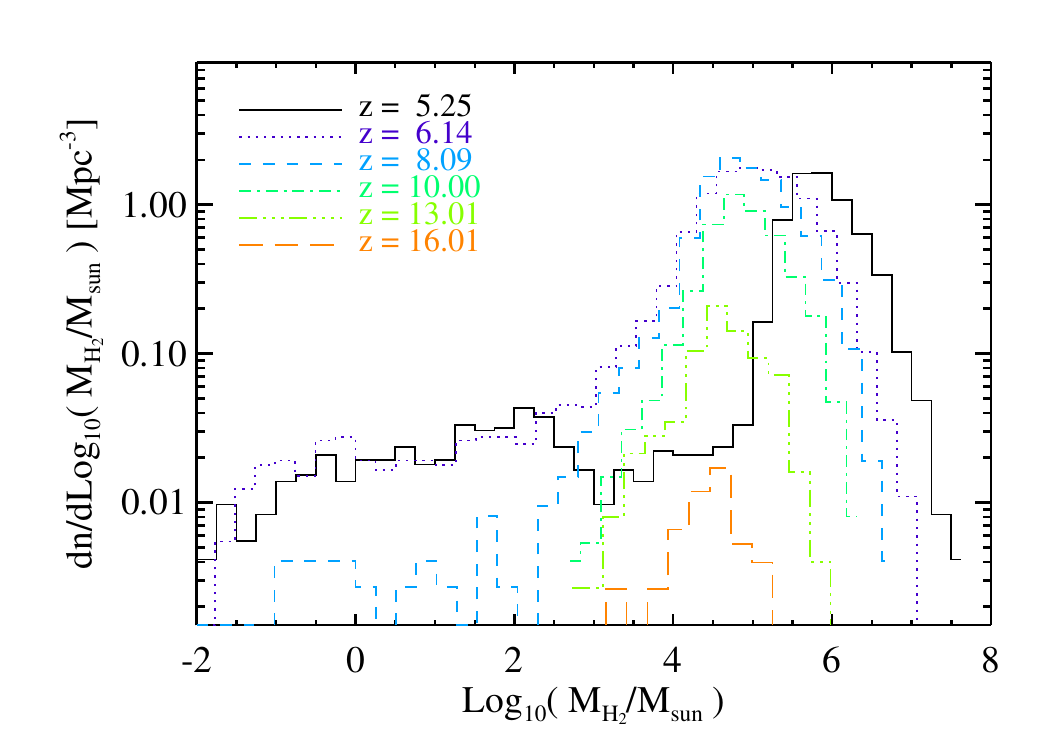}\\
 \vspace{-0.5cm}
 \includegraphics[width=0.5\textwidth] {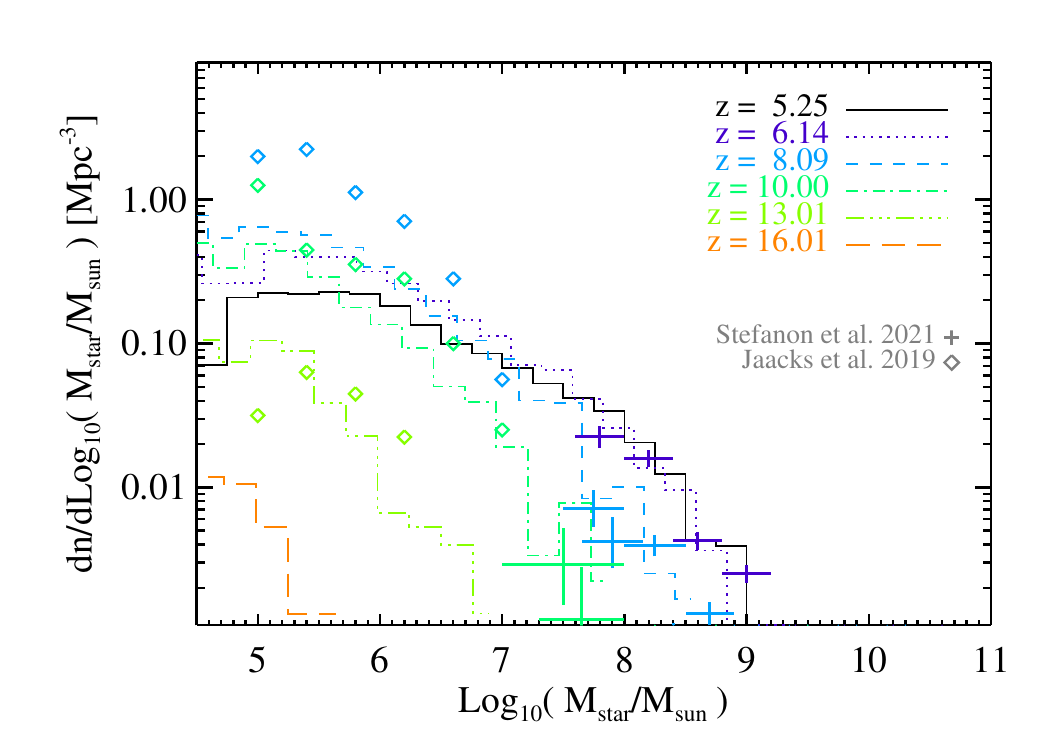}
 \caption{\small HI (top), H$_2$ (centre),and stellar (bottom) mass functions at different $z$ with a fit to $z=0$ ALFALFA HI data by \cite{Jones2018} (top panel, dotted grey line), $z=6$--10 HST/Spitzer masses by \cite{Stefanon2021} (bottom panel, coloured crosses) and expectations by \cite{Jaacks2019} (bottom panel, coloured diamonds). }
 \label{fig:phasemassfunctions2}
\end{figure}

\noindent
Fig.~\ref{fig:phasemassfunctions} shows mass functions for cold, warm, and hot gas at $z \simeq 5$--16.
The mass function is defined as the number of objects with a given (phase) mass per unit volume and per unit logarithmic mass bin.
In all three cases, the mass function evolves from narrow, poorly sampled distributions ($z\simeq16$) to broader and richer ones ($z\simeq 5$) with higher normalisation values towards lower redshifts.
That is a direct consequence of cosmic structure formation, which causes an increase over time of the number density of formed galaxies.
The cold and warm phases peak at $M_{\rm cold} \sim M_{\rm warm} \sim 10^7\,\rm M_\odot $ and the hot phase at $ M_ {\rm hot} \sim 10^5\,\rm M_\odot $ at roughly any $z$.
The high-mass end of the distributions show that the maximum values of cold, warm, and hot masses are very different, with $ M_{\rm cold} $ reaching $\sim 10^9\,\rm M_\odot$, but $ M_{\rm warm} \sim 10^{10}\,\rm M_\odot$ and $ M_{\rm hot} $ only $\sim 10^8\,\rm M_\odot$.
Larger objects are able to retain larger amounts of baryons, while smaller objects have shallower potential wells, and hence are more susceptible to feedback energetics.
A clear example is offered by the cold-phase mass distribution between $z\simeq6$ and 5, when reionisation heats up cold, diffuse gas and leads to a decrease in the peak mass and the low-mass tail.
At  $z\gtrsim 6$ the redshift evolution is quite neat and regular and results from the interplay between local gas cooling and stellar heating.
By comparing the masses of different gas phases at any fixed $z$, we find in general a minor role played by hot gas and a more relevant role played by cold and warm gas at most scales.
\\ 
\noindent 
Looking at the details of cold-mass distributions, in Fig.~\ref{fig:phasemassfunctions2} we make predictions for the HI and H$_2$ mass functions. 
The former closely follows the $M_{\rm cold}$ behaviour, as expected from its typical abundances (see Fig.~\ref{fig:omegas}); hence, the HI mass function is a reliable tracer of the cold-gas mass function.
In the upper panel, we also show a fit derived by ALFALFA data and valid at $z=0$ for $M_{\rm HI} > 10^6 \,\rm M_\odot$ \cite[][]{Jones2018}.
Clearly, the low-redshift trend is not a good description of the high-$z$ HI distribution, when mass regimes and gas thermal states are rather different.
In fact, the lack of early, grown, massive structures biases the resulting high-$z$ HI distributions towards values of $ M_{\rm HI} < 10^9 \,\rm M_\odot $.
Halo mass growth and/or mergers at later times are expected to shift HI masses up to $M_{\rm HI} \simeq 10^{10}$--$10^{11} \,\rm M_\odot$ and reionisation would lower the abundance of smaller (mostly unshielded) objects with $M_{\rm HI}\sim 10^6$--$10^8\,\rm M_\odot$.
The behaviour of the H$_2$ mass function is more convolved, because $M_{\rm H_2}$ can grow in a variety of temperature and density regimes and both stellar feedback and reionisation are efficient at dissociating unshielded H$_2$ mass.
In particular, the high-$M_{\rm H_2}$ end is driven by molecular formation in thick cooling gas, while the low-mass tail is determined by H$_2$ destruction due to stellar feedback ($z>6$) or reionisation ($z\lesssim 6$).
\\
The mass distribution of the young ($ t_{\rm age} < 100\,\rm Myr $) stellar component is plotted in the bottom panel of Fig.~\ref{fig:phasemassfunctions2} and compared with UV-based determinations for $z = 6$--10 HST and Spitzer galaxies \cite[][]{Stefanon2021}.
The resulting trend decreases with $M_{\rm star}$ and follows a Schechter form, while redshift evolution is evident in the normalisation and reflects the level of star formation activity from early to late times.
A plethora of previous works have shown stellar mass functions, such as those in the bottom panel of Fig.~\ref{fig:phasemassfunctions2}.
Nevertheless, the low-mass regime is still debated and more observational data of faint or normal star-forming galaxies at those epochs are needed to draw definitive conclusions.
Broadly speaking, we find a power-law trend down to $M_{\rm star} \sim 10^6\,\rm M_\odot$ and a flattening of the distribution at lower $M_{\rm star}$, while the high-mass regime is in line with the data by \cite{Stefanon2021}.
Independent theoretical results by \cite{Jaacks2019}, despite focusing on a smaller-mass range, find a behaviour that is qualitatively similar to ours (although steeper at $z\sim8$), with a mild flattening around $ M_{\rm star} \sim 10^5$-$10^6 \,\rm M_\odot$.
The existing discrepancies could be ascribed to the different stellar-population assumptions, selection effects, or numerical resolution.
As is noted therein, the flattening values correspond to the transition between (smaller) H$_2$-cooling haloes and (larger) H-cooling haloes, with implications for cooling efficiencies.
Thus, gas cooling is key for the establishment and shape of the early stellar mass function.

%********************************************************
%********************************************************
%********************************************************

\subsection{Baryon budget, galaxies and the IGM} \label{sect:relations}

We explore where different baryon phases are located and their relations with the hosting environment by looking at the mass density parameters in bound structures and in the IGM, as well as their possible correlations with the main galaxy properties.

%********************************************************
%********************************************************

\subsubsection{Mass phases in bound structures and IGM} \label{sect:omegas2}

\begin{figure*}
 \centering
 \includegraphics[width=0.49\textwidth]{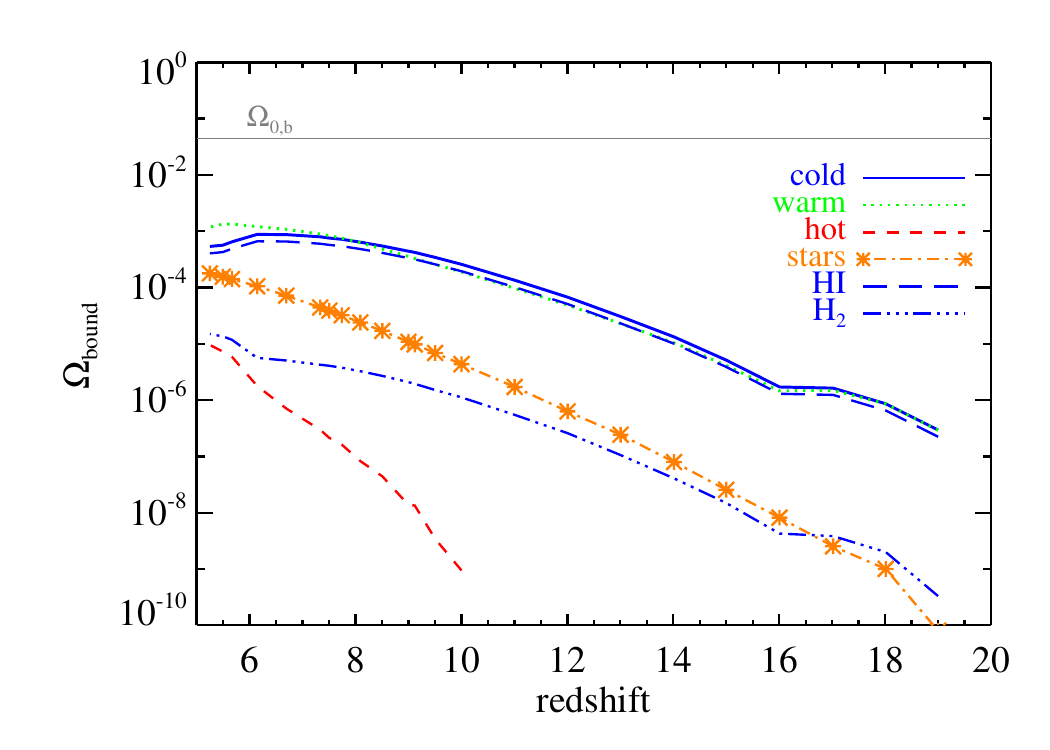}
 \includegraphics[width=0.49\textwidth]{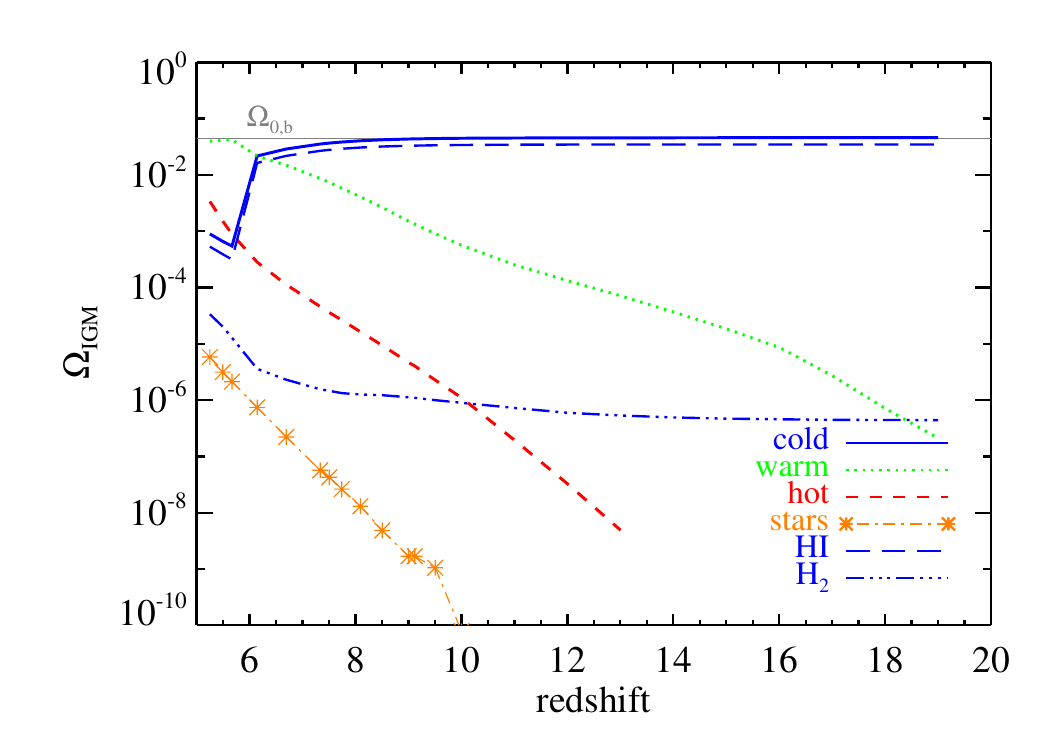}
 \caption{\small Baryon density parameters in bound objects (left) and the IGM (right) as function of redshift for cold (solid lines), warm (dotted lines), hot (short-dashed lines), stellar (dot-dashed lines with asterisks), HI (long-dashed lines), and H$_2$ (dot-dot-dot-dashed lines) phases. The horizontal grey line marks the value of $\Omega_{\rm 0,b}$. Stars form out of cold gas that dominates the baryon budget at $z \gtrsim 6$ and is overtaken by warm gas at later times. }
 \label{fig:omegas2}
\end{figure*}
Mass density parameters in bound structures and the IGM were estimated by selecting the total mass within a distance of $R < R_{\rm vir}$ from halo centres ($\Omega_{\rm bound}$) and the mass lying beyond any $ R_{\rm vir}$ ($\Omega_{\rm IGM}$) for each phase.
The selected bound mass therefore included both galaxies and their circumgalactic medium (CGM).
The share of cold, warm, hot, stellar, HI, and H$_2$ mass phases in bound structures ($ \Omega_{\rm bound, cold} $, $ \Omega_{\rm bound, warm} $, $ \Omega_{\rm bound, hot} $, $ \Omega_{\rm bound, stars} $, $ \Omega_{\rm bound, HI} $, $ \Omega_{\rm bound, H_2} $) and diffuse gas ($ \Omega_{\rm IGM, cold} $, $ \Omega_{\rm IGM, warm} $, $ \Omega_{\rm IGM, hot} $, $ \Omega_{\rm IGM, stars} $, $ \Omega_{\rm IGM, HI} $, $ \Omega_{\rm IGM, H_2} $) are presented in Fig.~\ref{fig:omegas2}, while corresponding bound-to-cosmic and IGM-to-cosmic mass density ratios are in Appendix~\ref{appendixGasFractions}.
\\
The trends for bound objects follow the halo growth during cosmic evolution with little amounts of mass in bound structures at early times and increasingly large amounts of bound mass at later epochs.
This is particularly evident for $ \Omega_{\rm bound, warm} $, $ \Omega_{\rm bound, hot} $, and $ \Omega_{\rm bound, stars} $ that increase monotonically with time and signal the build-up of gaseous and stellar content in galaxies.
The trend of $ \Omega_{\rm bound, cold} $ is affected by reionisation at $z\simeq 6$, when the UV background radiation ionises HI ($ \Omega_{\rm bound, HI} $ drops) and slightly boosts H$_2$ in the thin regions of halo peripheries ($\Omega_{\rm bound, H_2}$ gets enhanced).
This causes $ \Omega_{\rm bound, warm} $ to dominate the baryons in collapsed structures at $z < 6$.
When compared to the whole cosmic mass, the resulting $\Omega_{\rm bound}$ values are different from those in Fig.~\ref{fig:omegas} and discrepancies reach a few dex (consistently with the phase average evolution shown in Appendix~\ref{appendixGasFractions}, Fig.~\ref{fig:phaseevolution}).
\\
For the diffuse gas, the values of $ \Omega_{\rm IGM, cold} $, $ \Omega_{\rm IGM, warm} $, and $ \Omega_{\rm IGM, hot} $ tightly follow the corresponding $ \Omega_{\rm cold} $, $ \Omega_{\rm warm} $, and $ \Omega_{\rm hot} $ cosmic behaviour and the picture arising from Fig.~\ref{fig:omegas2} is similar to the one of Fig.~\ref{fig:omegas}.
In the IGM case, the stellar phase is negligible, as primordial stars are typically formed in bound structures (early galaxies).
While HI traces IGM cold gas well, H$_2$ features a resulting $ \Omega_{\rm IGM, H_2} \sim 10^{-6}$ at $z\simeq 18$ (in line with primordial abundances) and $ \Omega_{\rm IGM, H_2} \gtrsim 10^{-5}$ at $z<6$.

%********************************************************

\subsubsection{Phase relations} \label{sect:phaserelations}

\begin{figure*}
 \centering
 \includegraphics[width=0.35\textwidth] {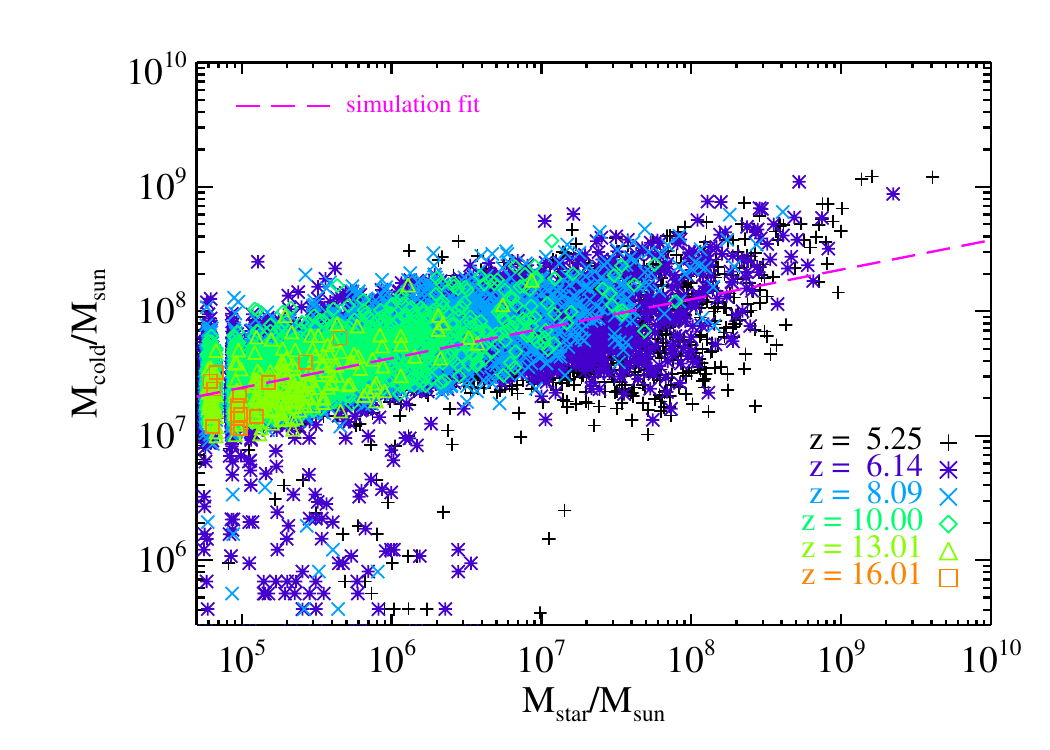}
 \hspace{-0.75cm}
 \includegraphics[width=0.35\textwidth] {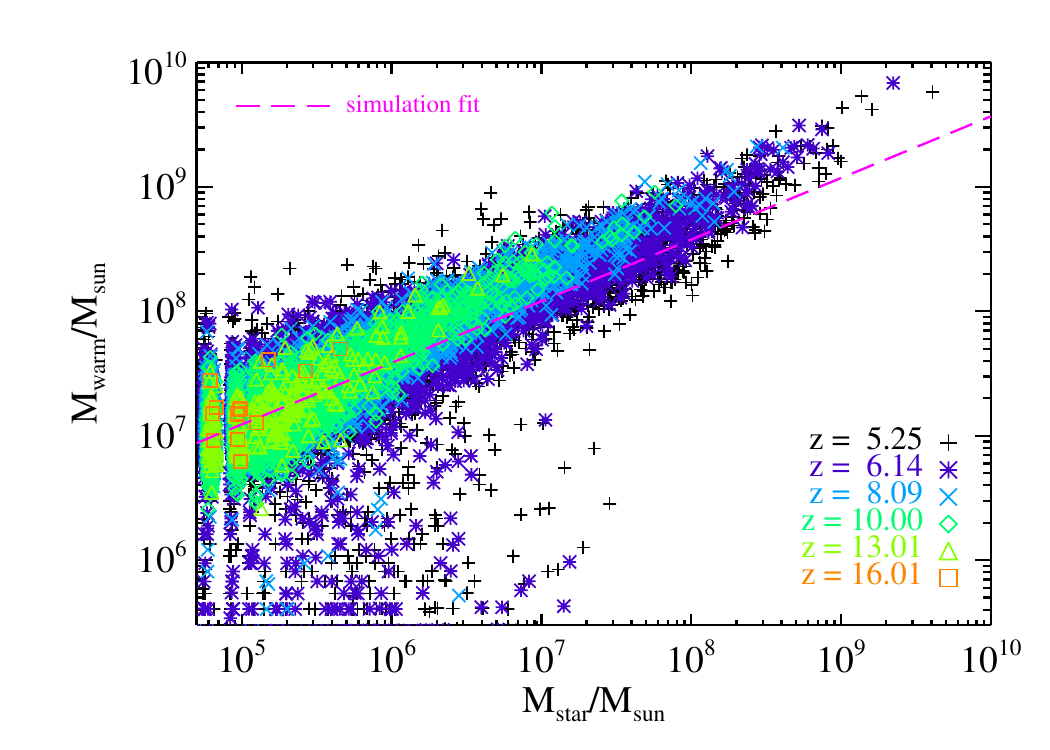}
 \hspace{-0.75cm}
 \includegraphics[width=0.35\textwidth] {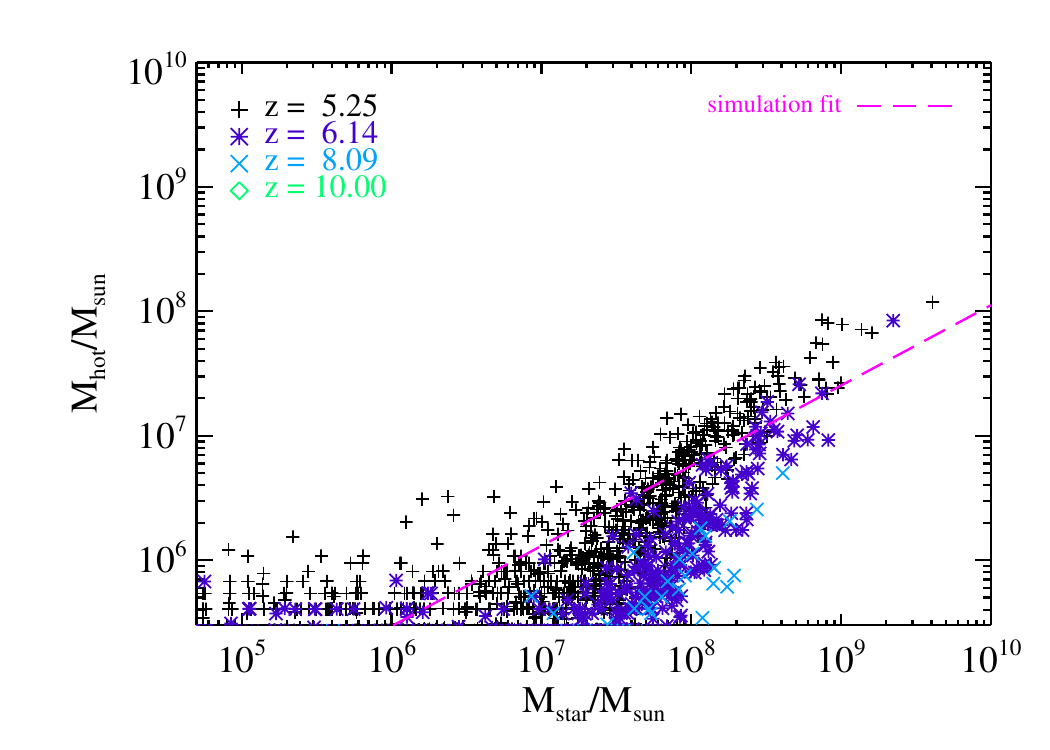}
 \\
 \vspace{-0.5cm}
 \includegraphics[width=0.35\textwidth] {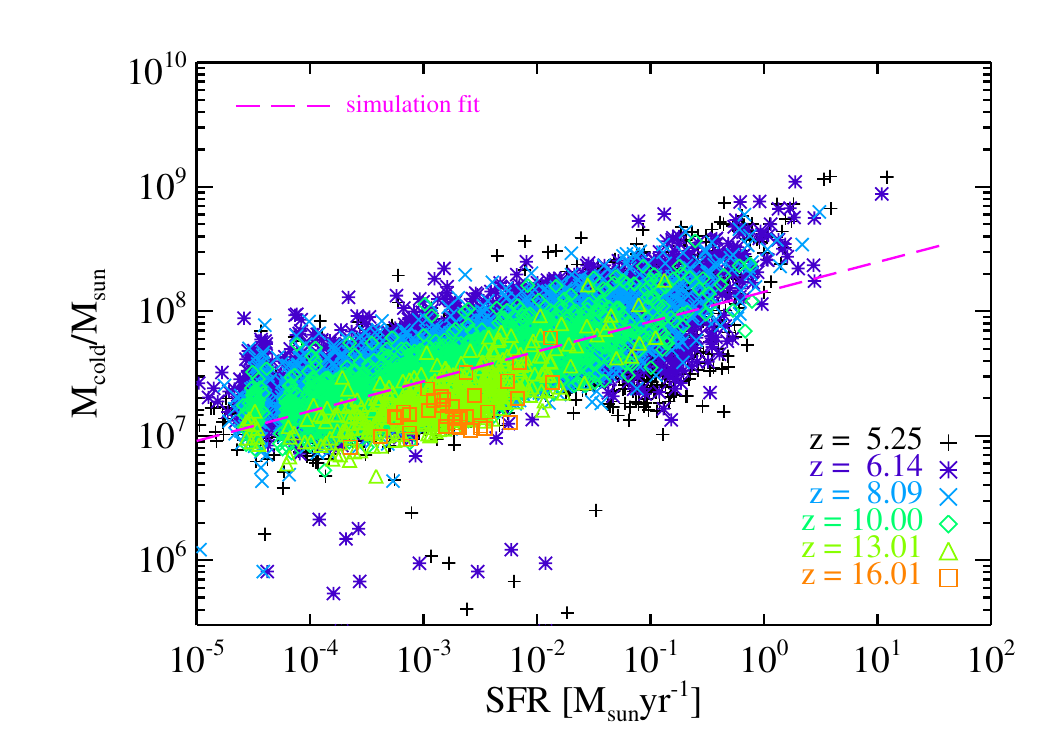}
 \hspace{-0.75cm}
 \includegraphics[width=0.35\textwidth] {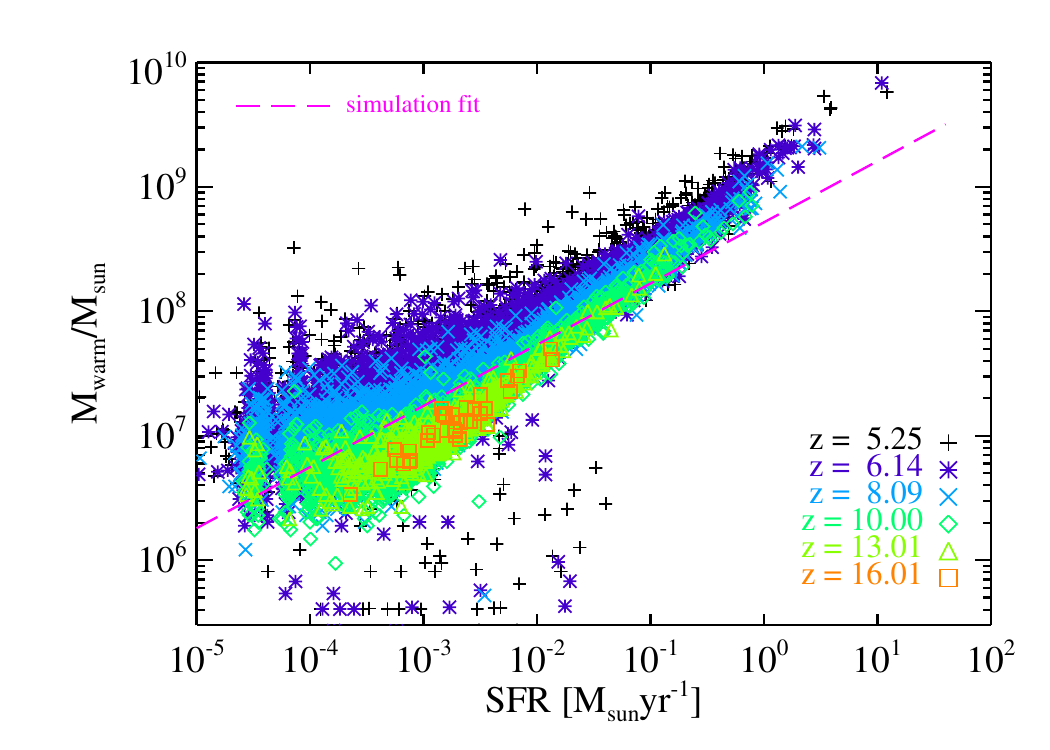}
 \hspace{-0.75cm}
 \includegraphics[width=0.35\textwidth] {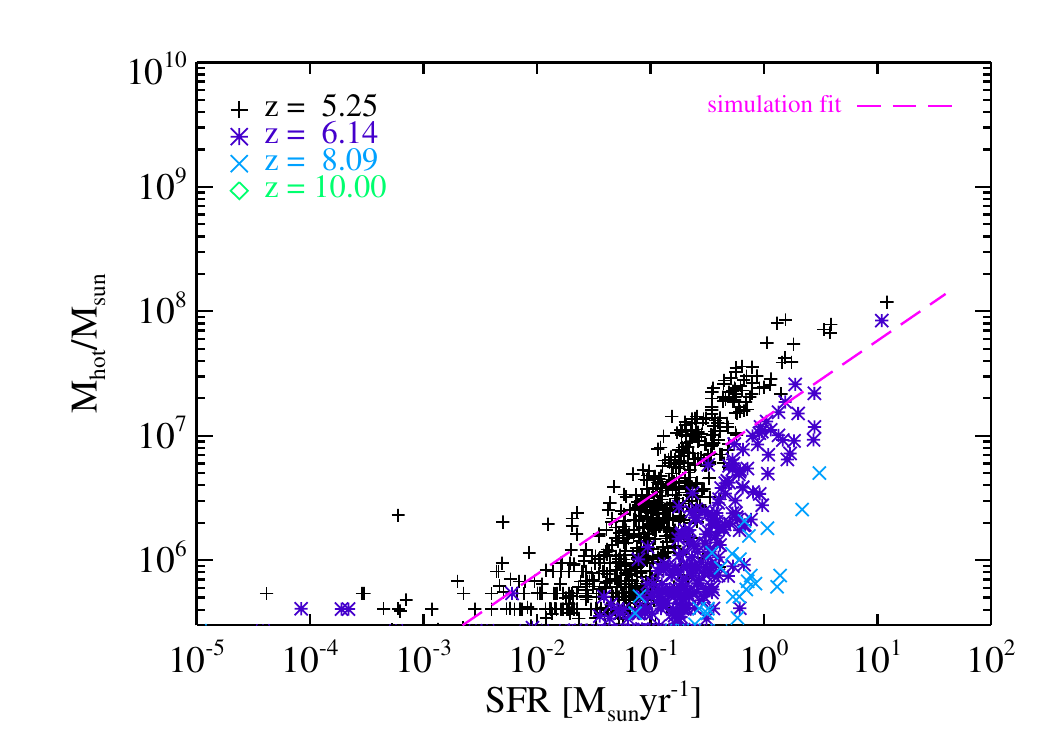}
 \caption{\small Cold (left column), warm (central column), and hot (right column) gas mass versus $M_{\rm star}$ (top row) and SFR (bottom row) for galaxies at redshift $z = $5.25, 6.14, 8.09, 10.00, 13.01, and 16.01, with simulation fits (dashed magenta lines) reported in Appendix~\ref{appendixFits}.
Different gas phases probe different $M_{\rm star}$ and SFR regimes with little redshift evolution, but a large environment and feedback-driven scatter.}
 \label{fig:phasemassrelations}
\end{figure*}
It is debated whether gas phases correlate with the main galaxy properties and if they could be used to infer unknown quantities.
\\
To address these questions, we check gas scaling relations in Fig.~\ref{fig:phasemassrelations}, where cold, warm and hot gas masses in bound structures are plotted against the hosting $M_{\rm star} $ (upper row) and SFR (lower row).
Corresponding mass fractions are shown in Appendix~\ref{appendixGasFractions}.
To better guide the eye, we overplot fits to simulation results, as reported in Appendix~\ref{appendixFits}.
Gas phase masses generally increase with $M_{\rm star}$ and SFR and the results in the figure show that the hot component is subdominant by a factor of $\sim 10^2$--$10^3$ at all $z$.
Warm and cold phases share comparable amounts of mass with slightly more cold gas in smaller systems  ($ M_{\rm star} \lesssim 10^7 \rm M_\odot$).
The trend is shallower for the cold material and steeper for the warm one (which dominates at $\rm SFR \gtrsim 0.01$--$0.1 \,\rm M_\odot$).
Simulation results suggest that cold, warm and hot gas masses increase with both $M_{\rm star}$ and SFR with slopes of  $\sim 0.2$, 0.5 and 0.6 (Tab.~\ref{tab:fits}).
\\
\begin{figure*}
 \centering
 \includegraphics[width=0.35\textwidth] {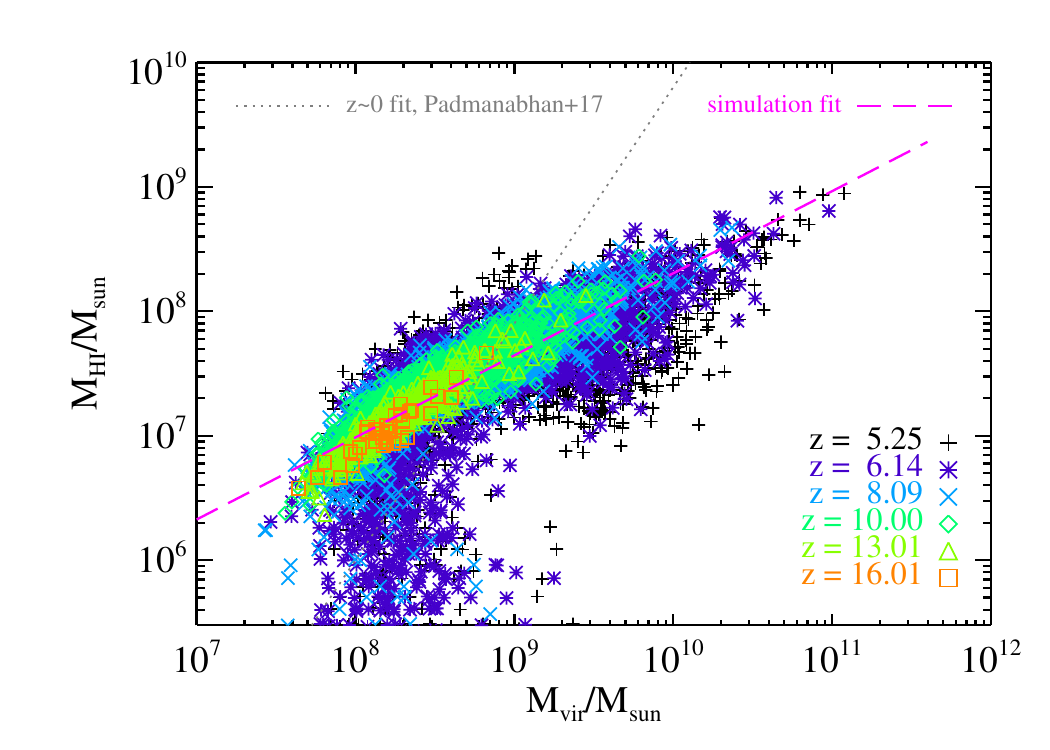}
 \hspace{-0.75cm}
 \includegraphics[width=0.35\textwidth] {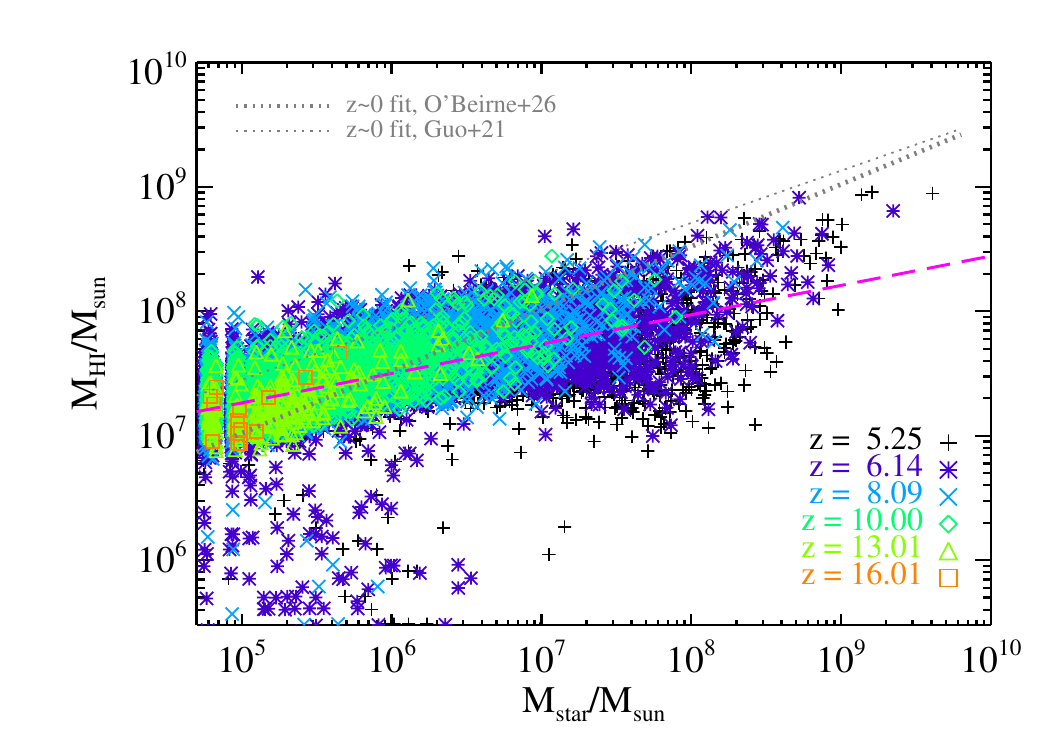}
 \hspace{-0.75cm}
 \includegraphics[width=0.35\textwidth] {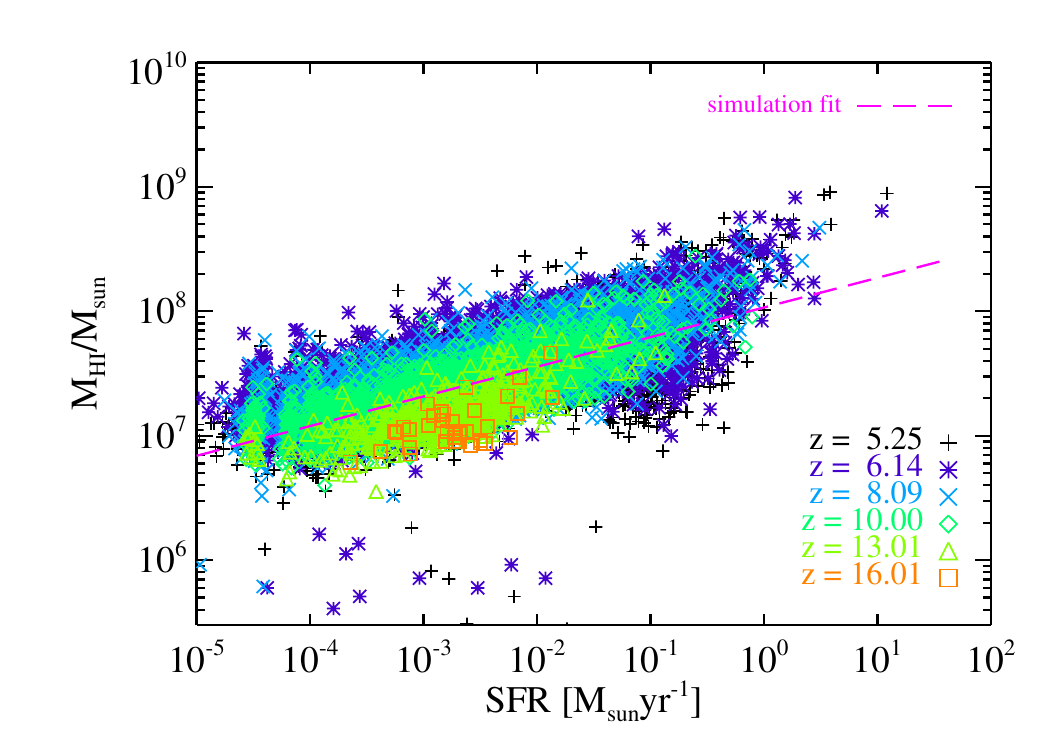}
\\
 \vspace{-0.5cm}
 \includegraphics[width=0.35\textwidth] {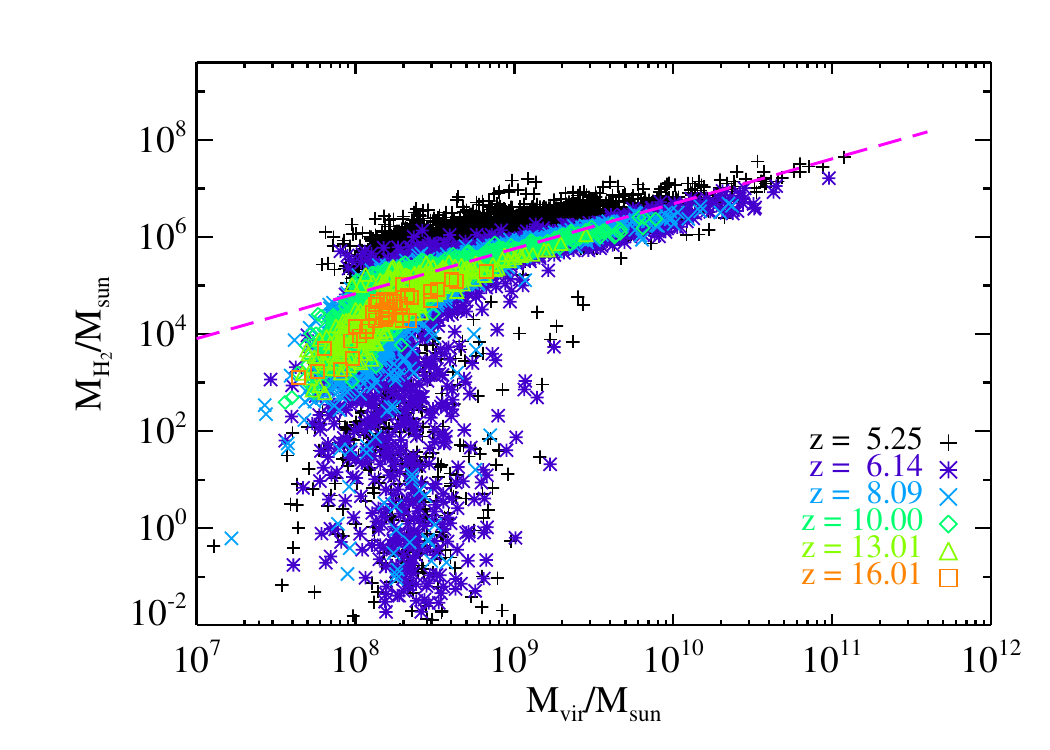}
 \hspace{-0.75cm}
 \includegraphics[width=0.35\textwidth] {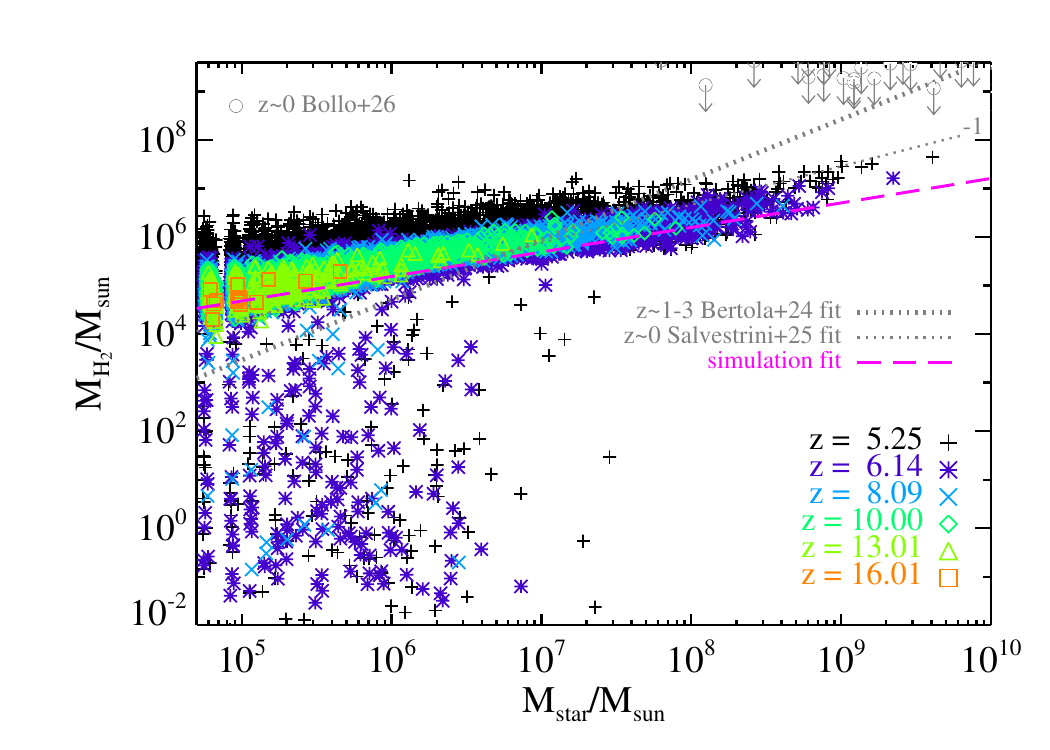}
 \hspace{-0.75cm}
 \includegraphics[width=0.35\textwidth] {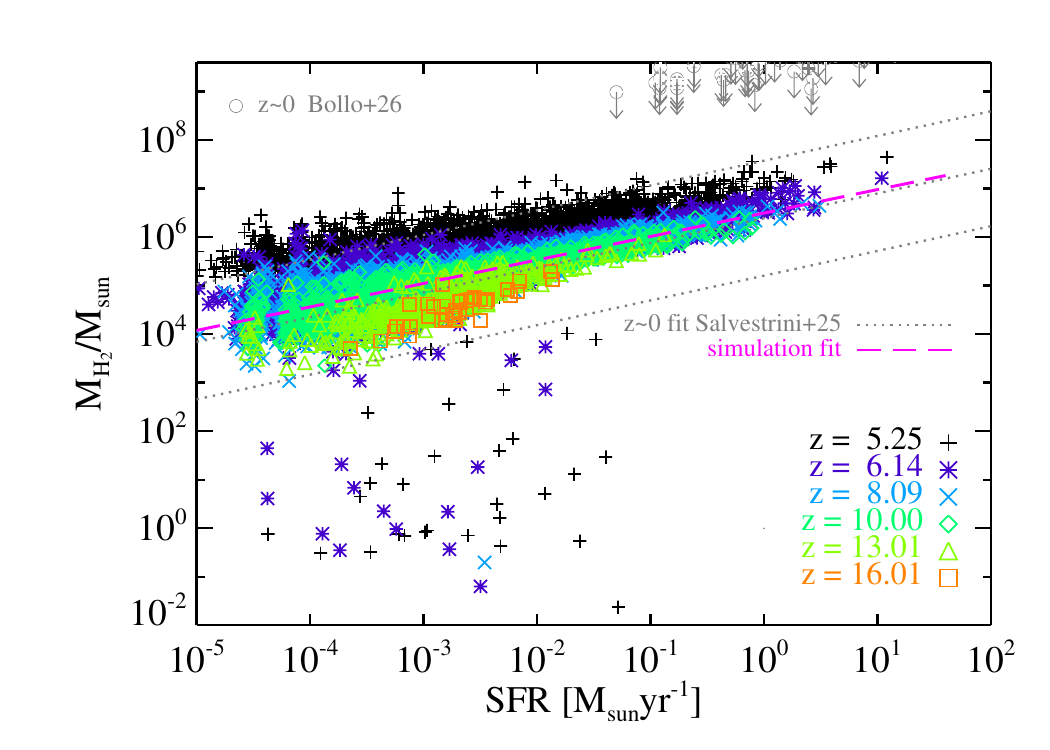}
 \caption{\small HI (top) and H$_2$ (bottom) masses versus $M_{\rm vir}$ (left), $M_{\rm star}$ (centre), and SFR (right). Observational fits for HI and H$_2$ data are shown by grey dotted lines and MUSE-ALMA Haloes H$_2$ upper limits \cite[][]{Bollo2026} by grey circles and arrows. The fit to the MUSE-ALMA Haloes $M_{\rm HI} $ vs. $M_{\rm star}$ is $ \rm Log(M_{\rm HI}/M_\odot ) = 0.50 \, Log(M_{\rm star}/M_\odot) + 4.52 $ (O'Beirne et al. 2026, priv. comm.) and is extrapolated from galaxy data at $M_{\rm star} \sim 10^8$-$10^{12} \, \rm M_\odot$. Multi-parametric fits by \cite{Salvestrini2025} are shown for $\rm SFR = 10^{-1} M_\odot\,yr^{-1}$ (lower central panel) and  $M_{\rm star} = 10^{8.5}, 10^{7} $, and $ 10^{5.5} \rm M_\odot$ (dotted lines from top to bottom in the lower right panel). Simulation fits (dashed magenta lines) are in Appendix~\ref{appendixFits}. Results are little dependent on redshift and feature a large environment-driven scatter.
 }
 \label{fig:phasemassrelations2}
\end{figure*}
\noindent
We further explore the trends of HI and H$_2$ masses as functions of $M_{\rm vir}$, $M_{\rm star}$, and SFR in Fig.~\ref{fig:phasemassrelations2}.
We compare our predictions with $z\sim 0$ HI data from HIPASS and ALFALFA surveys \cite[][]{PK2017, Guo2021} and with H$_2$ data from ALMA/NOEMA surveys of star-forming galaxies at $z\sim 1$--3 \cite[][]{Bertola2024}, nearby DustPedia galaxies \cite[][]{Salvestrini2025}, and MUSE-ALMA galaxies \cite[][]{Bollo2026}.
HI and H$_2$ mass relations feature little evolution and the scatter broadens due to both larger sample statistics and increasing feedback impacts at lower $z$.
Their scalings with $M_{\rm star}$ and SFR are similar, consistently with the main-sequence relation at this epoch, and are characterised by slopes of roughly 0.2 and 0.5 (Tab.~\ref{tab:mHIH2MvirMstarSFR}).
The H$_2$ behaviour is in agreement with recent MUSE-ALMA upper limits at $z\simeq 0.5$ by \cite{Bollo2026} and the fundamental plane by \cite{Salvestrini2025}, whose multi-parametric fit suggests an $M_{\rm H_2}$-SFR slope of 0.51 at $z\simeq 0$ \cite[see also][]{Sargent2014}.
Low-$z$ observations are qualitatively in line with our predictions, although they bear remarkable uncertainties \cite[][]{Chowdhury2022, Bianchetti2025}.
\\
Supplementary material about HI and H$_2$ total gas fractions $f_{\rm HI} $ and $f_{\rm H_2}$ versus $M_{\rm vir}$, $M_{\rm star}$, SFR, and $Z$ is given in Appendix~\ref{appendixGasFractions}.
Here, we only stress that, due to feedback effects, $f_{\rm HI} $ may reach values $ \lesssim 0.1$ of the whole gas mass (in objects dominated by warm or hot gas), while $f_{\rm H_2}$ is typically around a few percent, but reaches values larger than $10\%$ in cold, metal-enriched, star-forming sites.
Finally, we checked the relations between gas phases and metallicities and found that they are similar to the ones in Figs.~\ref{fig:phasemassrelations} and \ref{fig:phasemassrelations2} because of the mass-metallicity relation.

% *********************************************************************************
% *********************************************************************************

\vspace{-0.25cm}
\subsection{Baryon cycling: Connecting gas and stars} \label{sect:cycle}

Baryon phases are influenced by both the stellar activity -- that returns a fraction of stellar mass to the gaseous environment during stellar evolutionary stages (stellar return fraction; $R$) -- and the efficiency with which gas is converted into stars during a typical depletion time (star formation efficiency; SFE).
These two phenomena are signposts of the interaction between gas and stars and we quantify them in the next section.

\subsubsection{Stellar return fraction} \label{sect:R}

\begin{figure}[h]
 \centering
 \includegraphics[width=0.5\textwidth] {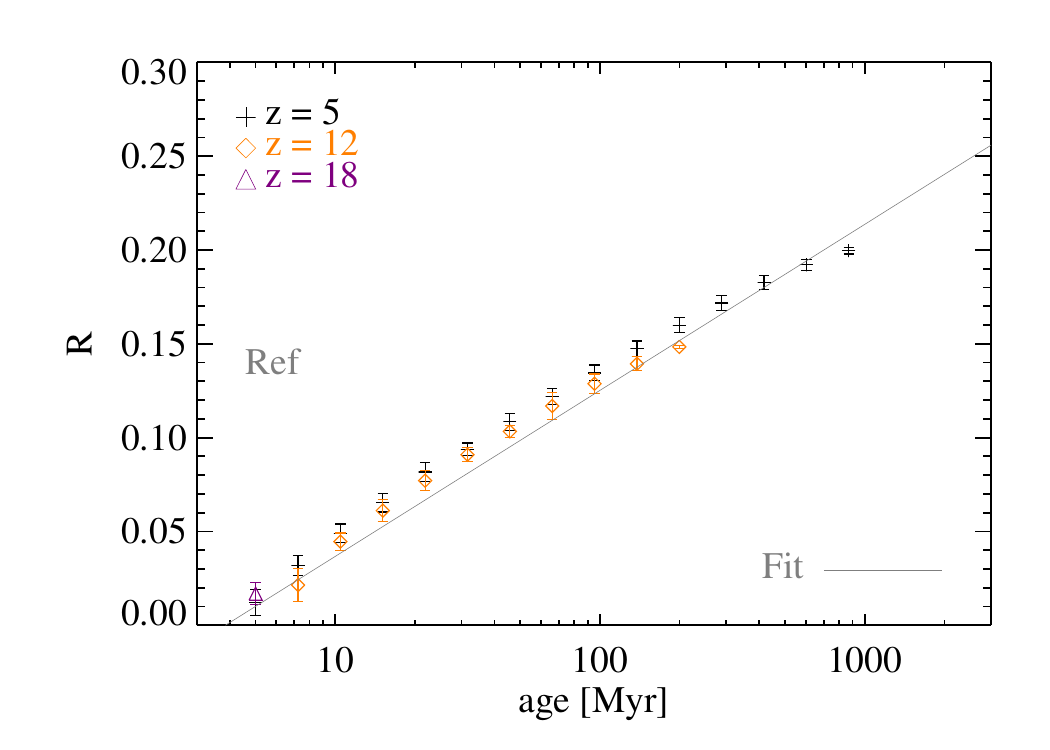}
 \caption{\small Stellar return fraction $R$ as a function of stellar age for simulated data at redshifts $z=5$ (crosses) , 12 (diamonds), and 18 (triangles) in the Ref run. Data points refer to median $R$ values with median absolute deviation, while the solid grey line is the corresponding fit (eq~\ref{eq:fit}).}
 \label{fig:Rage}
\end{figure}

\noindent
The stellar return fraction, $R$, depends on the assumed stellar IMF, metal yields, mass loss during SN explosions and AGB winds, stellar remnants, and stellar lifetimes.
Most authors rely on IMF-integrated averages of $R$, although this can vary for different populations and metallicities.
Since in our simulations we track the stellar age ($t_{\rm age}$) of each population and the respective chemical enrichment, here we improve on previous estimates by computing $R$ self-consistently at different times.
For each output, the average return fraction is $ R  = 1 -  \Omega_{\rm stars} / \Omega_{0,{\rm stars}} $, with $ \Omega_{\rm stars} $ and $ \Omega_{0,\rm stars} $ current and initial stellar mass density parameters.
\\
Fig.~\ref{fig:Rage} shows the resulting average trend for $R(t_{\rm age})$ at $z=5$, 12, and 18, fitted as a function of $t_{\rm age}$ as follows:
\begin{align}
\label{eq:fit}
R (t_{\rm age}) &= 0.0885 \,{\rm Log} \left( \frac{ t_{\rm age}}{ 3.853 \,{\rm Myr} }	\right),
\end{align}
if $ t_{\rm age} > 3.853\,\rm Myr$ and $R = 0$ otherwise.
\\
We note that, at $ t_{\rm age} \simeq 10\,\rm Gyr$, $R \simeq 0.3$, consistently with the expectations for a Salpeter IMF.
\\
\noindent
Our results imply that in high-$z$ galaxies the amount of stellar mass that is released into the surrounding gas is smaller than the amount inferred by common IMF-integrated estimates.
Our $z=5$ figures are around 0.15--0.20, i.e. half of the typically adopted values.
This is a consequence of the younger ages in high-$z$ populations that have had not enough time to evolve through all the stellar stages and eject large amount of mass back to the surrounding environment.
These considerations suggest that the recycled material affects the baryon budget by a fraction of $\lesssim 20$\% at $z>5$ and, since gas phases influenced by star formation feedback are predominantly warm, the corresponding $\Omega_{\rm warm}$ values (in e.g. Fig.~\ref{fig:omegas}) would share such recycled baryons at the time of their ejection.
On the other hand, early SN and AGB events (that take place on timescales of up to a few hundreds million years) seem efficient at boosting $R$ to levels that are only a factor of two away from late-time expectations.
These findings are not severely affected by metallicity effects (Appendix~\ref{appendixA}) and impact available high-$z$ SFRDs, which appear to suffer from $R$ mis-estimates on top of IMF uncertainties (in Tab.~\ref{tab:properties}).

\subsubsection{Gas depletion and star formation} \label{sect:tdepl}

\begin{figure}
 \centering
  \includegraphics[width=0.5\textwidth] {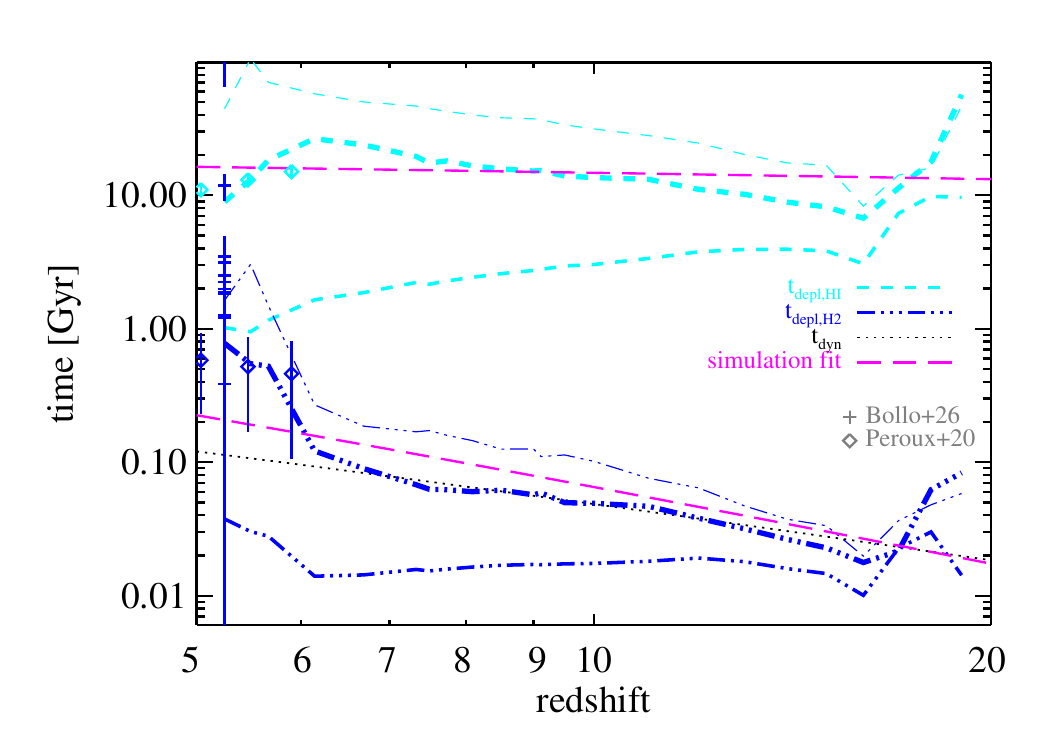}
 \caption{\small Redshift evolution of average (regular lines), mean (thin lines), and median (thick lines) HI and H$_2$ depletion times compared to dynamical time, $ t_{\rm dyn} $ (dotted black line), and observational expectations by \cite{Bollo2026} at low $z$ (crosses with 1$\sigma$ dispersion) and \cite{PH2020} at $z\simeq$5--6 (diamonds with 1$\sigma$ dispersion). Simulation fits (Appendix~\ref{appendixFits}) are  overplotted with long-dashed magenta lines.}
 \label{fig:tdeplredshift}
\end{figure}
\begin{figure}
 \centering
 \includegraphics[width=0.5\textwidth] {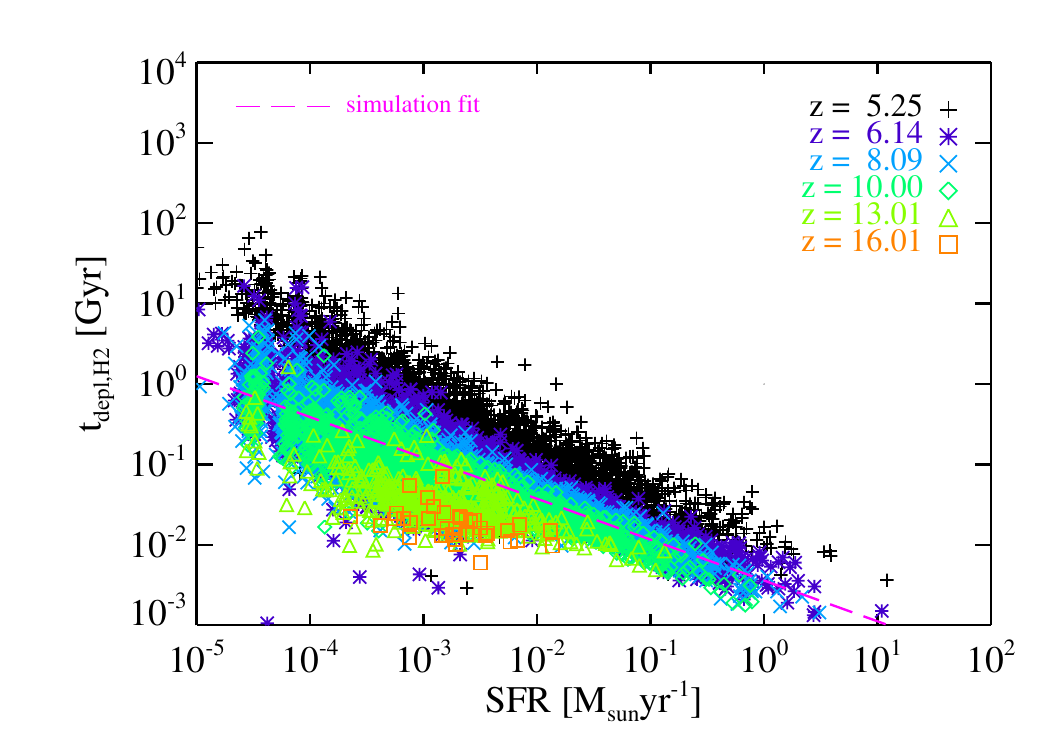}\\
 \vspace{-0.5cm}
 \includegraphics[width=0.5\textwidth] {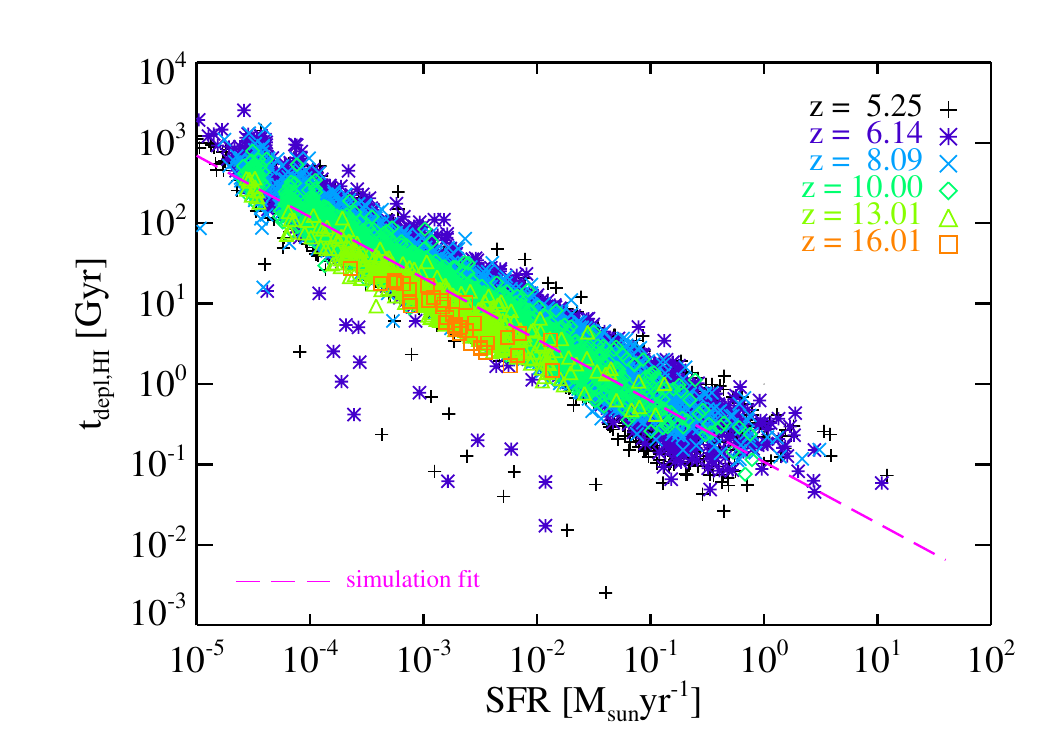}
 \caption{\small H$_2$ (top) and HI (bottom) depletion times versus SFR at $z = $5.25, 6.14, 8.09, 10.00, 13.01, and 16.01 with simulation fits (Appendix~\ref{appendixFits}). Further trends with galaxy properties are shown in Appendix~\ref{appendixDepletionTimes}.}
 \label{fig:tdepl}
\end{figure}
Depletion times effectively quantify the timescales needed to transfer mass from the cold HI and H$_2$ phases to the stellar phase and are defined as:
$ t_{\rm depl, HI} =  M_{\rm HI}/ \rm SFR$
and
$ t_{\rm depl, H_2} =  M_{\rm H_2}/ \rm SFR$.
\\
In Fig.~\ref{fig:tdeplredshift}, the average HI depletion time presents an evolution from $ t_{\rm depl, HI} \sim 10\,\rm Gyr $ at $z\sim 20$ to about 1~Gyr at $z\sim 5$. 
Mean and median values are larger, at variance with the typical Hubble times at those epochs.
The values for $ t_{\rm depl, H_2} $ lie around $\sim 0.01$--$1\,\rm Gyr$ at $z\sim 5$--20.
They are shorter than the Hubble time at the same epochs and support the formation of bursty objects at $z\sim 15$ and beyond.
Mean and median values may differ from simple averages by a factor of a few up to one dex, as is shown in the figure.
As a reference, the fitting expressions for the median HI and H$_2$ trends at $z\simeq $5--20 are
\begin{eqnarray}
\label{fit:tdeplHI}
{\rm Log} (t_{\rm depl,HI} / \rm Gyr) 	= 1.345 - 0.170\, {\rm Log} (1+z), \\
\label{fit:tdeplH2}
{\rm Log} (t_{\rm depl,H_2} / \rm Gyr) 	= 0.943 - 2.045\, {\rm Log} (1+z).
\end{eqnarray}
The latter roughly follows the dynamical time of density contrasts $ \Delta \simeq 500$, $t_{\rm dyn} = \pi t_{\rm H} / \sqrt{ 2 \Delta} \simeq  0.1 \, t_{\rm H}$, with $ t_{\rm H}$ Hubble time
(Fig.~\ref{fig:tdeplredshift}). Fits for average and mean evolution are in Appendix~\ref{appendixFits}.
\\
In Fig.~\ref{fig:tdepl}, $ t_{\rm depl, HI} $ and $ t_{\rm depl, H_2} $ are plotted as functions of SFR (further relations with $M_{\rm star}$, gas $Z$, and sSFR can be found in Appendix~\ref{appendixDepletionTimes}).
Their trend decreases as a result of ongoing star formation activity.
In particular, the bulk of $t_{\rm depl, H_2} $ values are about 0.01-0.05~Gyr at $z\simeq 16$ and about 0.1--1~Gyr at $z\simeq 6$.
Minimum values are reached in larger ($M_{\rm star} \gtrsim 10^8\,\rm M_\odot$), more star-forming ($ \rm SFR > 0.1\,\rm M_\odot yr^{-1} $) structures where typical gas metallicities are $Z > 0.01\,Z_\odot$.
Because of the H$_2$ sensitivity to star-forming processes, the $t_{\rm depl, H_2} $ scatter is broader than $t_{\rm depl, HI} $ and is due to environment and feedback episodes that alter gas molecular content and cooling capabilities in a non-trivial way.
HI trends at different redshifts feature resulting normalisations that are higher by one or two dex.
Fits to simulation results demonstrate that HI and H$_2$ depletion times decrease with SFR (stellar mass)
with slopes of $-0.7$ and $-0.5$ ($-0.6$ and $-0.3$), respectively (Tab.~\ref{tab:depletion}).
\\
\begin{figure}
 \flushleft
 \includegraphics[width=0.5\textwidth] {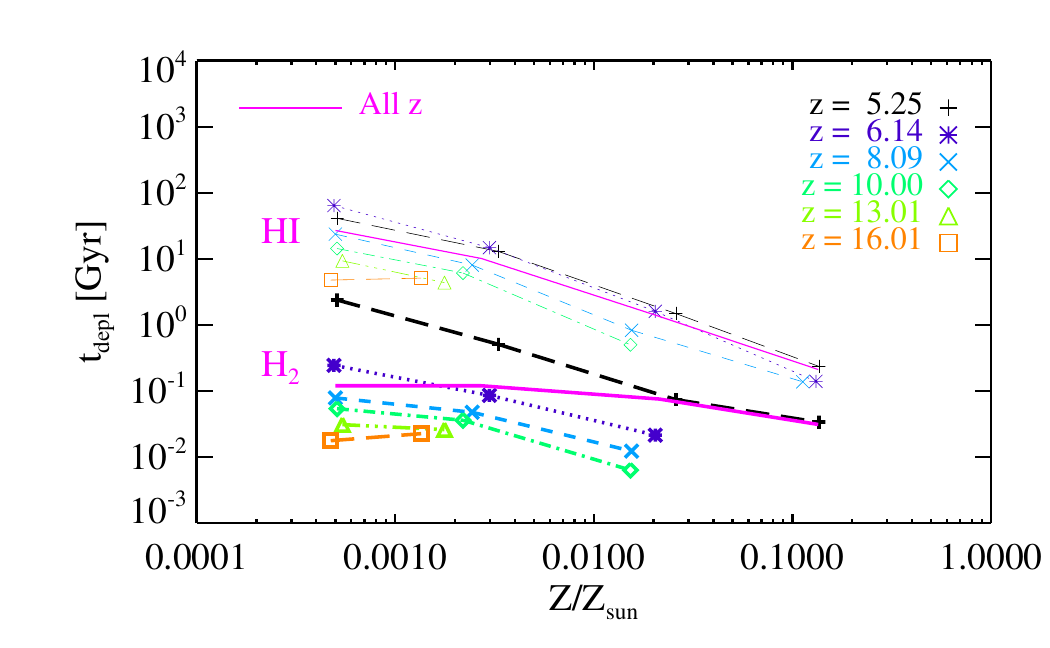}
 \caption{\small Median HI (thin symbols) and H$_2$ (thick symbols) depletion times vs $Z$ at different $z$ (coloured symbols) and for the whole $z$ range (solid magenta lines). Gas metallicity affects depletion times at all $z$.}
 \label{fig:tdeplmetallicity}
\end{figure}
In Fig.~\ref{fig:tdeplmetallicity} the median HI and H$_2$ depletion times versus gas metallicity are displayed for the whole redshift range as well as for the different $z$ considered in one-dex metallicity bins.
The figures show neatly that average trends are decreasing with gas metallicity as a consequence of more efficient star formation in metal-rich environments.
The difference between the average $t_{\rm depl, HI} $ and $t_{\rm depl, H_2} $ is almost three dex at $Z\lesssim 0.001 \,\rm Z_\odot$ and one dex at $Z\gtrsim  0.1 \,\rm Z_\odot$.
This suggests that metals are efficient in lowering $t_{\rm depl, HI} $, although they are less crucial for $t_{\rm depl, H_2} $.
\\
\begin{figure}[t]
\centering
\includegraphics[width=0.5\textwidth]{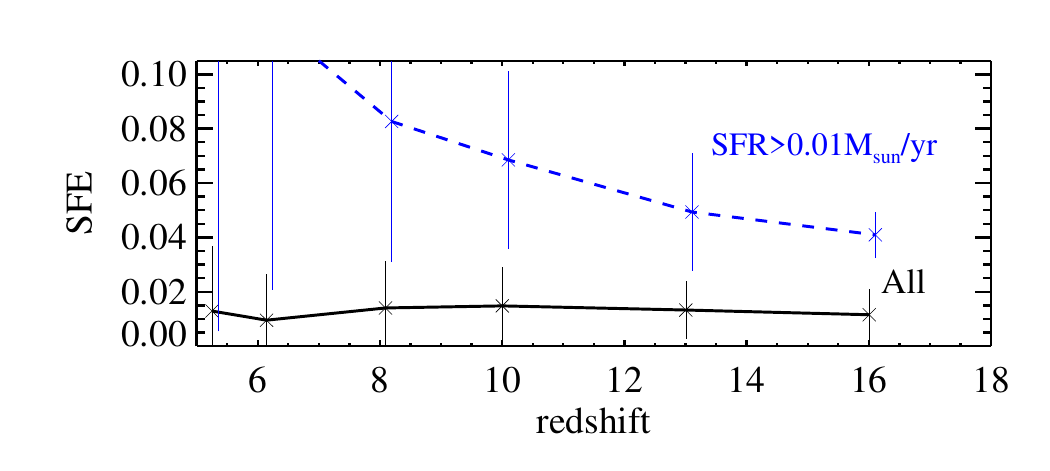}\\
\vspace{-0.7cm}
\includegraphics[width=0.5\textwidth]{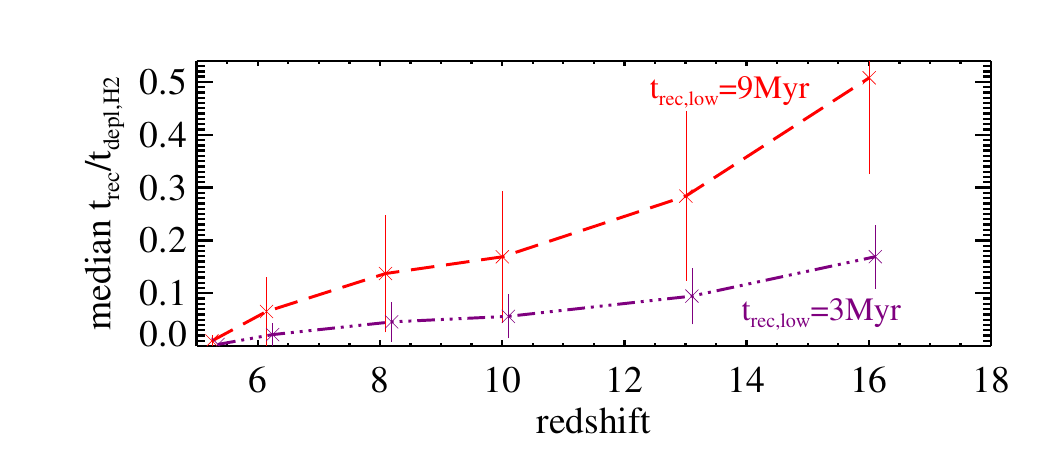}
\caption{\small Top: Median SFE redshift evolution with 1$\sigma$ spread for all (solid line) and
$\rm SFR > 0.01 \, M_\odot yr^{-1}$ (short-dashed line) objects. Bottom: Median $t_{\rm rec} / t_{\rm depl,H_2} $ ratios with 1$\sigma$ spread for $t_{\rm rec, low} = 9 \,\rm Myr$ (long-dashed line) and 3~Myr (dot-dashed line). }
\label{fig:epsilon}
\end{figure}
Gas timescales allow us to infer the SFE of a given object, $ t_{\rm ff} / t_{\rm depl} $, with $t_{\rm ff}$ and $t_{\rm depl}$ gas free-fall time and depletion time (Fig.~\ref{fig:epsilon}, top panel).
We find a median $ {\rm SFE} \sim 0.01$--0.04, and values of around $ 0.1$ for more star-forming objects ($\rm SFR >0.01\,\rm M_\odot yr^{-1}$).
This is in agreement with  low-$z$ data \cite[][]{Ni2025} and the lack of evidence for increased SFE at early times \cite[][]{Zavala2019, Donnan2025}.\\
Another interesting timescale is the recovery time, $t_{\rm rec}$, which gives the time needed for gas to cool again after being heated by stellar feedback during the initial stages of star formation. We check its role in describing star-forming events in the lower panel of Fig.~\ref{fig:epsilon}.
The values of $t_{\rm rec} $ are functions of SN mass or, equivalently, lifetime ($t_{\rm age}$), and in Appendix~\ref{appendixFits} we provide fits to the results by \cite{Jeon2014} in the range of $t_{\rm rec} \sim 9$--$300\,\rm Myr$ for 15--$200\, \rm M_\odot$ SN explosions hosted in a halo with mass $5 \times 10^5 \, \rm M_\odot $.
We adopt the mentioned $t_{\rm rec} $-$t_{\rm age}$ fit in Appendix~\ref{appendixFits} and estimate median values of $ t_{\rm rec}/t_{\rm depl, H_2} $ for the simulated galaxy population.
To bracket uncertainties about $t_{\rm rec} $ lower values ($t_{\rm rec, low}$), we consider both $ 9\, \rm Myr$ and $ 3\, \rm Myr$ for $t_{\rm rec, low}$ \cite[in line with Tab.~1 in][]{Jeon2014} when using our fit.
We find that $t_{\rm rec}$ behaves as a good tracer of the initial episodes of star formation in molecular gas, when it accounts for roughly 20-50\% of the H$_2$ depletion time ($z\simeq 16$); however, ongoing feedback effects make its contribution drop down to 1-10\% at $z\simeq 5$-6.

%*****************************************************************************

\section{Discussion} \label{sect:discussion}

%*****************************************************************************

This paper presents an analysis of the baryon phases at the epoch of reionisation and, as for any research work, the obtained results may depend on the model assumptions performed, but also give hints as to how to interpret observational findings.

\subsection{Model assumptions}
We considered star particles as simple stellar populations described by a Salpeter IMF, although different choices are possible due to uncertainties in the sampling of low-mass, long-lived stars (see Tab.~\ref{tab:properties}).
We included energy ejection and mass loss from type-II SNe, low-mass stars and intermediate-mass stars.
We accounted for type-Ia SNe by considering the classical (favoured) single-degenerate scenario \cite[][]{WhelanIben1973}, in which they form in binary systems after a delay of $\sim 0.1$~Gyr and about 50\% of them explode within 1~Gyr, irrespectively of the exact delay time distribution \cite[][]{IbenTutukov1994, Greggio2005, Cappellaro2015, Maoz2017}.
The properties of the resulting population might depend on the inclusion of stellar rotation and post-zero-age-main-sequence evolution at high $z$, though \cite[][]{Hassan2025}.
\\
Gas cooling takes place via resonant, fine-structure, and molecular lines and is responsible for the transition from hot to warm and cold phases \cite[][]{Maio2007, Maio2010}, while winds are generated within the smoothing kernel of star-forming particles with a constant velocity of 350~km/s and are decoupled from hydro calculations after their ejection in overdense regions \cite[][]{Maio2011}.
Different wind implementations could potentially affect the gas located near star-forming sites, but the impacts should be limited as long as wind velocities are kept in realistic ranges, of around a few $10^2\,\rm km/s$, and mass loading factors do not exceed extreme values of $\sim 5$ \cite[][]{Perrotta2024}.
\\
We checked resolution effects and found that there can be differences in the warm, hot, and stellar phases, but the resulting trends are similar within one dex or less (Appendix~\ref{appendixA}).
This finding is related to the optimal gas resolutions adopted here which allow us to resolve gas phases at scales from megaparsecs to tens of parsecs at $z\gtrsim 10$.
This is consistent with previous works \cite[][]{Creasey2011} that demonstrated that numerical overcooling can severely affect results for particle gas masses higher than $\sim$10$^6\,\rm M_\odot$.
We selected objects by means of their virial mass and number of constituting particles in order to exclude tiny, poorly resolved structures.
Different selection criteria (explored in Appendix~\ref{appendixSelection}) might lead to slightly different results; however, the general trends should be preserved.
\\ 
The temperature thresholds to distinguish cold, warm, and hot gas phases are somewhat arbitrary; however, most literature studies follow the choices we perform here.
A typical threshold of $10^4\,\rm K$ is common in observational investigations to study neutral and HI gas masses at different redshifts \cite[][]{Nelson2020}, while the threshold between warm and hot gas is debated and typically ranges between a few $10^6\,\rm K$ and $10^7\,\rm K$.
Here we chose 10$^7$~K and note that the impacts of adopting slightly lower values would be limited, due to the large amount of warm mass that dominates over hot gas.
\\
We identified bound and IGM baryons by means of the virial radius.
Alternative criteria might impact the final outcomes.
\\
Our results for cosmic mass density parameters demonstrate that cold gas is the main baryon phase until reionisation, when its contribution drops down to $ \Omega_{\rm cold} \sim 10^{-3} $ and warm gas takes over.
Of course, the details of this transition may depend on the reionisation history, the UV-background model adopted, and assumptions about heating and chemistry rates \cite[][]{Maio2022}.
Radiative-transfer effects of the radiation that escapes primordial star-forming regions might have an impact on the actual atomic and molecular abundances, since UV and Lyman-Werner photons might dissociate them in unshielded regions.
Due to their rarity, primordial, solar-like, or OB-star sources do not induce dramatic variations in the average HI and H$_2$ content over megaparsec scales \cite[as quantified in][]{MaioPetkova2016}.\\
The sSFR trends we find are in line with observational data (Appendix~\ref{appendixMStdepl}); however, caution must be taken when comparing numerical models with observations that may rely on different assumptions about the galaxy main sequence.
We neglect primordial black-hole feedback, since in typical conditions its effects are expected to be minor due to the small galaxy masses at these epochs \cite[see further discussion in e.g. ][]{Maio2019}.
\\
With these assumptions, our star formation efficiencies are of the order of a few percent, while average stellar return fractions are $R\lesssim 0.2$ at the epoch of interest here (i.e. up to 20\% of the initial stellar mass is converted into warm gas).
As a consequence, any small fraction of cold gas that forms stars competes with increasing fractions of stellar mass turned back into gas via direct mass loss.
These quantities affect the mass transfer from a phase to another; hence, different modelling or theoretical uncertainties of such processes might impact our findings.
\\
We stress that the main thing responsible for the transition of gas from cold to warm or hot phases is not the mass transfer itself, which is quantitatively rather small, but the heating energy produced and injected into the surroundings after such small amounts of gas have formed stars.
This happens via SN explosions and feedback, and the differentials of the curves in Figs.~\ref{fig:omegas} and \ref{fig:omegas2} give a quantitative measure of the actual mass transfer.
These physical processes lead the mass flows among different phases, determining the complete baryon cycle.

\subsection{Model variations and implications for the early SFRD}
The primordial IMF is particularly debated in the literature \cite[e.g.][and references therein]{Palla1983, Silk1983, Bromm2001th, ABN2002, Tornatore2007, Maio2010, Bromm2013, Liu2020, KG2023} and a top-heavy shape is commonly adopted for PopIII regimes, following prescriptions by \cite{Larson1998}.
His argument was based on the higher Jeans (Bonnor-Ebert) masses expected at higher redshift, due to the increasing cosmic-background temperature. Subsequent studies have highlighted the role of further physical processes, such as local gas shielding, partial recombination, magnetic fields and binary or cluster formation, which would bring proto-stellar seed masses down to values comparable to solar and following a power-law (unbiased) distribution \cite[][]{Yoshida2007, Clark2011, Greif2011, Hirano2014, Stacy2022, Chon2022, Machida2025, Eda2026}.
Direct  evidence is not available yet, and thus the PopIII IMF remains unknown.\\
\begin{figure}
\centering
\includegraphics[width=0.5\textwidth]{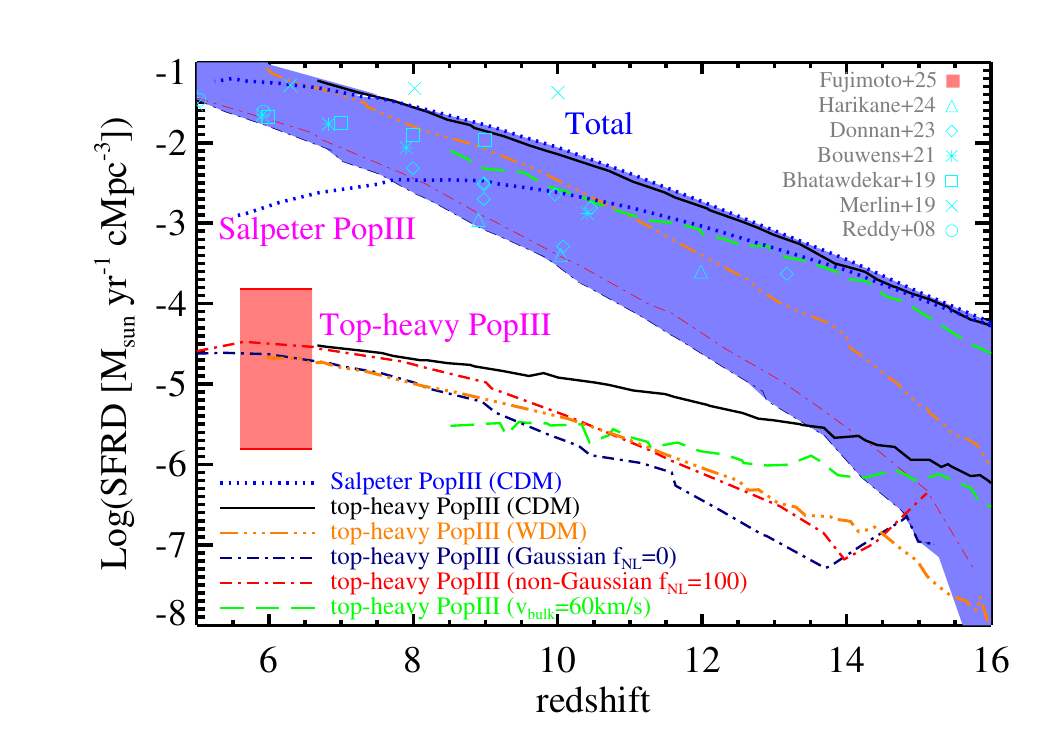}
\caption{\small Observed (symbols) and simulated (lines) comoving SFRDs for Salpeter (this work) and top-heavy (previous works) PopIII IMFs and model variations including: CDM and WDM \cite[][]{Maio2015, Maio2023}, non-Gaussianities \cite[$f_{\rm NL} = 0$, 100;][]{MaioIannuzzi2011}, and primordial bulk flows at 60~km/s \cite[][]{Maio2011vb}. Total SFRDs for different models and resolutions (blue shade) and JWST PopIII determinations (red shade) are also shown. Observational SFRDs are from \citealt{Reddy2008}, \citealt{Merlin2019}, \citealt{Bhatawdekar2019}, \citealt{Bouwens2021}, \citealt{Donnan2023a,Donnan2023b}, \citealt{Harikane2024}, and \citealt{Fujimoto2025}. 
The tabulated data for the theoretical predictions are available at 
\url{https://adlibitum.oats.inaf.it/umberto.maio/predictionsECSF.html}.
}
\label{fig:popIIIcomparison}
\end{figure}
\begin{figure}[]
\centering
\includegraphics[width=0.35\textwidth, angle=270]{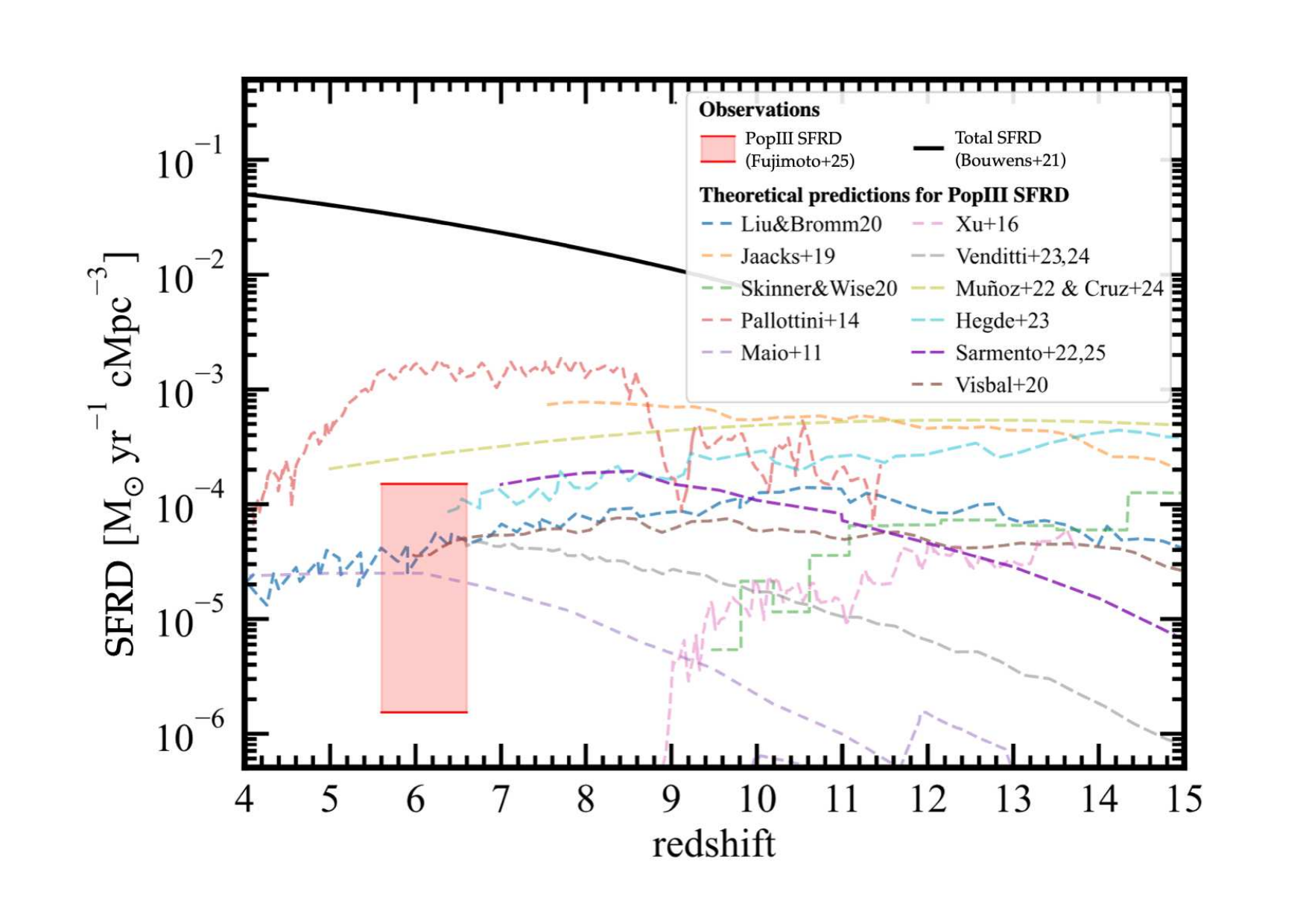}
\caption{\small Top-heavy PopIII SFRDs (broken lines) from simulations (\citealt{Maio2010, Maio2011}, \citealt{Pallottini2014}, \citealt{Xu2016}, \citealt{Jaacks2019}, \citealt{Liu2020}, \citealt{Skinner2020}, \citealt{Venditti2023}) and semi-analytical models (\citealt{Visbal2020}, \citealt{Munoz2022}, \citealt{Hegde2023}, \citealt{Sarmento2025}) compared to total SFRD data (solid line) and JWST PopIII determinations (shaded area) taken from \cite{Fujimoto2025}. }
\label{fig:popIII}
\end{figure}
To test its impacts on the early SFRD, in Fig.~\ref{fig:popIIIcomparison} we collect our expectations for different model variations, including Salpeter and top-heavy PopIII IMFs in our Ref box, and changes in the background matter density field in different cosmic volumes ($\rm L \simeq 1$-$142\,\rm Mpc$), such as: warm darm matter (WDM), non-Gaussianities (as parameterised by the $f_{\rm NL}$ parameter, where  $f_{\rm NL} = 0 $ is the Gaussian case), and high-order corrections to linear theory inducing coherent megaparsec-scale gas bulk valocities of $v_{\rm bulk} = 60\,\rm km/s$ at 2$\sigma$ levels \cite[][]{Maio2011vb, MaioIannuzzi2011, Maio2015, Maio2023}.
While the total SFRDs are similar and suffer mostly from uncertainties in volume sampling or resolution (blue shade), different PopIII IMFs neatly determine different PopIII SFRDs and affect the timing of metal enrichment, photon production, and massive-black-hole formation \cite[][]{Maio2011, MaioPetkova2016, Maio2019, Ma2017}.
Changes in the background (WDM, $f_{\rm NL} = 100$ or $v_{\rm bulk}$ cases) produce limited impacts on PopIII SFRDs, whose trends are still consistent with the latest JWST PopIII constraints at $z\simeq 6$ \cite[][]{Fujimoto2025}.
This demonstrates that the effects of alternative cosmological scenarios are subdominant with respect to baryon modelling \cite[][]{Maio2011, Maio2011vb, Maio2012grb, Maio2015, Maio2023, Pillepich2018, Villaescusa2021, DSouza2022, Zang2024neutrino, Barberi2024, Despali2025, Acharya2025, DEramo2025}.
\\
Fig.~\ref{fig:popIII} summarises the behaviour of the PopIII SFRD for additional models in the literature.
The heterogeneity of theoretical results suggests that current uncertainties or degeneracies are still remarkable, almost two dex ($z\simeq 6$) or larger ($z > 6$), and a definitive consensus must still be reached.

\subsection{Comparison with the literature}
Drawing precise conclusions based on direct comparisons with observations at the epochs of interest here is not trivial and quantification of early gas phases is still an open problem.
Observational determinations of stellar masses are difficult (one-dex error), as well as HI mass estimates, while H$_2$ values rely on indirect (CO) low-$z$ calibrations.
Detections of kiloparsec-scale [CII] emission up to $z\simeq 7$ are consistent with the existence of significant amounts of cold gas in primordial sources \cite[][]{Bischetti2024, Zanella2024, DecarliDiaz2025, Algera2025, Narita2026, Vizgan2026, Chiang2026}, in line with our conclusions.
There seems to be growing evidence of $z > 6$ cold outflows detected via OH$^{+}~290\,\mu\rm m$ and OH~$119\,\mu\rm m$ absorption lines tracing atomic diffuse gas and dense molecular gas, respectively (e.g. ALMA detections by \citealt{Shao2022}, \citealt{Salak2024}, and \citealt{Spilker2025}).
\\
Qualitatively speaking, given the remarkable differences we find between the trends of the cosmic $\Omega$ parameters and the $\Omega_{\rm bound}$ ones, one must conclude that estimates of the baryon budget based on the mass of bound structures could be biased and recent $\Omega_{\rm HI}$ results by \cite{Oyarzun2025} go in this direction.
\\
In large numerical simulations, H$_2$ fractions estimated at post-processing seem to be almost insensitive to the sub-grid models adopted; however, in such analysis the cold and molecular phases cannot be directly traced due to the poor ISM physics \cite[][and references therein]{Ward2022}.
H$_2$ can be further explored through semi-analytical prescriptions calibrated with the present-day observed galaxy populations \cite[][]{Lagos2015, Diemer2018, Dave2019, Szakacs2022}, but this approach tends to underestimate higher-$z$ molecular masses by a factor of a few.
\\
During the cosmic stellar-mass build-up, the amount of the mass returned to the gas phases evolves non-linearly, since the formation of stars is accompanied by a stellar return fraction, $R$, that grows with stellar age (Fig.~\ref{fig:Rage}).
Although consistent with IMF and stellar-mass uncertainties (Tab.~\ref{tab:properties}), dependences of $R$ on stellar metallicity (Appendix~\ref{appendixA}) are less severe than what is derived from analytic calculations assuming fixed metallicity values \cite[][]{ Vincenzo2016}.
More detailed analyses based on simulations that include stellar evolution, time-dependent metal spreading, and feedback mechanisms (such as the one we perform here) are still lacking in the literature.
\\
Baryon mass functions at different $z$ show a variety of trends with typically decreasing high-mass tails.
This is somehow reminiscent of the Schechter formalism, but with more complex behaviour at the low-mass end.
When comparing our results to observations of the HI mass function at $z\simeq 0$ by \cite{Jones2018}, we see that low-$z$ determinations are smoother and extend towards higher HI masses.
This highlights a clear redshift evolution of gas mass distributions.
\\
Baryon phases in bound objects are linked to the main galaxy properties, such as mass, SFR, and metallicity, while dependencies of our results on the specific SFR are usually weak, consistently with recent theoretical findings about early fine-structure line emission \cite[][]{Bisbas2022, Casavecchia2024, Casavecchia2025, ME2024,Nyhagen2025, Nakazato2026}.
In general, cold, warm, and hot gas masses increase with hosting $M_{\rm star}$, SFR, and gas $Z$ and the found relations bare a scatter of one or two dex, as a consequence of star formation stochasticity \cite[][]{Katz1996}.
Thus, tight observational correlations might be sample-biased and capture only the trends of more powerful sources, mostly at high redshift.
\\
The broad scatter in the gas phases is reflected in the depletion times.
Observational determinations for star-forming galaxies at $z\sim 0$ suggest gas depletion times of $\sim 2$--8~Gyr \cite[][]{Tumlinson2017} decreasing within $M_{\rm star} \sim 10^{8.5}$--$10^{11.5} \,\rm M_\odot$.
First surveys of molecular gas in PHIBSS cosmic-noon star-forming galaxies \cite[][]{Tacconi2013, Tacconi2018} found $ t_{\rm depl, H_2} \propto (1+z)^{-0.6} $, well approximated by $ t_{\rm dyn} \simeq t_{\rm H} / 10$.
Recent estimates inferred by QSO and main-sequence galaxy observations at $ z \simeq 6-7$ suggest $ t_{\rm depl,H_2} \sim 0.01$--$0.1 \,\rm Gyr$ \cite[][]{Zavala2022, Salvestrini2025}, in line with our results.
We find that depletion times depend on redshift, but, when averaged over the $z\gtrsim 5$ range, median $t_{\rm depl, H_2}$ values are weakly dependent on gas metallicity and lie around 100~Myr as a result of feedback-regulated star formation.
Such short H$_2$ timescales make the formation of Milky Way progenitors possible at $z \gtrsim 6$ \cite[][]{Rusta2024}.
Corresponding star formation efficiencies are around $ 0.01 $ and do not evolve significantly.
In this respect, recent findings in isolated-galaxy simulations by \cite{Polzin2024} are in line with ours.
\\
Overall, the theoretical timescales we find in primordial epochs help us understand the first phases of cosmic star formation and the role of feedback and environment in shaping the early baryon cycle.
Furthermore, the impact of star formation heating explains why absorption-selected galaxies \cite[e.g.][]{Bollo2026} are often inefficient at converting cold gas into stars, despite their sizeable molecular reservoir.

\subsection{Observational insights}
From an observational point of view, measurements of $M_{\rm H_2}$ and $M_{\rm star}$ can be inferred from millimetre and optical observations, while $M_{\rm HI}$ is becoming available at cosmological distances thanks to 21~cm detections.
For this reason, the following definitions of observational HI and H$_2$ gas-to-star mass fractions and baryon mass fraction are commonly adopted: $ \mu_{\rm HI} = M_{\rm HI} / M_{\rm star} $, $ \mu_{\rm H_2} = M_{\rm H_2} / M_{\rm star} $ and $ F_{\rm gas} = M_{\rm H_2} / ( M_{\rm H_2} + M_{\rm star}) = \mu_{\rm H_2} / ( \mu_{\rm H_2} + 1) $.
\\
\begin{figure}
 \centering
 \includegraphics[width=0.5\textwidth] {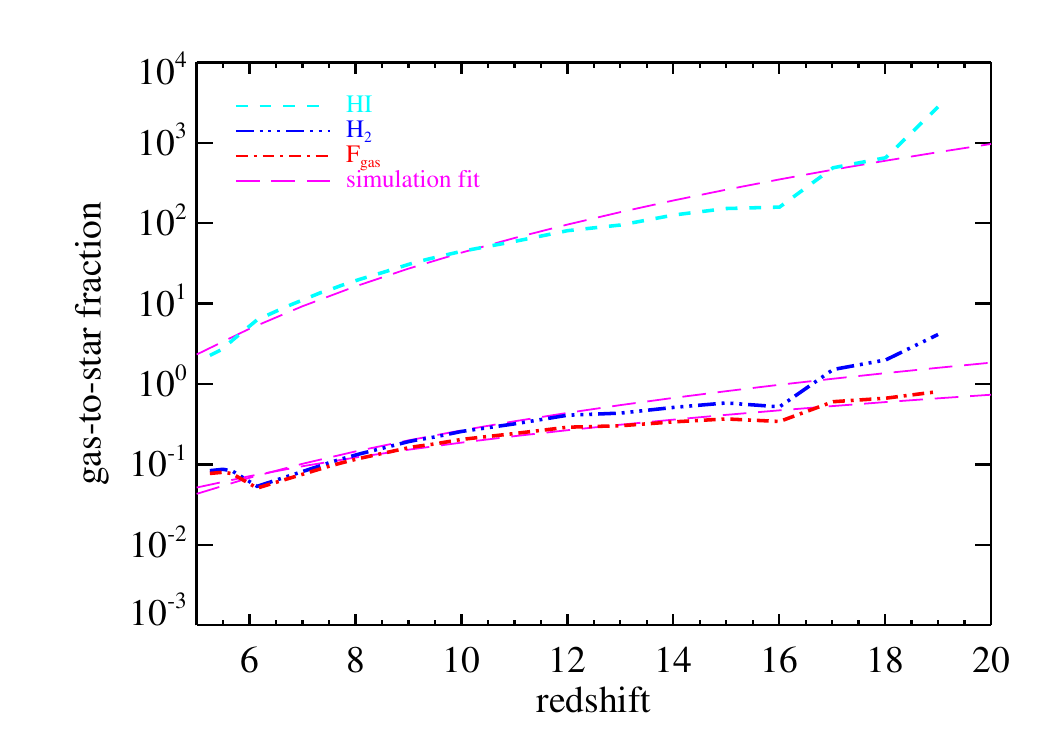}
 \caption{\small Average $\mu_{\rm HI} = M_{\rm HI} / M_{\rm star} $ (short-dashed line), $\mu_{\rm H_2} = M_{\rm H_2} / M_{\rm star} $ (dot-dot-dot-dashed line), and $ F_{\rm gas} = \mu_{\rm H_2}  / (\mu_{\rm H_2}  + 1) $ (dot-dashed line) redshift evolution with simulation fits (long-dashed lines; Appendix~\ref{appendixFits}). }
 \label{fig:Fobslikeredshift}
\end{figure}
Fig.~\ref{fig:Fobslikeredshift} displays the evolution of the average $\mu_{\rm HI}$, $\mu_{\rm H_2}$, and $ F_{\rm gas}$ in the first gigayear.
HI gas-to-star fractions decrease monotonically from $ \mu_{\rm HI} \sim 10^3$ at $z \gtrsim 20$ down to unity at $z\simeq 5$, while corresponding $\mu_{\rm H_2}$ values range from a few to 0.1.
These trends are due to the tiny amounts of stars formed in early epochs in comparison to the dominant cold-gas mass.
Subsequent stellar-mass growth and the associated feedback mechanisms cause a shallower evolution of cold-gas phases and a regular decrease in both $\mu_{\rm HI}$ and $\mu_{\rm H_2}$.
The $F_{\rm gas}$ behaviour is analogous to $\mu_{\rm H_2}$ with values ranging between $ \sim 0.1 $ ($z\simeq 5$) and 1 ($z\simeq 20$).
\\
In Fig.~\ref{fig:Fobslike} we present the dependencies of $\mu_{\rm HI} $, $\mu_{\rm H_2} $, and $F_{\rm gas} $ with $M_{\rm star}$ and corresponding fitting relations for the whole simulated data sample (Tab.~\ref{tab:mHIH2MvirMstarSFR}).
\begin{figure*}
 \centering
 \includegraphics[width=0.33\textwidth] {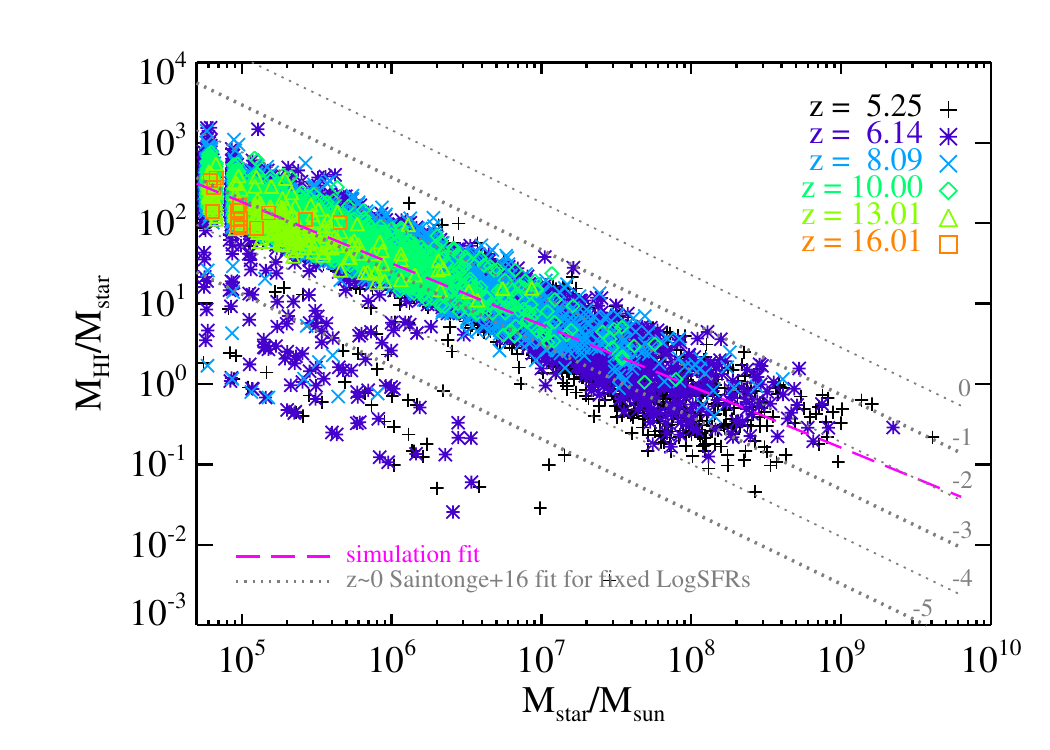}
 \includegraphics[width=0.33\textwidth] {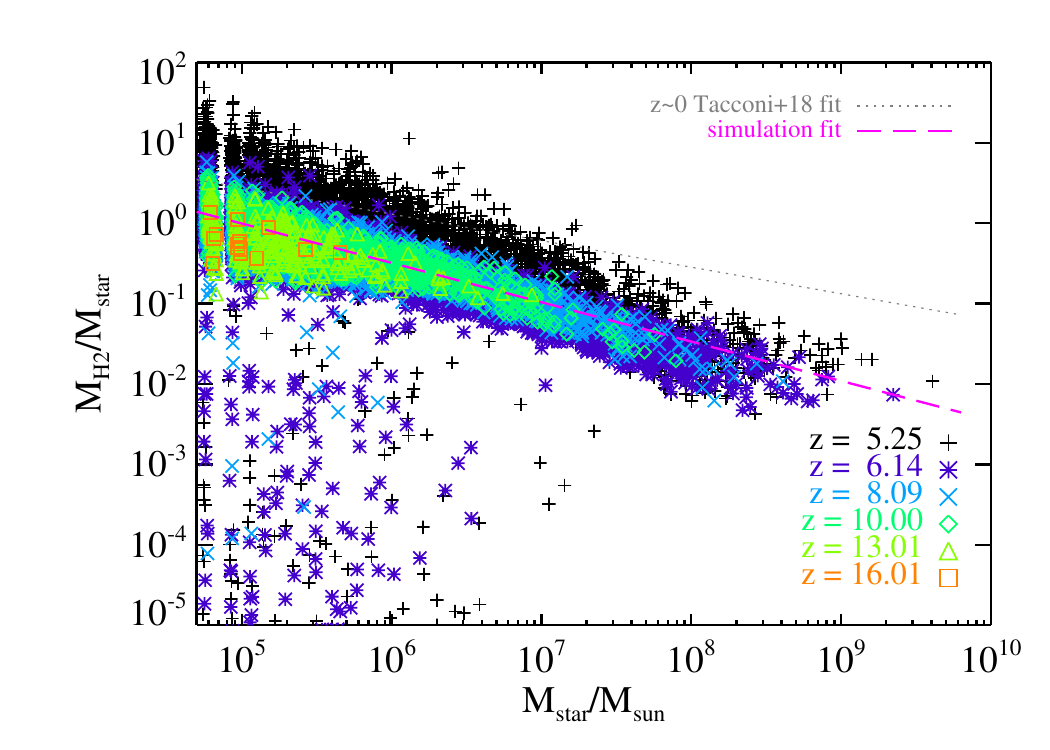}
 \includegraphics[width=0.33\textwidth] {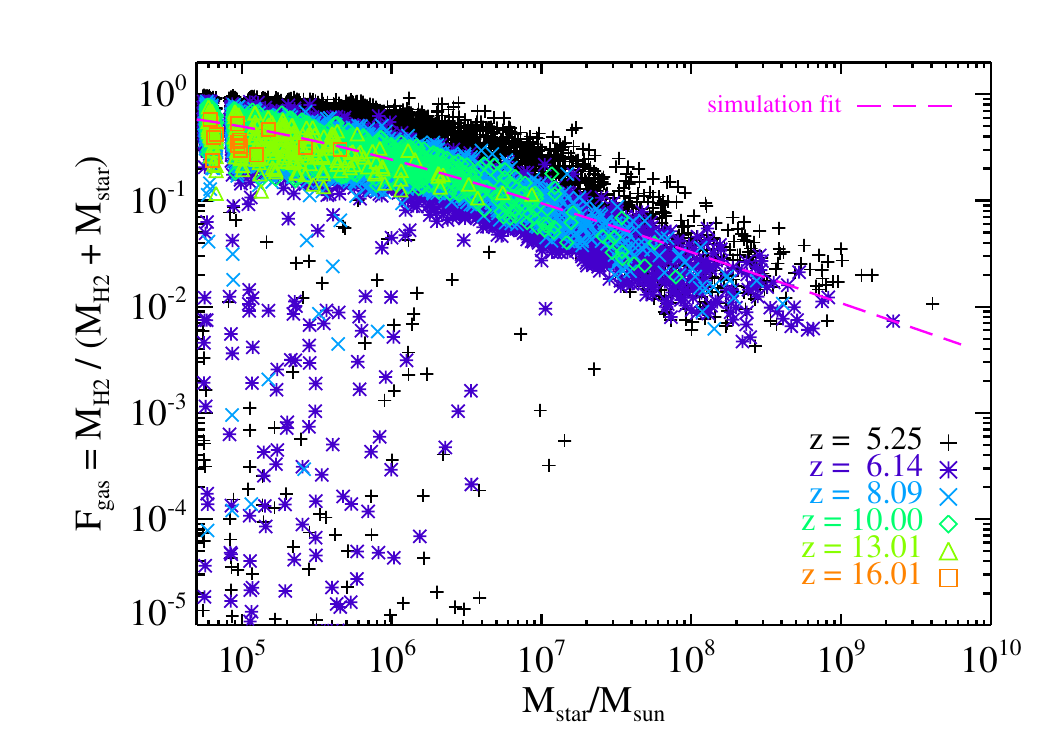}
 \caption{\small Gas-to-star fractions, $\mu_{\rm HI}$ (left), $\mu_{\rm H_2}$ (centre), and $ F_{\rm gas} = \mu_{\rm H_2} / (1 +  \mu_{\rm H_2}) $ (right) as functions of $M_{\rm star}$ compared to $z\simeq 0$ observational fit functions (dotted lines) for main-sequence galaxies \cite[][]{Saintonge2016, Saintonge2017, Tacconi2018}. Different dotted lines in the left panel correspond to fixed $\rm Log(SFR/M_\odot yr^{-1}) = 0$, -1, -2, -3, -4, and -5, from top to bottom. Dashed magenta lines refer to simulation fits (Appendix~\ref{appendixFits}). }
 \label{fig:Fobslike}
\end{figure*}
Observational fractions decrease with $M_{\rm star}$.
In particular, the relation between $\mu_{\rm HI} $ and $M_{\rm star}$ is fitted by $ \mu_{\rm HI} = 10^{6.08} (M_{\rm star}/M_\odot )^{-0.76} $, and the resulting normalisation and slope are roughly constant in $z$.
HI measurements from ALFALFA, GASS, COLDGASS, and xGASS data \cite[][]{ Brown2015, Saintonge2016, Catinella2018, Guo2021} show similar slopes, of between $-0.85$ and $-0.90$.
However, optically dark sources with a low surface brightness are also detected in HI and lie off the relation at $M_{\rm star} \lesssim 10^7 \,\rm M_\odot$ \cite[see recent WALLABY results by][]{OBeirne2025}.
\\
Our expectations for $\mu_{\rm H_2}$ feature a slope ranging between $-0.3$ (at $z\simeq 16$) and $-0.7$ (at $z\simeq5$) and give $ \mu_{\rm H_2} \propto M_{\rm star} ^{ -0.49} $ as average scaling.
That broadly agrees with low-$z$ observational results from, for example, xCOLDGASS field galaxies \cite[][]{Saintonge2017}, EDGE-CALIFA spiral galaxies \cite[][]{Bolatto2017}, SDSS star-forming galaxies \cite[][]{Popesso2020}, GASP jellyfish galaxies \cite[][]{Moretti2023}, and PHIBBS, ASPECS, and ALMA/NOEMA galaxies and cosmic-noon AGNs \cite[][]{Bertola2024} that similarly span a range of a few dex and feature a decreasing $\mu_{\rm H_2}$.
The exact slope of the $\mu_{\rm H_2}$-$M_{\rm star}$ relation is still debated.
For example, the analysis by \cite{Tacconi2018} is compatible with a slope of $-0.33$ at $z\sim 0$, while \cite{Bertola2024} find a slope of $-1$ at cosmic noon, without strong differences between AGN hosts and non-active galaxies.
It is unclear whether this is due to possible redshift effects; however, \cite{Papastergis2012} and \cite{Peeples2014} find trends analogous to ours, with $z\sim 0$ 
% HI+H$_2$ 
gas-to-star ratios in the range 0.1--100 and a slope between $-0.48$ and $-0.43$ using measurements by \cite{McGaugh2005, McGaugh2012}, \cite{Leroy2008} and \cite{Saintonge2011}.
\\
Our predictions for $F_{\rm gas}$ suggest typical values of between 0.1 and 0.8 for most of the sampled stellar masses and a drop below 10\% at more massive scales.
The trend has an average slope of roughly $-0.3$; however, it is better fitted by the $\mu_{\rm H_2}$-based relation $ F_{\rm gas } =  [1 + 0.0036 \, (M_{\rm star}/{\rm M_\odot})^{0.49}]^{-1}$.
In the figure, scattered data points below $\mu_{\rm HI}$, $\mu_{\rm H2}$, and $F_{\rm gas}$ trends represent galaxies that are either inactive or that have suffered outflow-driven HI and H$_2$ mass removal.
The $F_{\rm gas}$ bending reflects environment or feedback-induced molecular suppression and might be at the origin of the SFR-$M_{\rm star}$ bending at later time \cite[][]{Popesso2019, Perez2025}.
Estimates for nearby resolved galaxies observed with Hershel agree with typical $F_{\rm gas } \simeq 0.04$ and range between 0.02 and 0.6 depending on metallicity and dust content \cite[][]{Casasola2022}.
NOEMA determinations of $z < 1$ H$_2$-cooling bright central galaxies yield $F_{\rm gas} \simeq 0.1$--0.5 \cite[][]{Castignani2025}.
At $z\simeq 1$--5, observations of star-forming gas have been reporting routinely $F_{\rm gas} \simeq 0.3$--0.8 \cite[][]{Daddi2010, Harris2012, Magnelli2012, Saintonge2011, Tacconi2013, Bothwell2013, Scoville2016} and studies of resolved molecular gas in $z\simeq 2$--3 lensed active galaxies suggest $F_{\rm gas} \simeq 0.6$ \cite[][]{Spingola2020}.
Recent analysis of ALMA/SDSS $z\simeq 2$ quasars \cite[][]{Molyneux2025} agree with $ F_{\rm gas} \simeq 0.02$--0.24 ($ \mu_{\rm H_2} \simeq 0.02$--0.32).
Values from $z\sim 4$--8 primordial galaxies show $F_{\rm gas} \simeq 0.3$--0.9 \cite[][]{DZ2020, Heintz2023, Aravena2024, Fudamoto2025}.
At higher $z$ there is no convergence, yet, and upper limits at $ z \simeq 14 $ range between 0.5 \cite[][]{Helton2025} and 0.8  \cite[][]{Schouws2025OIII, Carniani2025}.
\\
All these observational results, together with the expected large theoretical scatter, are in line with the idea that high-$z$ star-forming galaxies might be more gas-rich than present-day ones due to their more limited evolution.
Environment or dynamical conditions (mass, SFR, $Z$, mergers) do play a relevant role and induce significant variations in the actual abundances \cite[as observationally suggested at lower $z$ by e.g.][]{Kanekar2018, Donevski2020, Zanella2023, Zanella2025}; hence, HI and H$_2$ do not necessarily fill up the whole galaxy gas budget.
The warm component can account for more than 50\% of the gas content in bound structures and HI and H$_2$ mean fractions can reach values as low as 0.2 and 10$^{-3}$, respectively (Appendix~\ref{appendixGasFractions}).
Hints about the role of warm cosmic gas are also suggested by low-$z$ H Ly$\alpha$ studies \cite[][]{ Viel2017, Scarlata2025, Gelli2025, Rowland2025} and partially ionised atomic species, such as OVI or CIV \cite[][]{Peeples2014, Werk2019, Lochhaas2025}.
Also MgII, NII, and OIII lines are diagnostics of mildly warm gas and have been recently detected at high $z$ \cite[][]{Hennawi2021, Kolupuri2025, Perna2025, Belladitta2025, Nakazato2025}, but a reliable quantification has proven to be difficult.
For hot gas, the situation is much more uncertain, as $T>10^7\,\rm K$ gas is mostly detected in low-$z$ galaxy groups and clusters via X-ray emission.
While hot-gas fractions can be related to total mass in both simulations and observations at $z < 1$ \cite[][]{Comparat2023, Zhang2024erositaI, Zhang2024erositaII, Rasia2025, Marini2025, Biffi2025}, the high-$z$ regime still deserves investigations.
In addition, the spread among available observations is $\sim$dex, due to the different data (XMM-Newton, Chandra, SZ, gravitational lensing), selections (X-ray or optical), or assumptions (hydrostatic equilibrium, scaling relations) adopted to retrieve the hot mass.

%*****************************************************************************

\section{Conclusions} \label{sect:conclusions}

%*****************************************************************************

\noindent
In this work we have exploited the up-to-date {\sc ColdSIM} numerical simulations to make predictions about the baryon budget of early galaxies and explore correlations that may be used to infer unknown quantities.
We implemented detailed non-equilibrium chemistry of several chemical species, which guaranteed that HI and H$_2$ abundances were traced consistently with metal-dependent cooling, heating, and feedback processes during cosmic structure evolution.
This allowed us to explicitly follow the cold, warm, and hot gas thermal phases without relying on poorly constrained sub-grid assumptions.
Our findings can be summarised as follows.
\begin{itemize}
	\item[--]
	Baryon mass density parameters at $z > 5$ are dominated by gas, both in bound structures (galaxies/CGM) and in the IGM, and bound objects contain only a small fraction of the entire baryonic mass, around $10^{-6}$--$10^{-2}$.
	\item[--]
	The cold phase dominates the epoch of reionisation, but warm gas is present by $z\sim 20$ and increases with time, taking over the cosmic baryon budget at $z\lesssim6$.
	\item[--]
	Due to uncertainties in the adopted IMF, the stellar mass density parameter can vary within 0.3~dex.
	\item[--]
	Phase mass functions evolve with $z$ and show feedback-driven trends at low masses and a more regular Schechter-like shape at the high-mass end.
	\item[--]
	Gas phases correlate with the host galaxy $M_{\rm star}$ and SFR, and feature a more scattered behaviour with gas $Z$.
	\item[--]
	HI traces the whole cold phase fairly well, while H$_2$ gas is tightly linked to star formation with $ t_{\rm depl, H_2} \sim 0.01$--$0.1\,\rm Gyr$.
	\item[--]
	Typical $ t_{\rm depl, HI} $ and  $ t_{\rm depl, H_2} $ decrease with $M_{\rm star}$ and SFR and correlate weakly with $Z$.
	\item[--]
	Median star formation efficiencies are around a few percent and do not evolve significantly with redshift.
	\item[--]
	The stellar return fraction increases up to $R\simeq$0.2 for $t_{\rm age} \simeq 1\,\rm Gyr$ (a factor of two lower than the values commonly adopted) and this impacts high-$z$ SFRDs.
	\item[--]
	Observation-based $\mu_{\rm HI}$,  $\mu_{\rm H_2}$, and $F_{\rm gas}$ fractions show decreasing trends with cosmic time, $M_{\rm star}$, and SFR, but cannot be considered as fair proxies of the whole galaxy gas content.
	\item[--]
	Matter phases and depletion times are related to host galaxy properties via simple scaling relations (see the appendix).
\end{itemize}

% =============================================================================

\begin{acknowledgements}
We thank the anonymous referee for constructive comments that allowed us to improve the original manuscript.
We are grateful to V.~Bollo and T.~O'Beirne for useful discussions and for kindly sharing MUSE-ALMA Haloes data and to S.~Fujimoto and R.~P.~Naidu for granting leave to reproduce their PopIII data collection.
UM expresses gratitude to the Italian National Institute of Astrophysics for financial support provided through Theory Grant no.~1.05.23.06.13 ``FIRST – First Galaxies in the Cosmic Dawn and the Epoch of Reionisation with High Resolution Numerical Simulations'' and Travel Grant no.~1.05.23.04.01.
UM also acknowledges warm hospitality at the European Southern Observatory during the completion of this work and inspiring interactions with Camilla~Maio.
We finally acknowledge the NASA Astrophysics Data System and the JSTOR archive for their bibliographic tools.
\end{acknowledgements}

% =============================================================================
\bibliographystyle{aa1}
\bibliography{bibl}
% =============================================================================

% ===========================================================================
\appendix
% ===========================================================================

\section{Numerical effects}  \label{appendixA}

\begin{figure}
\centering
\includegraphics[width=0.5\textwidth]{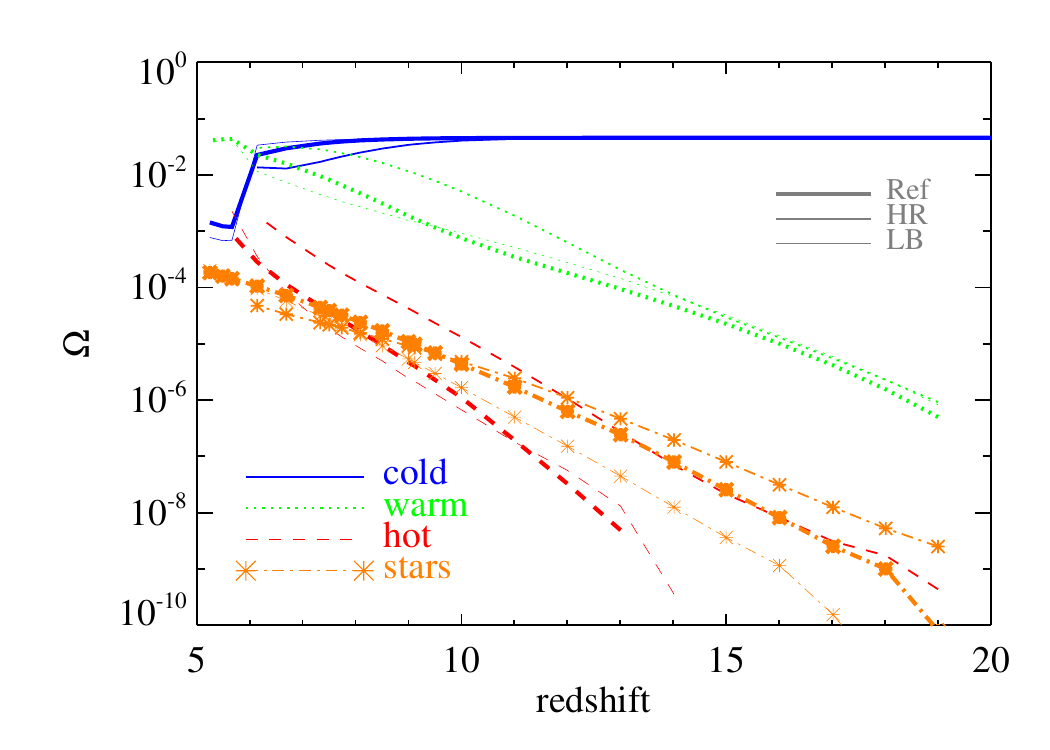}
\caption{\small Redshift evolution of the mass density parameters for the cold (solid lines), warm (dotted lines), hot (dashed lines) and stellar (dot-dashed lines) components in the Ref (thick lines), HR (regular lines) and LB (thin lines) simulations.}
\label{fig:resolution}
\end{figure}

\begin{figure}
 \centering
 \includegraphics[width=0.5\textwidth] {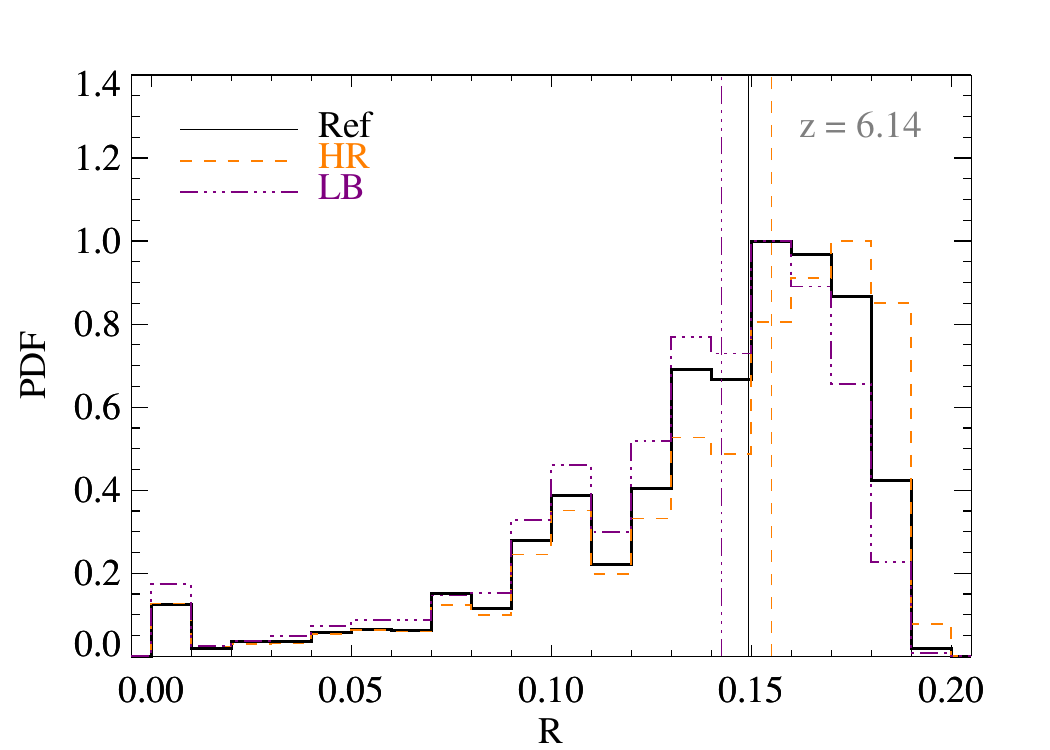} 
 \caption{\small Probability distributions (histograms) and median values (vertical lines) of $R$ for the Ref (solid black lines), HR (dashed orange lines) and LB (dot-dot-dot-dashed purple lines) simulations at $z\simeq 6$. }
 \label{fig:Rpdf_021}
\end{figure}

\begin{figure}
 \centering
  \includegraphics[width=0.5\textwidth]{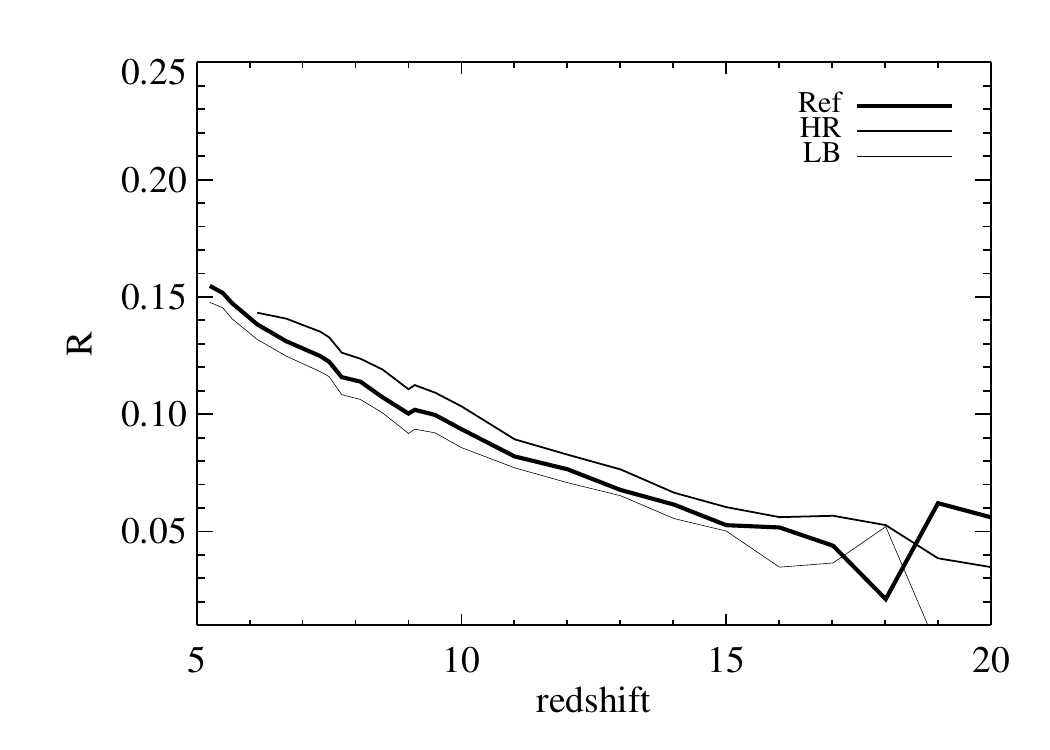} \\	
  \vspace{-0.5cm} 
  \includegraphics[width=0.5\textwidth] {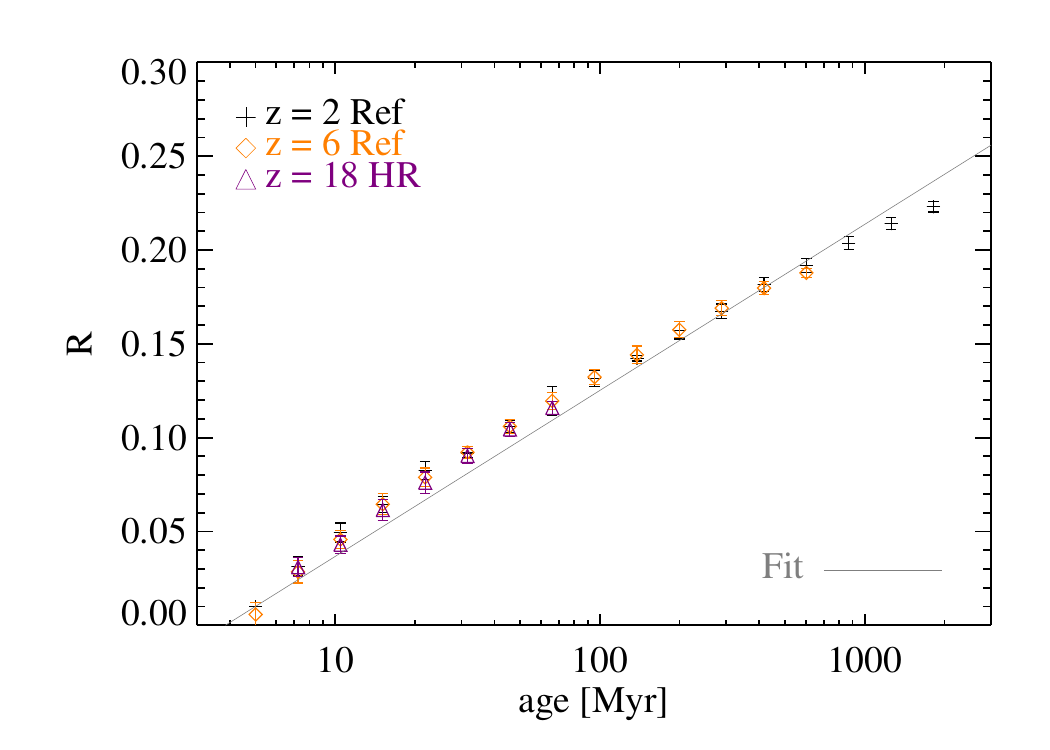}
 \caption{\small Top: Redshift evolution of the average $R$ at different redshifts for the Ref (solid black lines), HR (dashed orange lines) and LB (dot-dot-dot-dashed purple lines) simulations. Bottom: Dependence of $R$ on stellar age at $z\simeq 2$ (crosses), 6 (diamonds) and 18 (triangles), plotted together with the fitting function of eq.~\ref{eq:fit} (solid line). }
 \label{fig:Rrel}
\end{figure}

\begin{figure}
\centering
\includegraphics[width=0.5\textwidth]{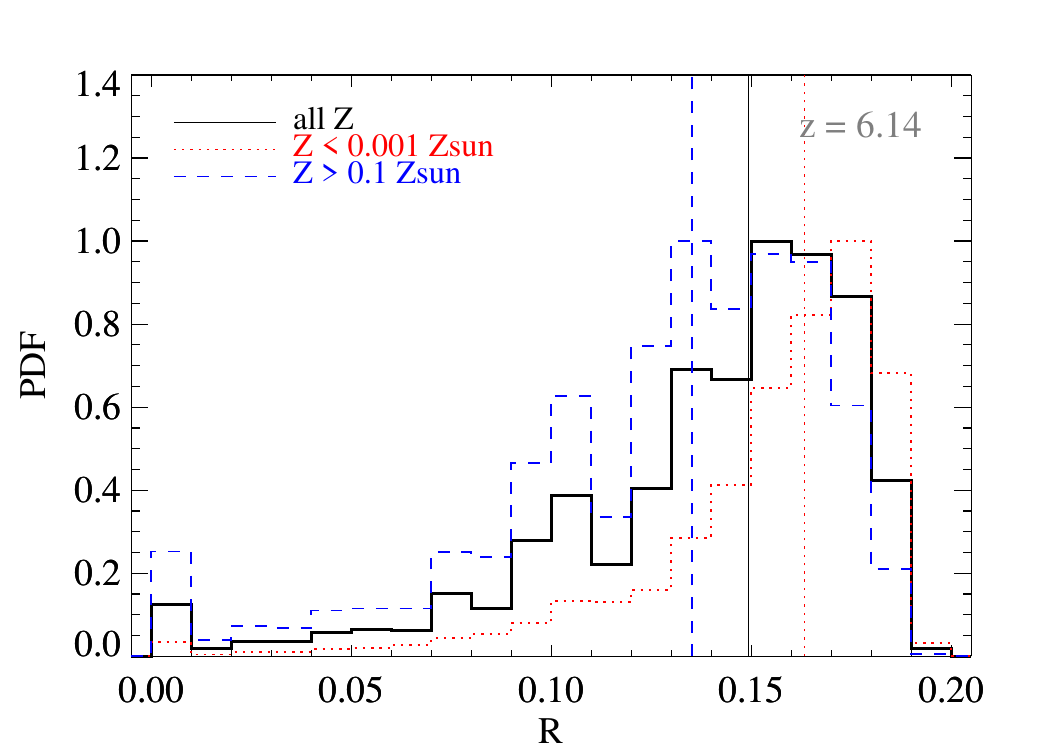}\\
\vspace{-0.5cm} 
\includegraphics[width=0.5\textwidth]{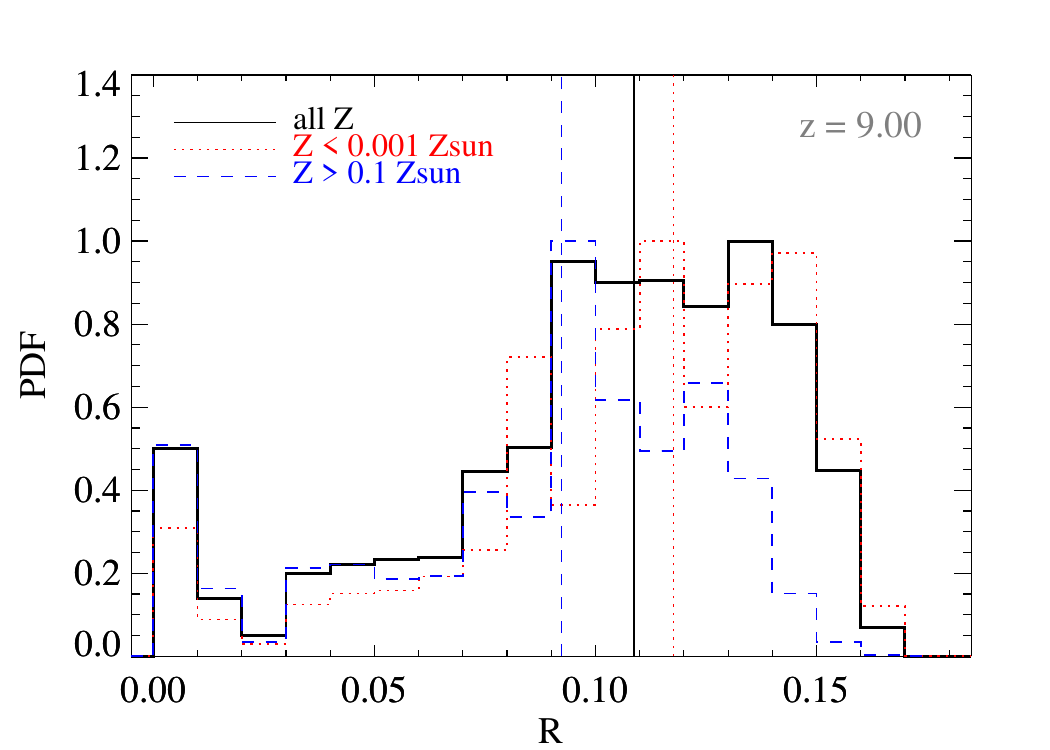}\\
\vspace{-0.5cm} 
\includegraphics[width=0.5\textwidth]{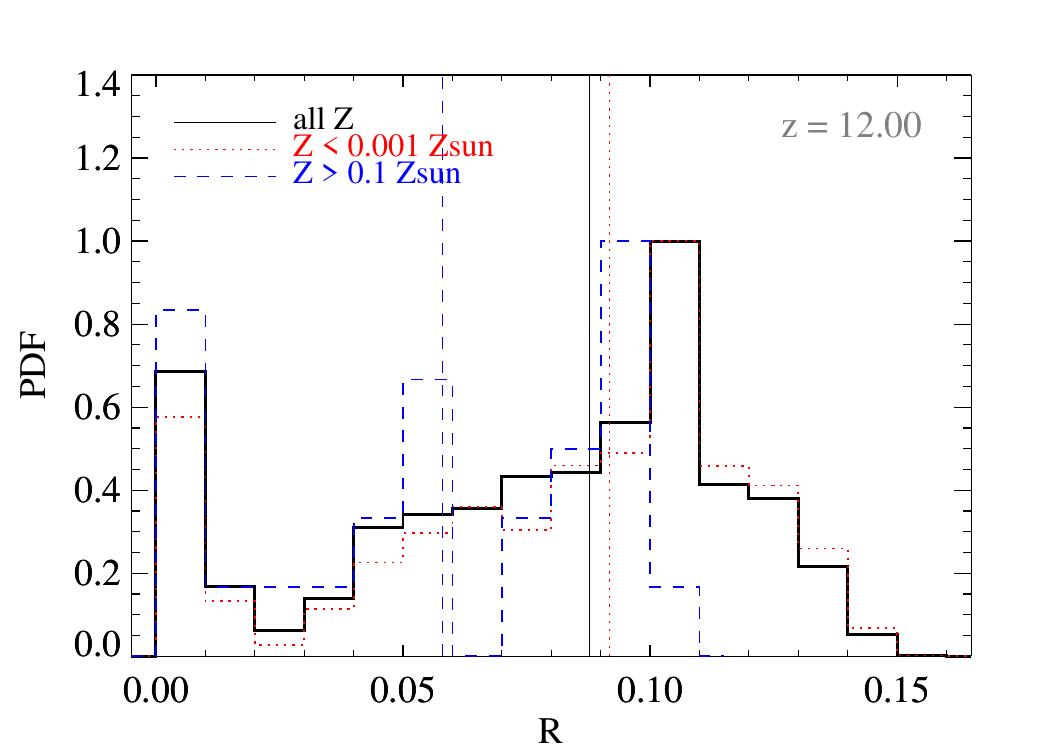}
\caption{\small Probability distribution functions (histograms) and median values (vertical lines) of $R$ computed for the Ref run at redshift $z\simeq 6$, 9 and 12 for different stellar metallicities: all $Z$ (solid black lines), $ Z < 0.001 Z_\odot $ (dotted red lines) and  $ Z > 0.1 Z_\odot  $ (dashed blue lines). }
\label{fig:RZ}
\end{figure}

%*****************************************************************************
%
In Fig.~\ref{fig:resolution} the impact of different box resolutions and sizes on the evolution of the matter density parameters is shown for cold, warm, hot and stellar phases  in the Ref, HR, and LB runs. 
Results converge to the same behaviour in all the boxes.
\\
Figs.~\ref{fig:Rpdf_021} and \ref{fig:Rrel} show the resulting distributions for the stellar return fraction $R$ at $z\simeq 6$, and its evolution as function of redshift and stellar age, as described by the fitting expression of eq.~\ref{eq:fit}:
$ R (t_{\rm age}) = 0.0885 \, {\rm Log} (t_{\rm age } / 3.853 \rm Myr) $.
$R$ values are obtained from self-consistent stellar-evolution calculations including time-dependent metal enrichment for a Salpeter IMF.
\\
Further numerical effects are discussed about stellar-metallicity dependences of the stellar return fraction $R$ in Fig.~\ref{fig:RZ}.
We find metallicity-driven variations from a few up to some ten percent with stronger impacts at high redshift, as demonstrated by the $R$ distributions for metallicity-selected structures ($Z < 0.001 \, \rm Z_\odot$ and $ Z > 0.1 \, \rm Z_\odot$).
At $z\simeq 6$, 9 and 12, sample peak and median are both around $R\simeq 0.15$, 0.11 and 0.09.
Metal-poor populations, that are older and more evolved, feature corresponding values of $R\sim 0.17$, 0.12 and 0.09, while younger metal-rich populations have $ R \sim 0.14$, 0.09 and 0.06.

\section{Selection effects}  \label{appendixSelection}
%*****************************************************************************
%
\begin{figure}
 \centering
 \hspace{-0.7cm}
 \includegraphics[width=0.25\textwidth] {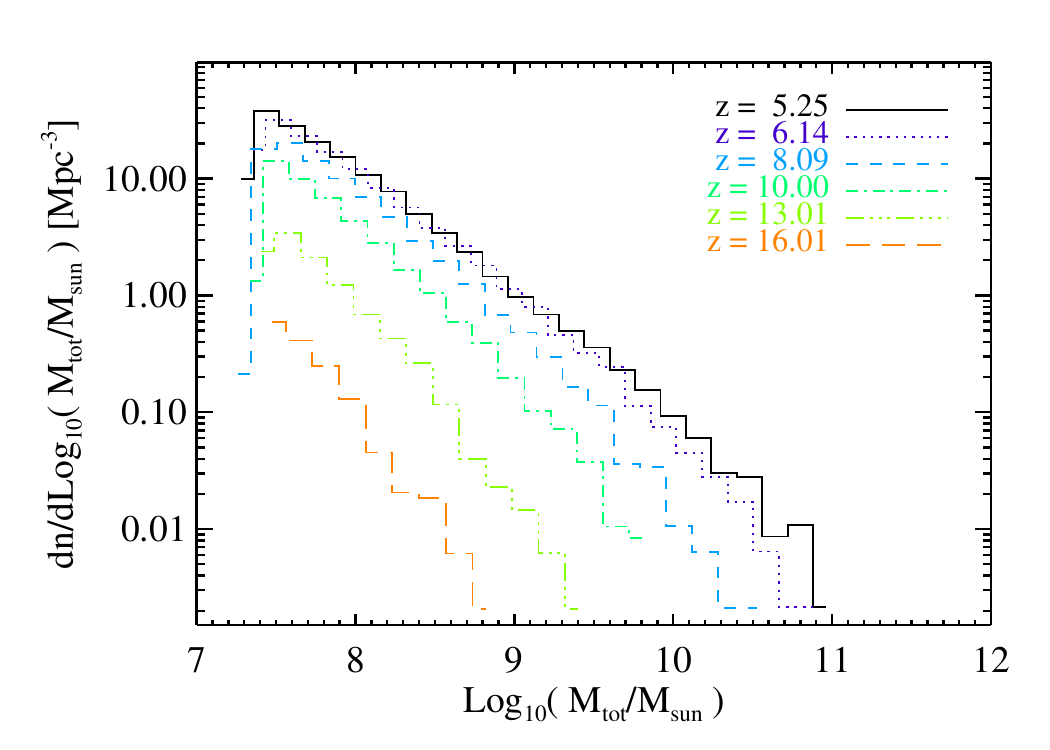}
 \hspace{-0.5cm}
 \includegraphics[width=0.25\textwidth] {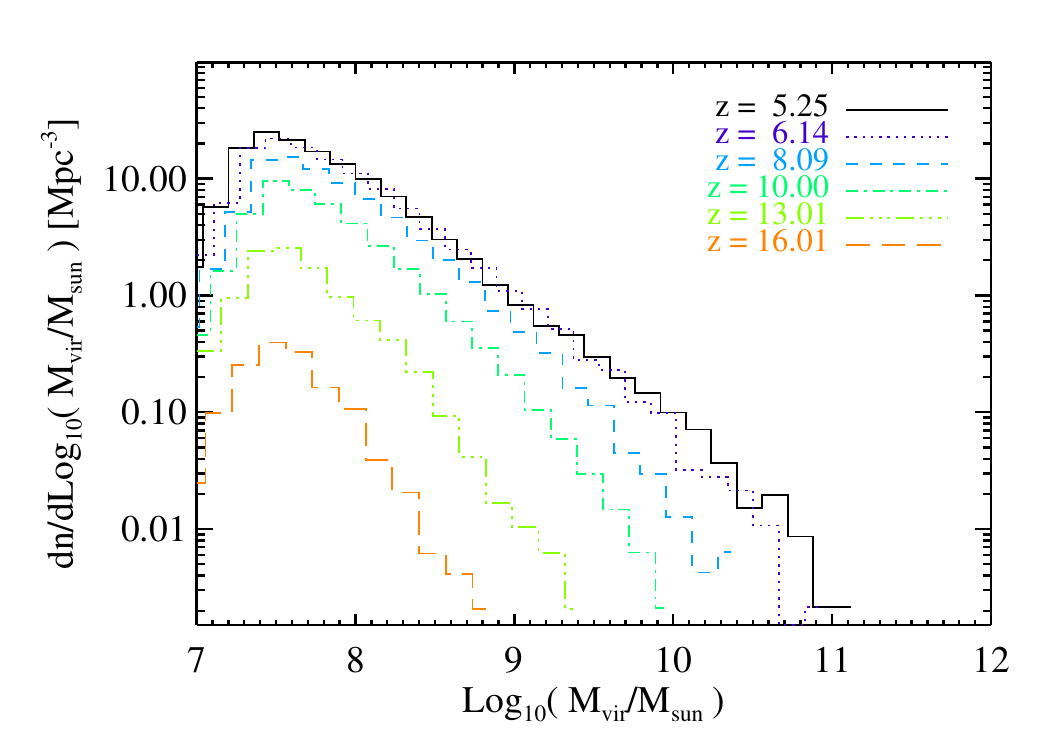}
 \\
 \vspace{-0.3cm} 
 \hspace{-0.7cm}
 \includegraphics[width=0.25\textwidth] {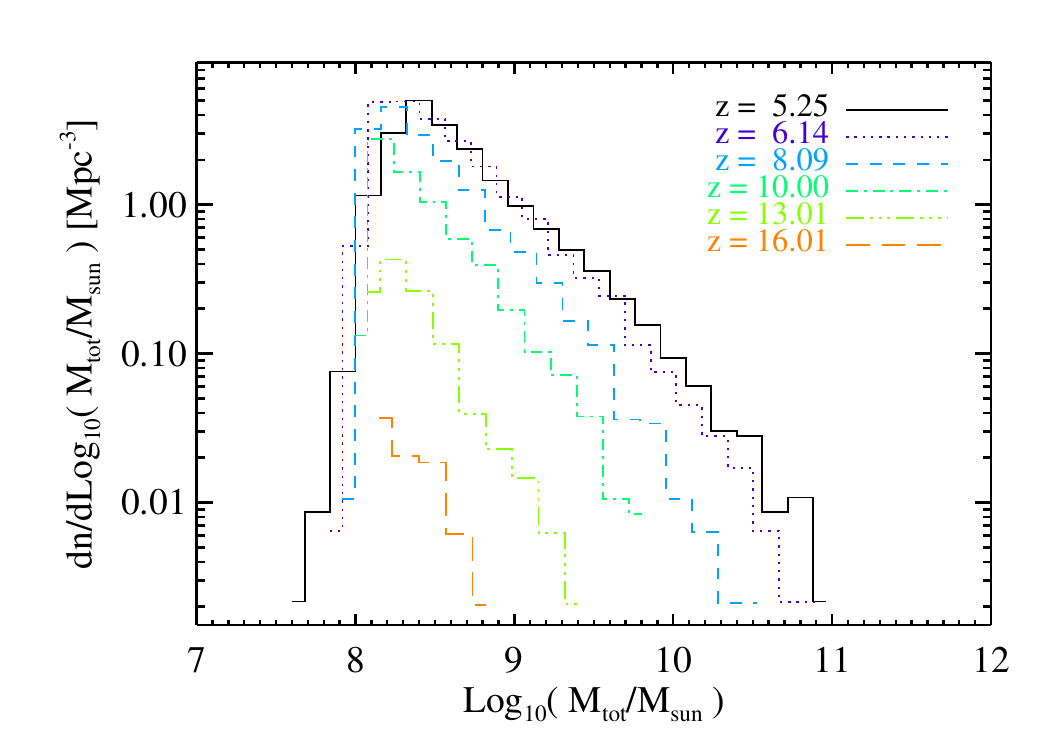}
 \hspace{-0.5cm}
 \includegraphics[width=0.25\textwidth] {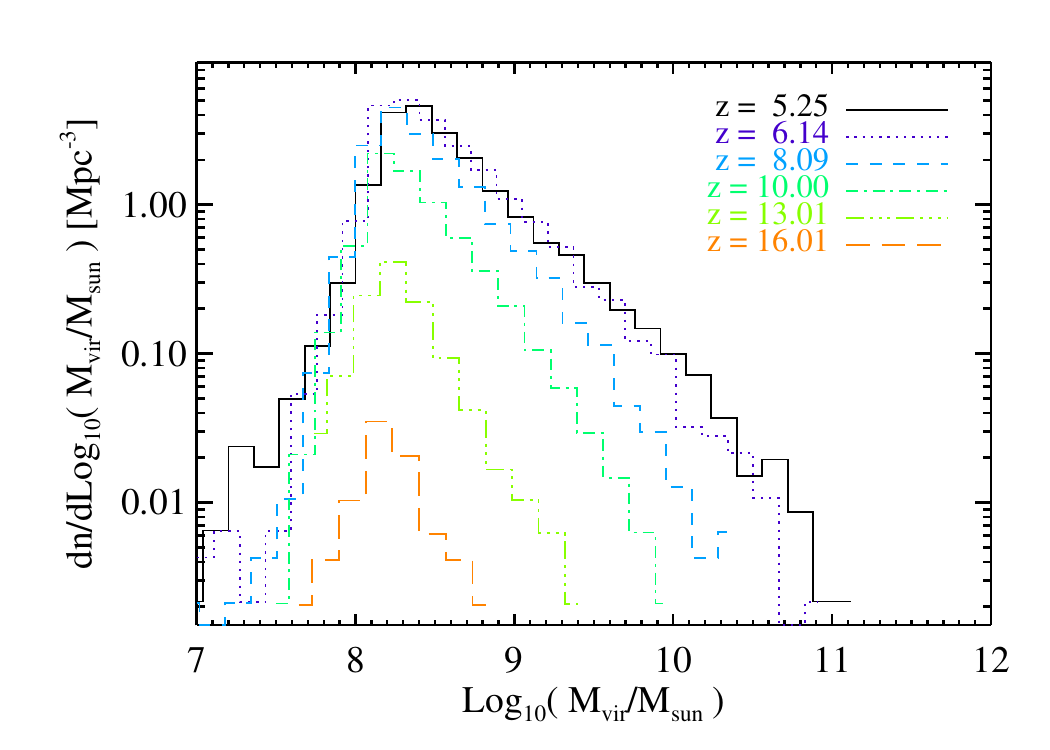}
 \caption{\small Top row: Halo total-mass functions (left) and virial-mass functions (right) at $z = 5.25$, 6.14, 8.09, 10.00, 13.01, 16.00. Bottom row: Corresponding functions for haloes with more than 300 particles.}
 \label{fig:haloselection}
\end{figure}
\begin{figure}
 \centering
 \hspace{-0.7cm}
 \includegraphics[width=0.25\textwidth] {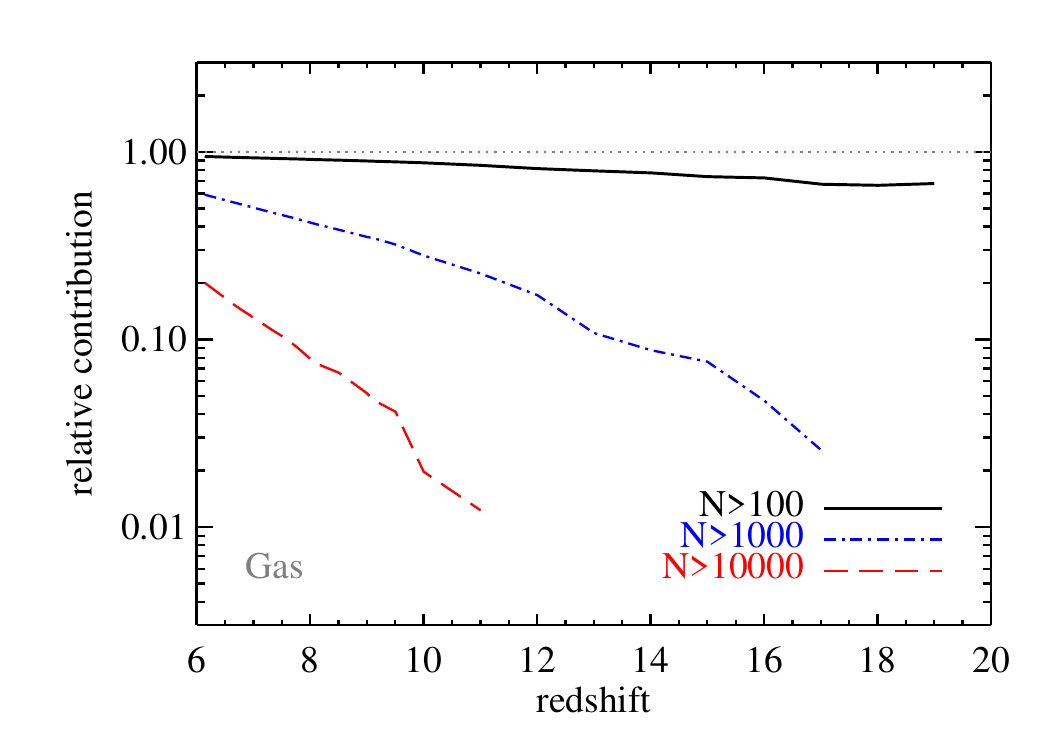}
 \hspace{-0.5cm}
 \includegraphics[width=0.25\textwidth] {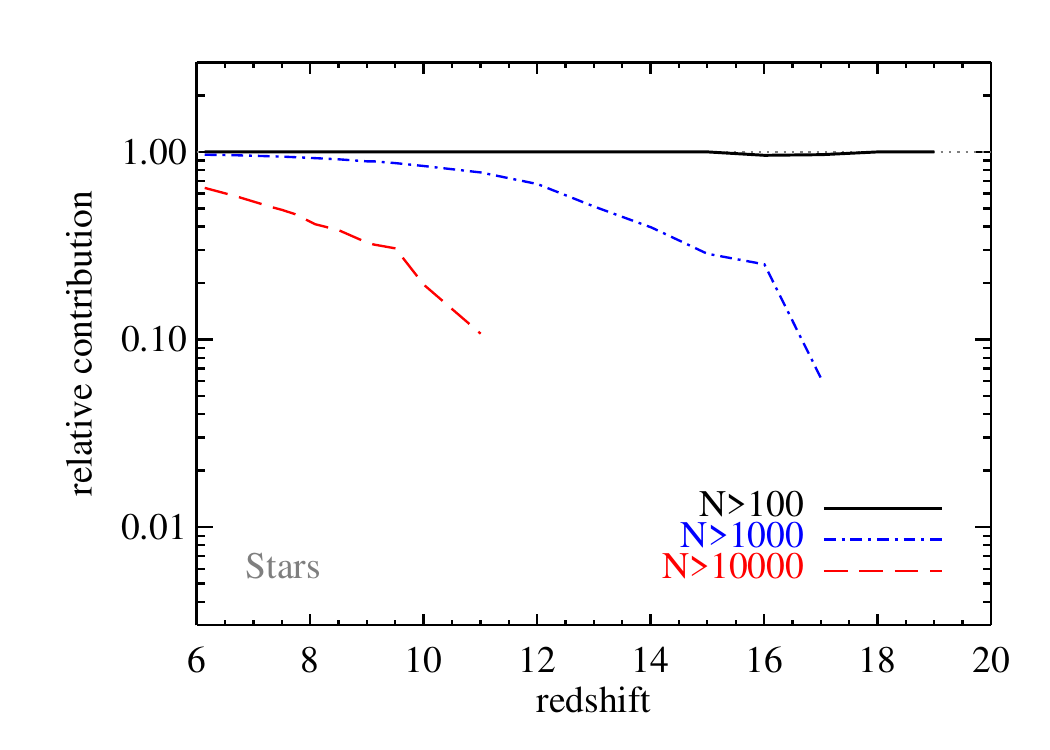}
 \\
 \vspace{-0.3cm} 
 \hspace{-0.7cm}
 \includegraphics[width=0.25\textwidth] {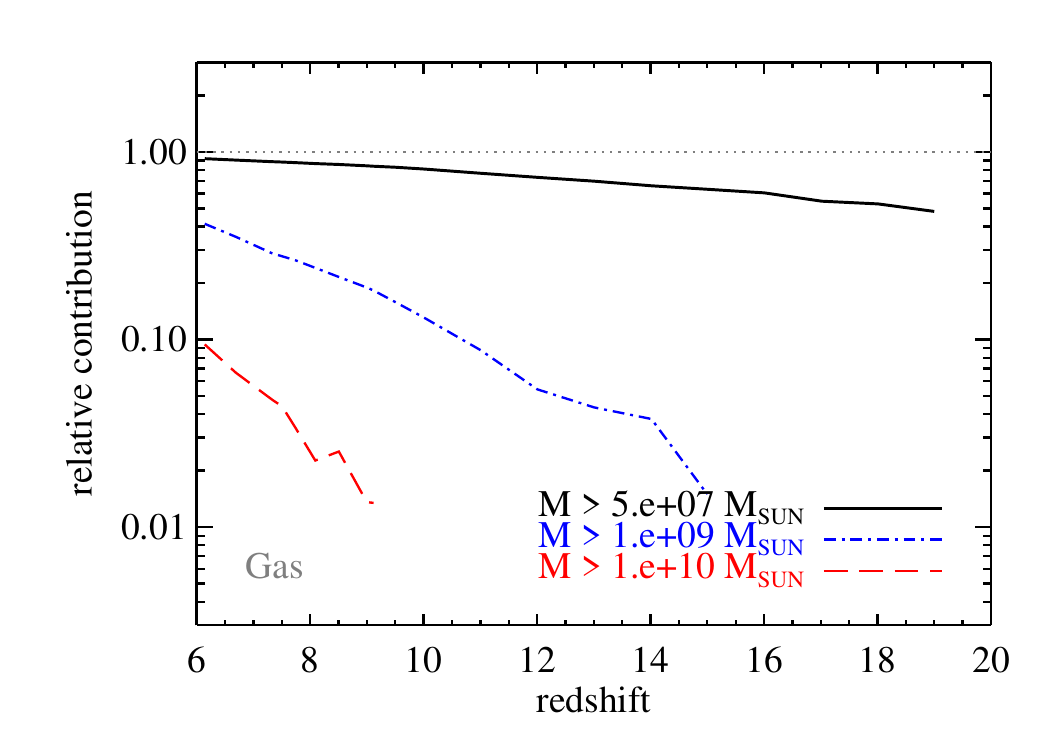}
 \hspace{-0.5cm}
 \includegraphics[width=0.25\textwidth] {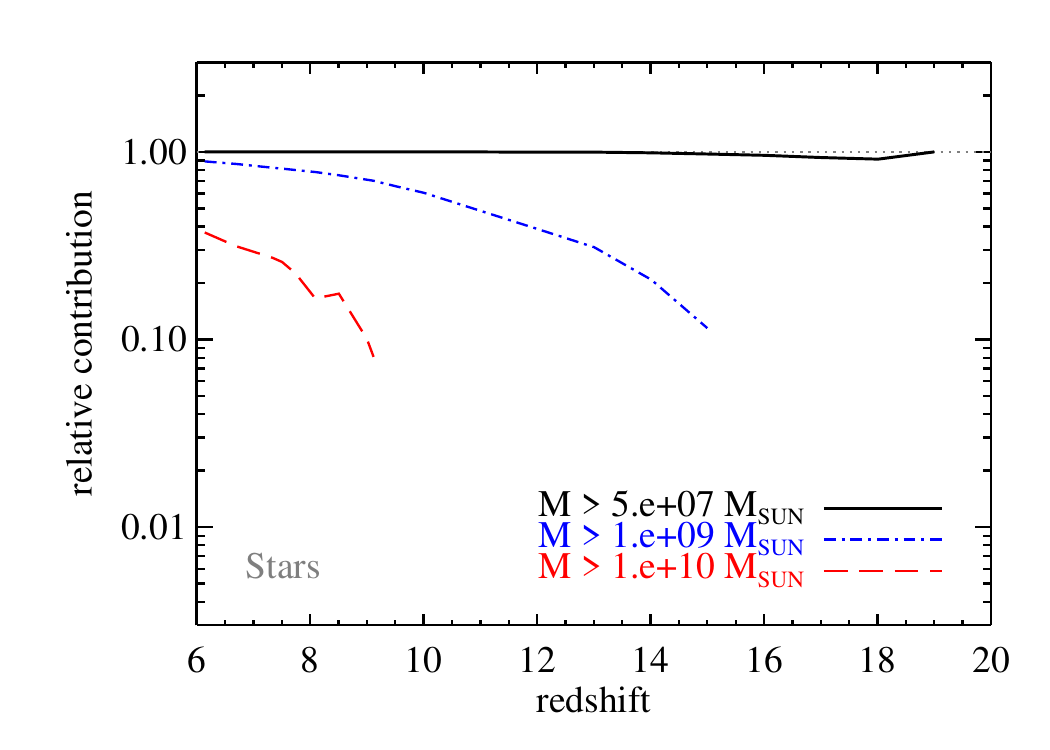}
 \caption{\small Top row: Contribution to the total gas mass (left) and stellar mass (right) of haloes with a number of particles $N>100$ (solid lines), 1~000 (dot-dashed lines), 10~000 (dashed lines) as a function of redshift.
Bottom row: Contribution of haloes with total mass $ M > 5 \times 10^7$ (solid lines), $ 10^9$ (dot-dashed lines) and $ 10^{10} \,\rm M_\odot$ (dashed lines). }
 \label{fig:fofratios}
\end{figure}
\noindent
In Fig.~\ref{fig:haloselection}, we show halo mass functions obtained with different selection criteria.
Total-mass functions and virial-mass functions feature similar trends, with slight differences in the low-mass tail.
When haloes with more than 300 member particles are considered, the low-mass tail is statistically less populated as tiny objects, presumably not well resolved, are excluded.
\\
In Fig.~\ref{fig:fofratios} we show the impacts of selection effects in terms of the relative contribution to the total mass locked in bound objects when different mass or number thresholds are applied.

\section{Main sequence}  \label{appendixMStdepl}
%*****************************************************************************

\begin{figure}
 \centering
 \includegraphics[width=0.5\textwidth] {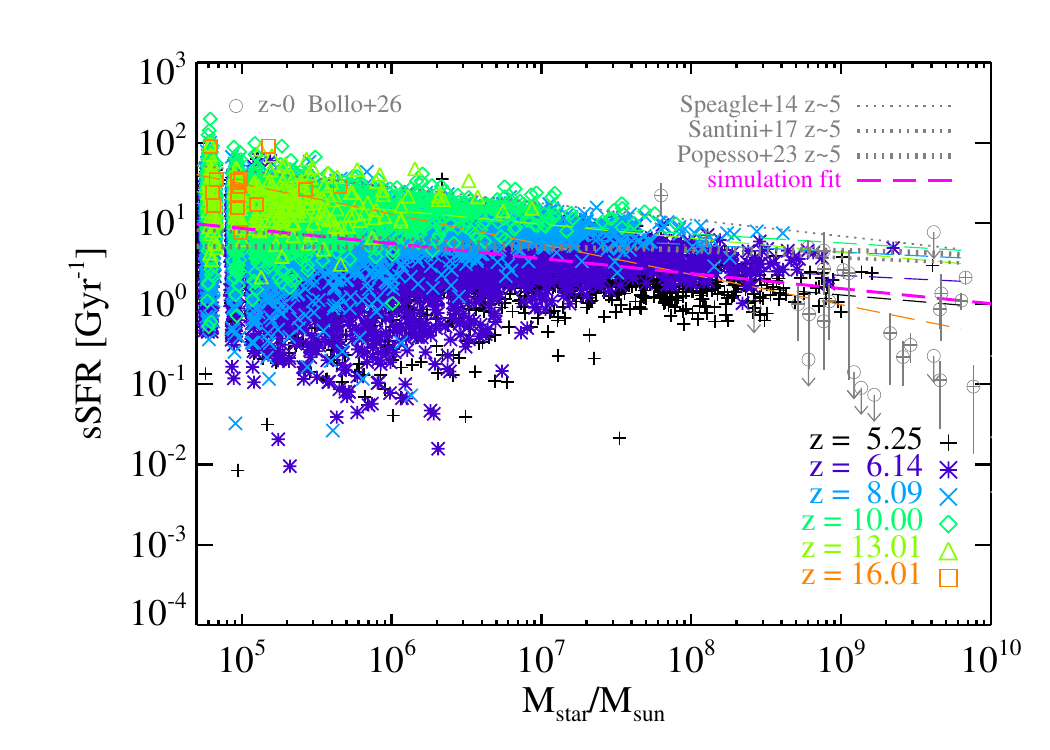} \\
 \vspace{-0.5cm} 
 \includegraphics[width=0.5\textwidth] {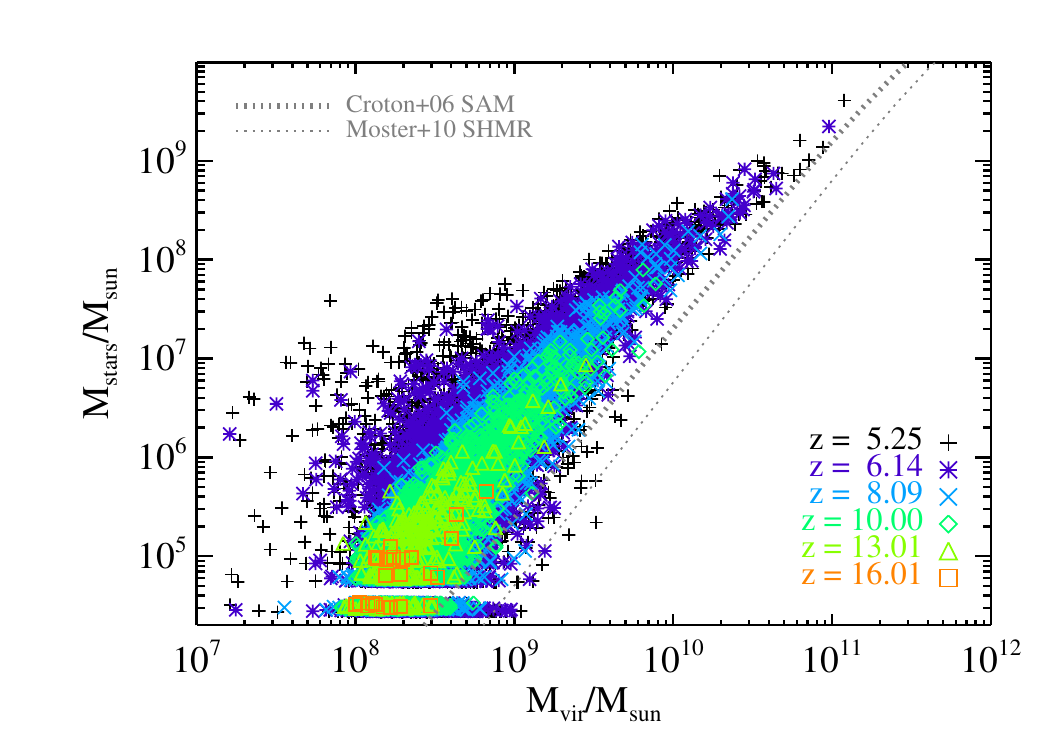}
 \caption{\small  Top: Simulated sSFR and $M_{\rm star}$ at different $z$ (coloured symbols), $z\simeq 5$ observational fits by \cite{Speagle2014}, \cite{Santini2017} and \cite{Popesso2023} (thin, regular and thick dotted lines) and $z\sim 0.5$ MUSE-ALMA determinations (circles) with upper limits (arrows) by \cite{Bollo2026}.
Fits to simulation data at each $z$ (thin dashed lines) are displayed together with the fit to the whole simulated sample (thick dashed magenta line), as per eq.~\ref{eq:MSfit} and Tab.~\ref{tab:fits}: $ {\rm Log} ({ \rm sSFR/Gyr^{-1} })  = 1.871 - 0.187 \, {\rm Log} (M_{\rm star}/{\rm M_\odot})$. Bottom: Simulated $M_{\rm star}$ and $M_{\rm vir}$ and fits from low-$z$ semi-analytical model \cite[][SAM, thick dotted line]{Croton2006} and the stellar-to-halo mass relation \cite[][SHMR, thin dotted line]{Moster2010}. }
 \label{fig:basicmstar}
\end{figure}

With the selection performed in this work (more than 300 particles to resolve the virial mass of haloes), we retrieve the basic relations involving SFR, stellar mass and halo mass at different epochs, as plotted in Fig.~\ref{fig:basicmstar}.
The simulated main sequence at different $z$ (upper panel) can be fitted by eq.~\ref{eq:MSfit} with the values of the parameters $a$ and $b$ listed in Tab.~\ref{tab:fits}.
The values we find are in good agreement with $z\simeq 5$ observational determinations
\cite[][]{Rodighiero2010, Rodighiero2014, Kashino2013, Speagle2014, Schreiber2015, Popesso2023, Herrera2025}.
The simulated relation between stellar mass and halo mass (lower panel) features a generally increasing trend at all $z$.
Recent literature works adopting heterogeneous numerical techniques are in line with the broad scatter we find at low masses \cite[][]{Ceverino2017, Rosdahl2018, Tacchella2018, Behroozi2019, Zhu2020, Stefanon2021}.

\section{Mass fractions and phases}  \label{appendixGasFractions}
%*****************************************************************************

%
\begin{figure*}
 \centering
 \includegraphics[width=0.49\textwidth]{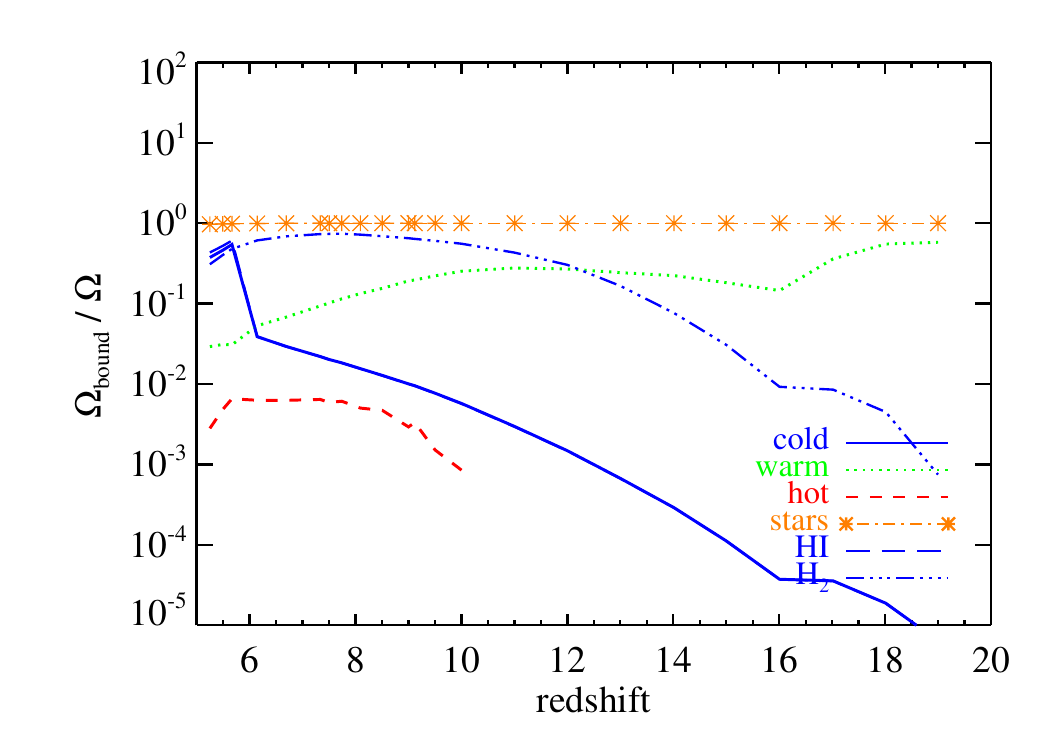}
 \includegraphics[width=0.49\textwidth]{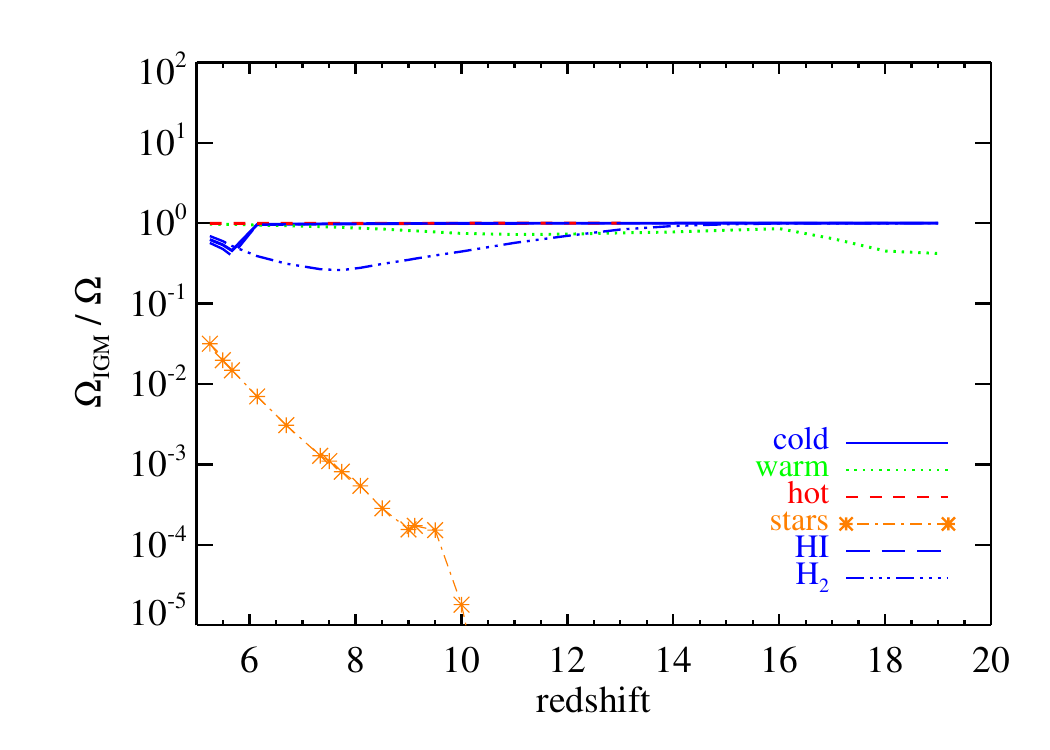} 
 \caption{\small Bound-to-cosmic (left) and IGM-to-cosmic (right) mass density ratio for the cold, warm, hot, stellar, HI and H$_2$ baryon phases. }
 \label{fig:omegas2ratios}
\end{figure*}
\begin{figure}
 \centering
 \includegraphics[width=0.5\textwidth]{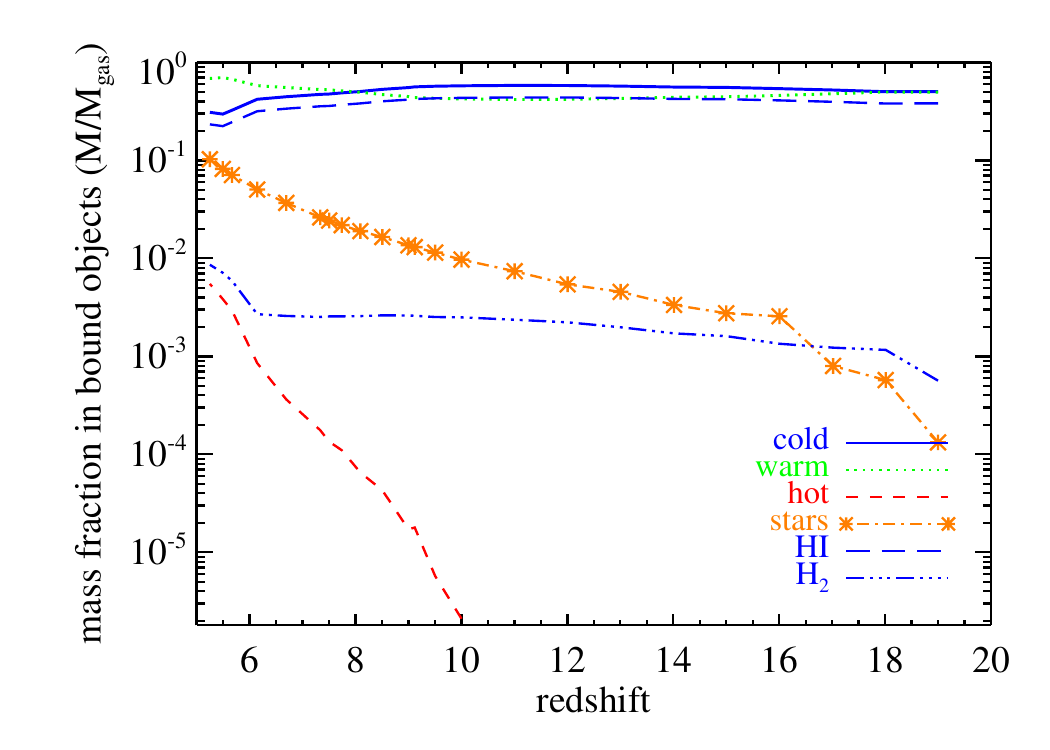}
 \caption{\small Average mass fraction of each baryon phase (cold, warm, hot, stars, HI, H$_2$) identified in bound objects at different redshifts.}
 \label{fig:phaseevolution}
\end{figure}
\begin{figure*}
 \centering
 \hspace{-0.3cm}
 \includegraphics[width=0.25\textwidth] {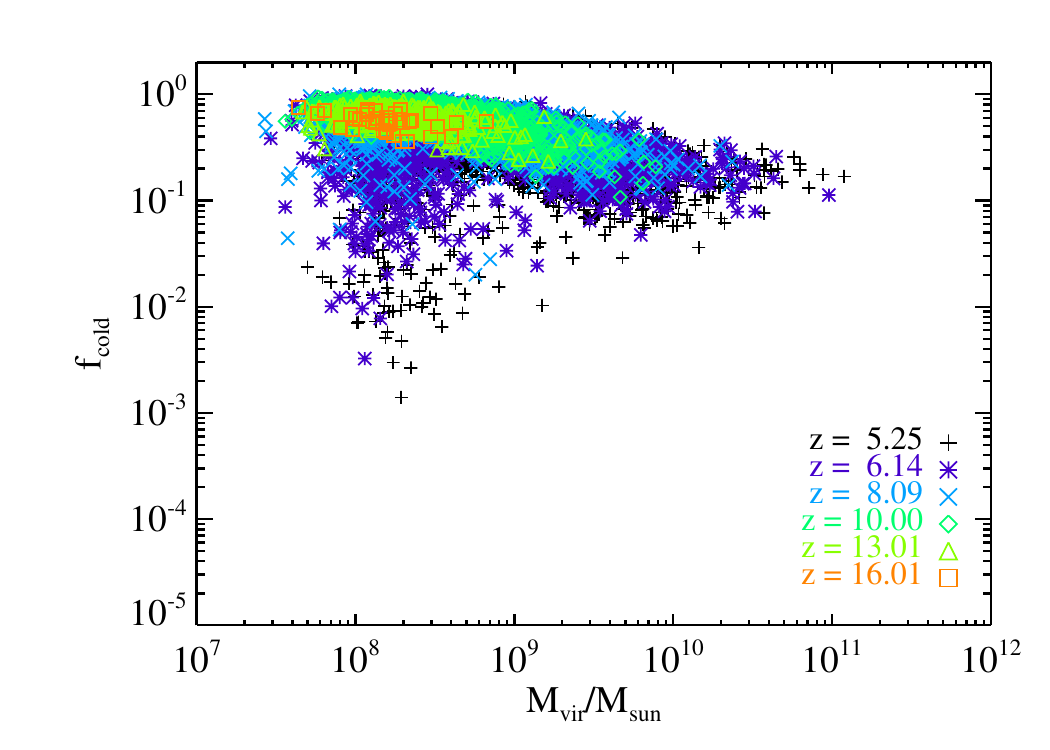}
 \hspace{-0.7cm}
 \includegraphics[width=0.25\textwidth] {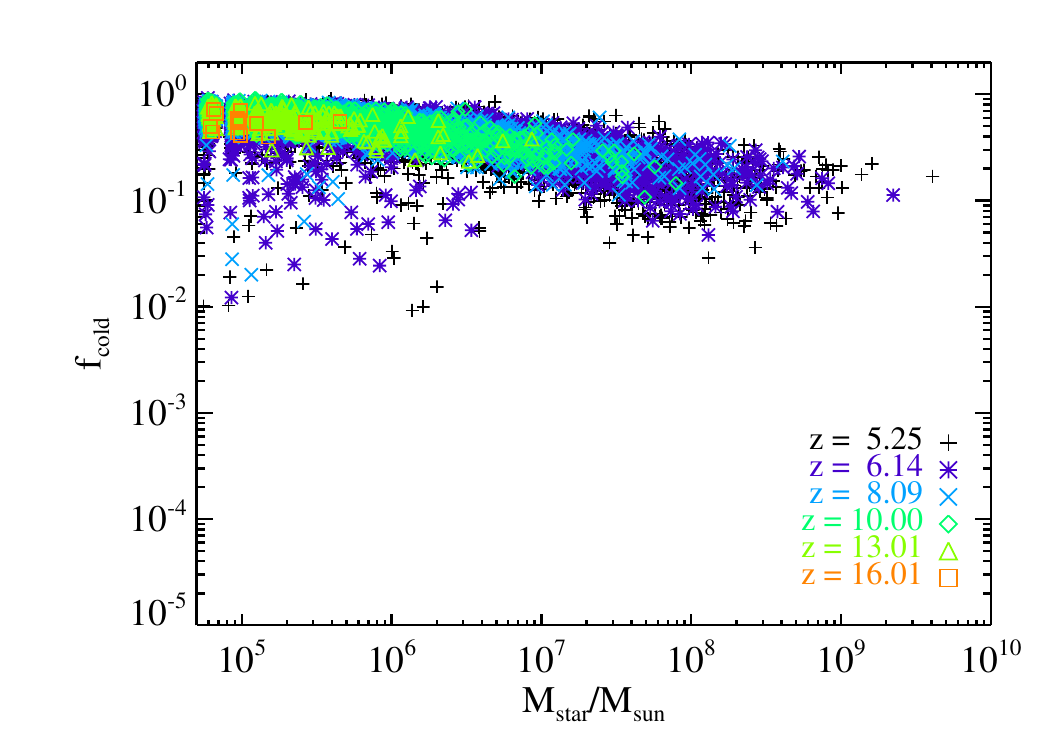}
 \hspace{-0.7cm}
 \includegraphics[width=0.25\textwidth] {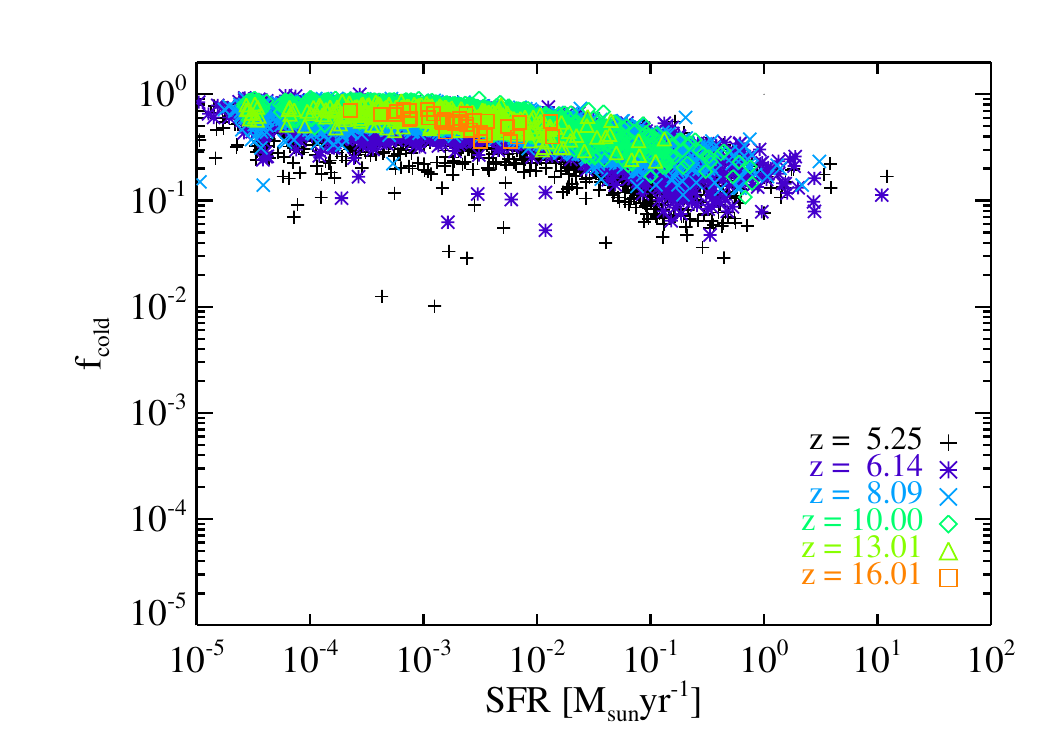}
 \hspace{-0.7cm}
 \includegraphics[width=0.25\textwidth] {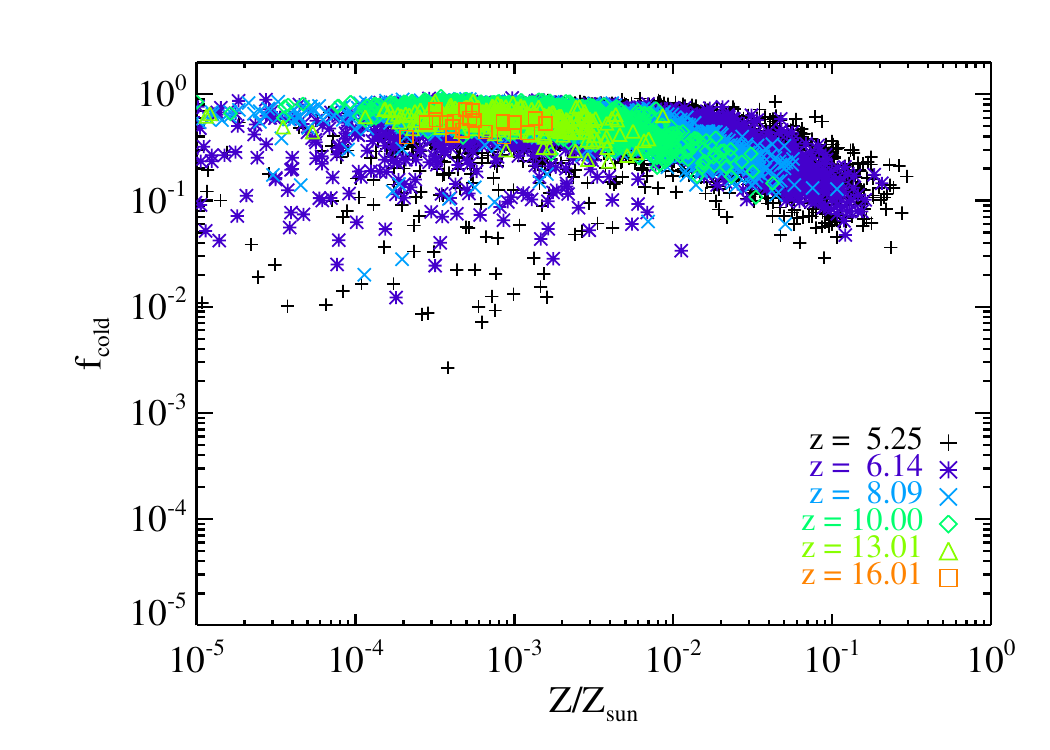}
 \\
 \vspace{-0.3cm} 
 \hspace{-0.5cm}
 \includegraphics[width=0.25\textwidth] {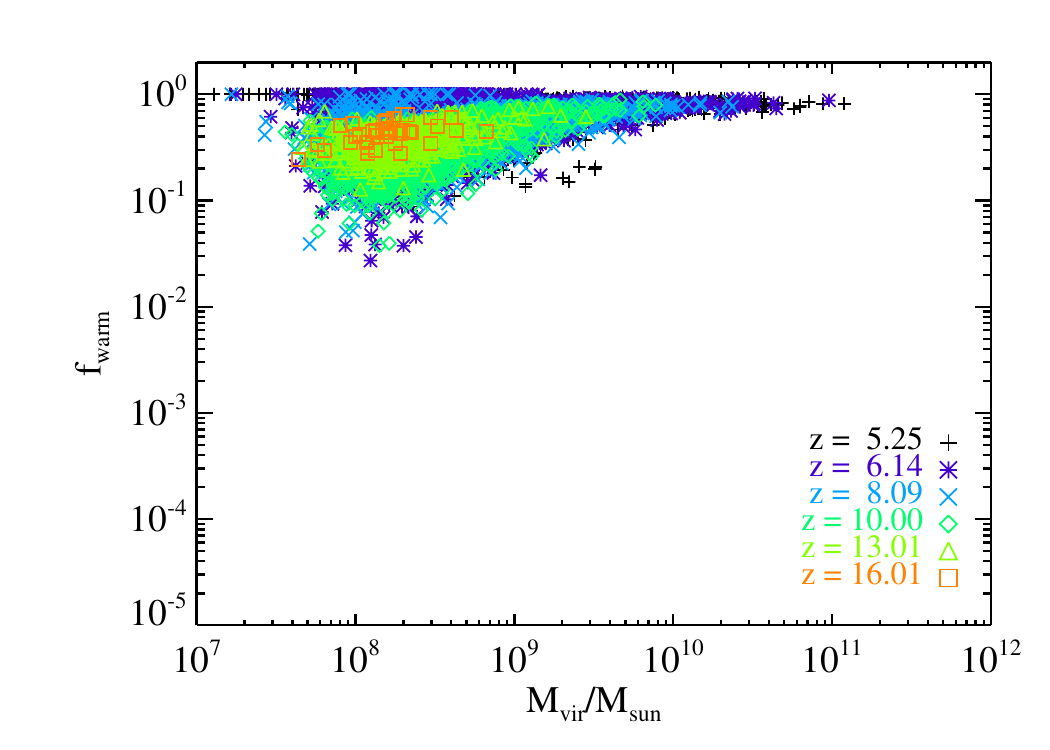}
 \hspace{-0.7cm}
 \includegraphics[width=0.25\textwidth] {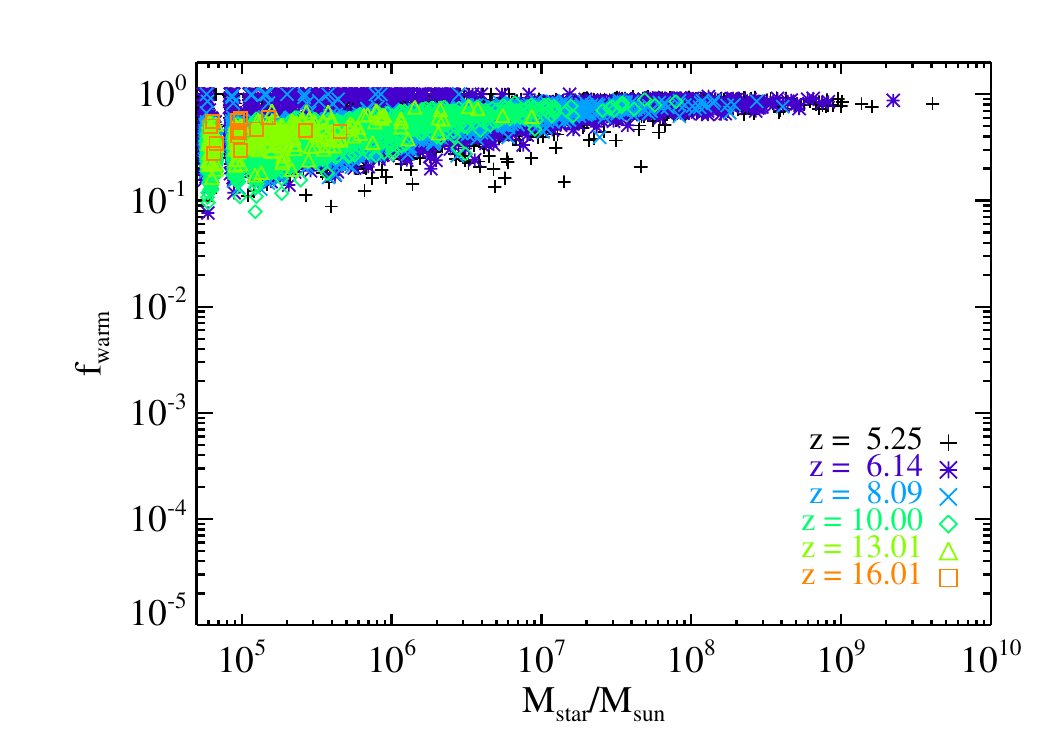}
 \hspace{-0.7cm}
 \includegraphics[width=0.25\textwidth] {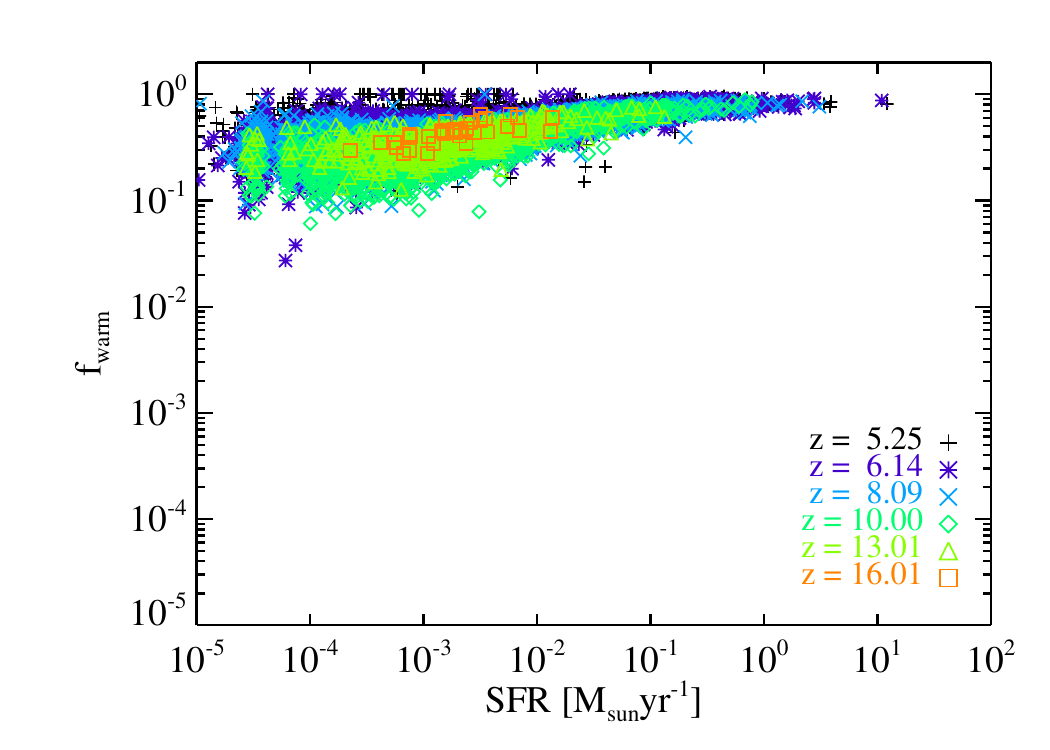}
 \hspace{-0.7cm}
 \includegraphics[width=0.25\textwidth] {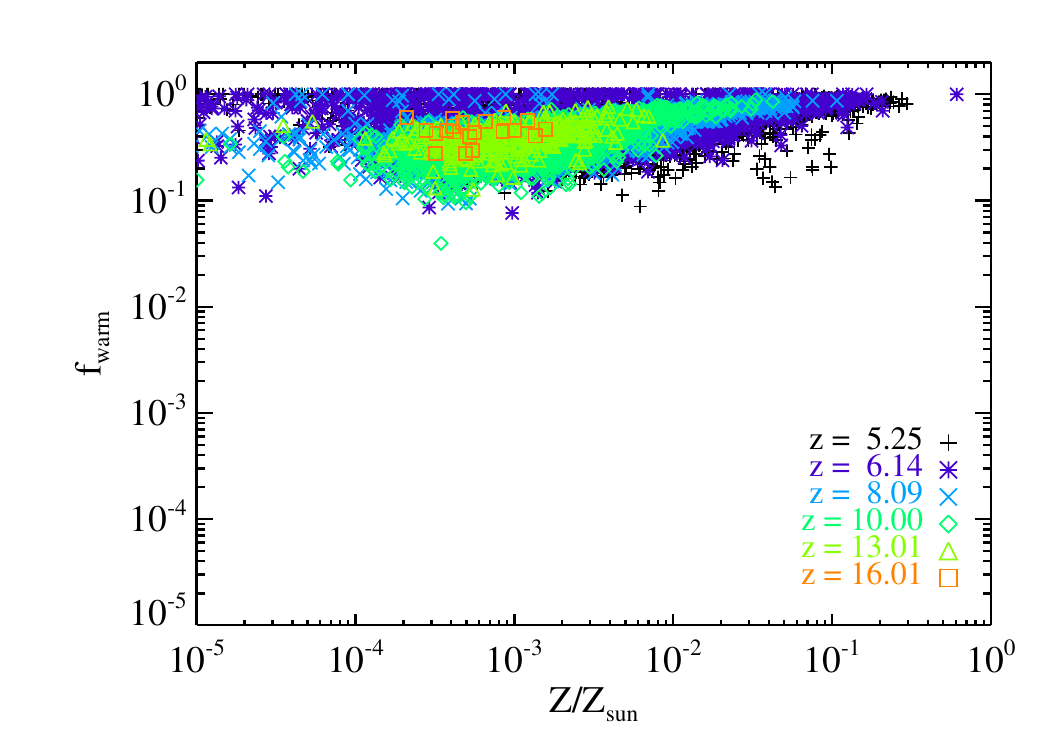}
 \\
 \vspace{-0.3cm} 
 \hspace{-0.5cm}
 \includegraphics[width=0.25\textwidth] {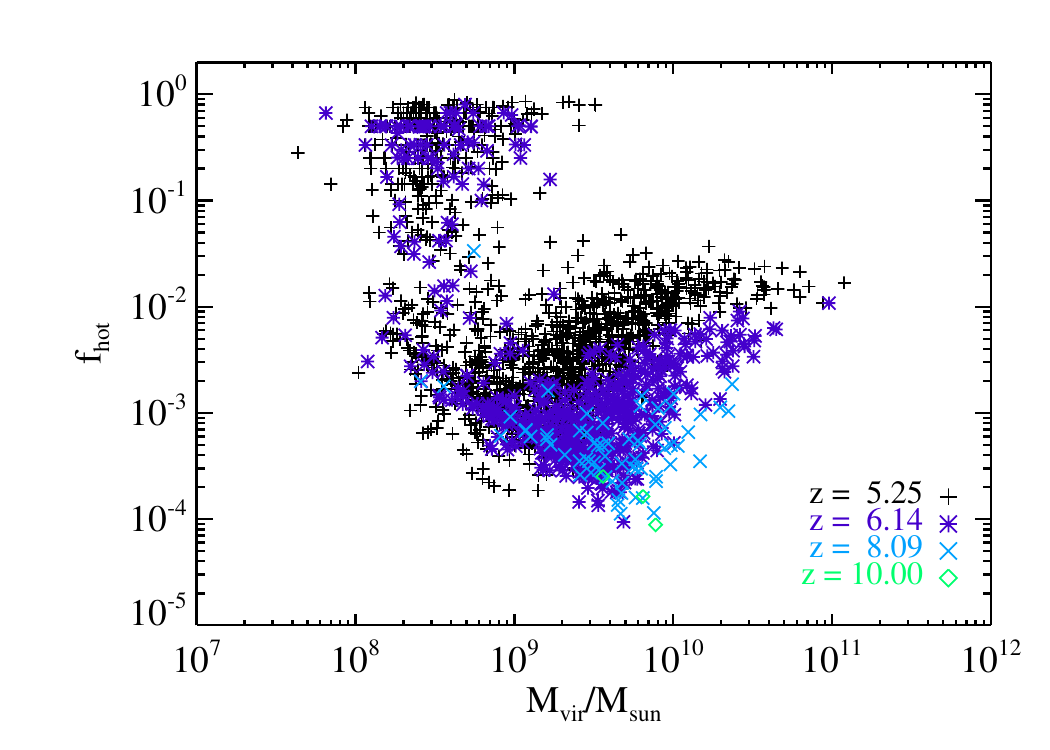}
 \hspace{-0.7cm}
 \includegraphics[width=0.25\textwidth] {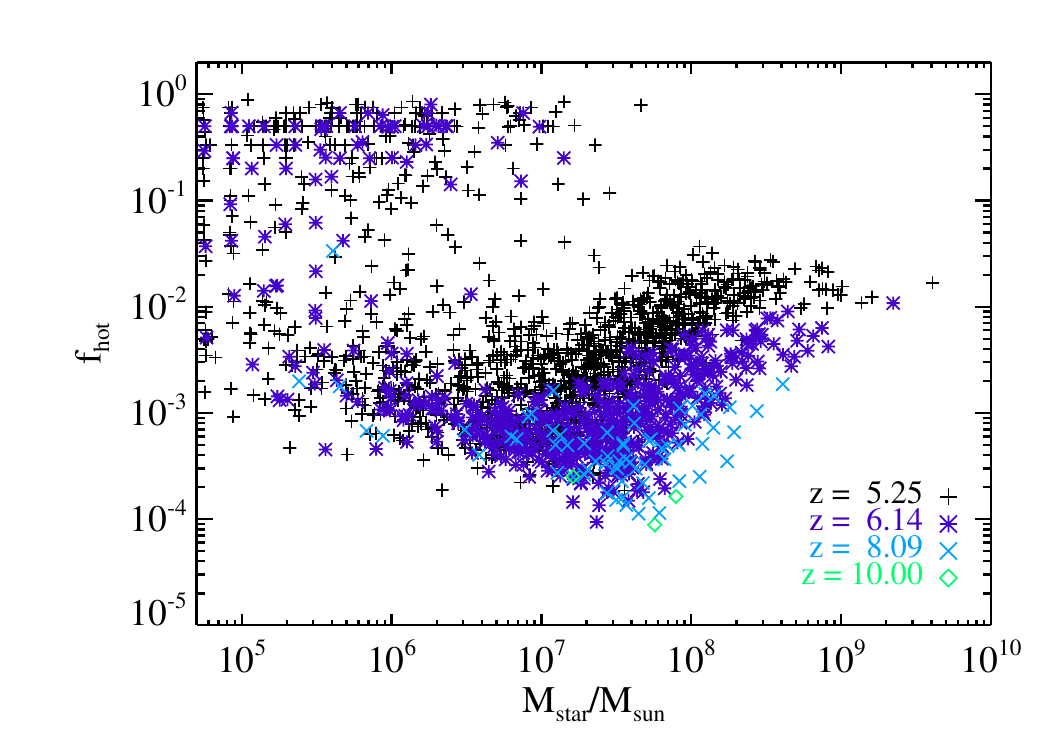}
 \hspace{-0.7cm}
 \includegraphics[width=0.25\textwidth] {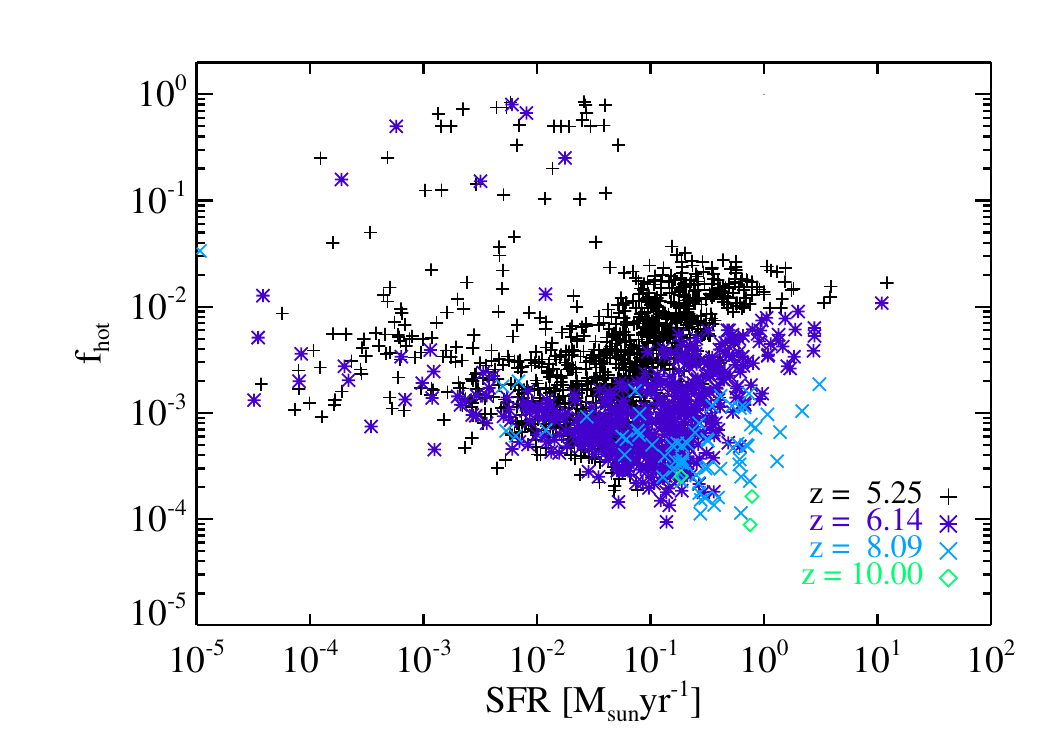} 
 \hspace{-0.7cm}
 \includegraphics[width=0.25\textwidth] {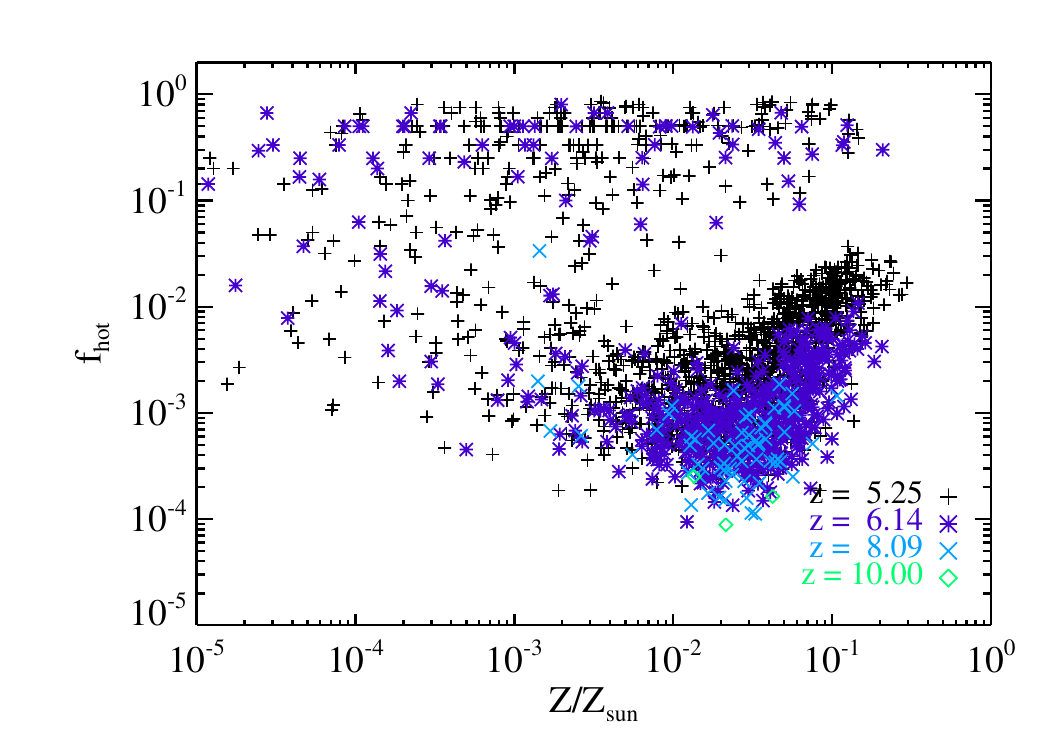} 
 \caption{\small Cold (top row), warm (central row) and hot (bottom row) gas fraction versus 
 virial mass (first column), stellar mass (second column), SFR (third column) and gas metallicity (fourth column) at $z = $5.25, 6.14, 8.09, 10.00, 13.01, and 16.01. }
 \label{fig:phasefraction}
\end{figure*}
\begin{figure*}
 \centering
 \hspace{-0.5cm}
 \includegraphics[width=0.25\textwidth] {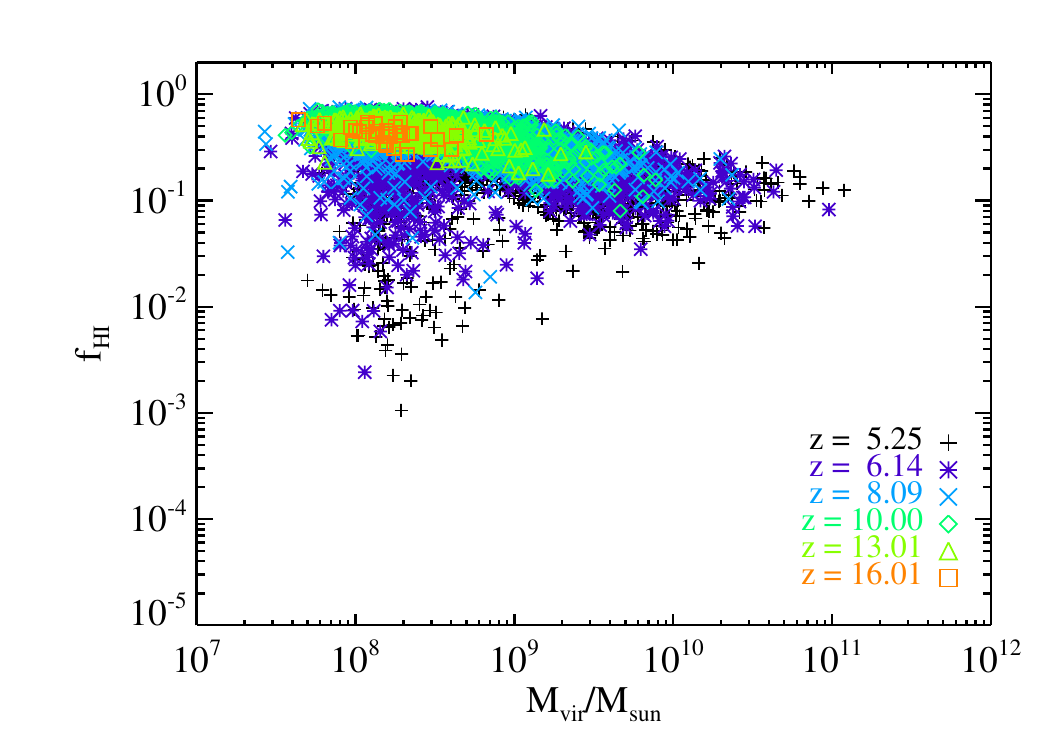}
 \hspace{-0.7cm}
 \includegraphics[width=0.25\textwidth] {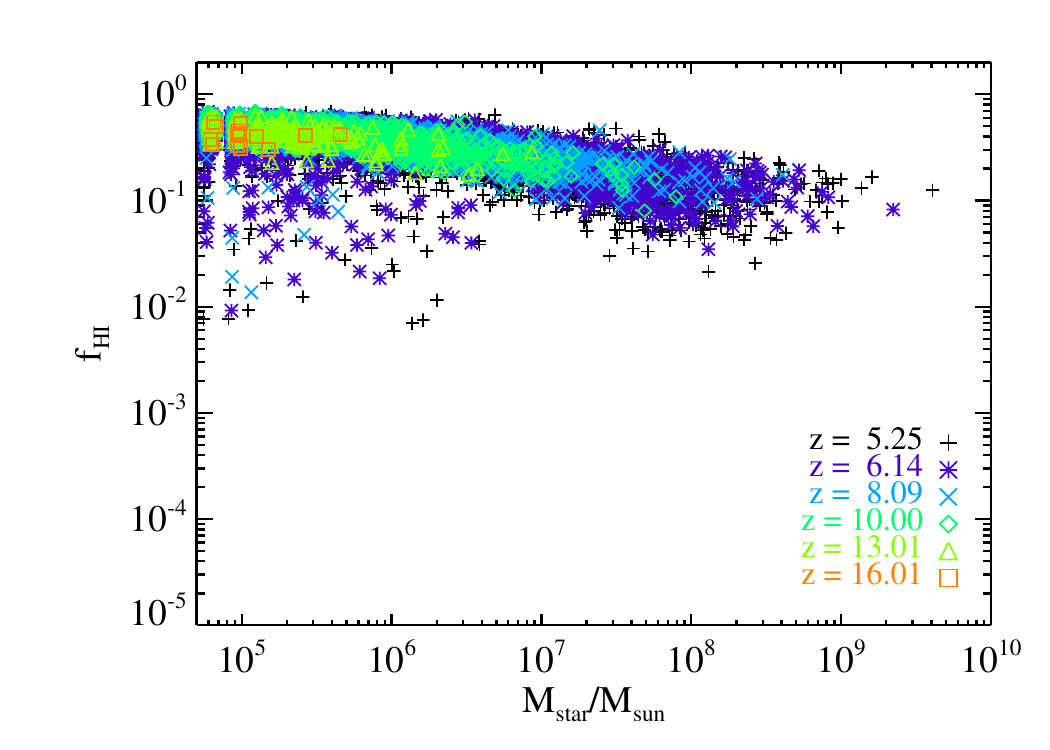}
 \hspace{-0.7cm}
 \includegraphics[width=0.25\textwidth] {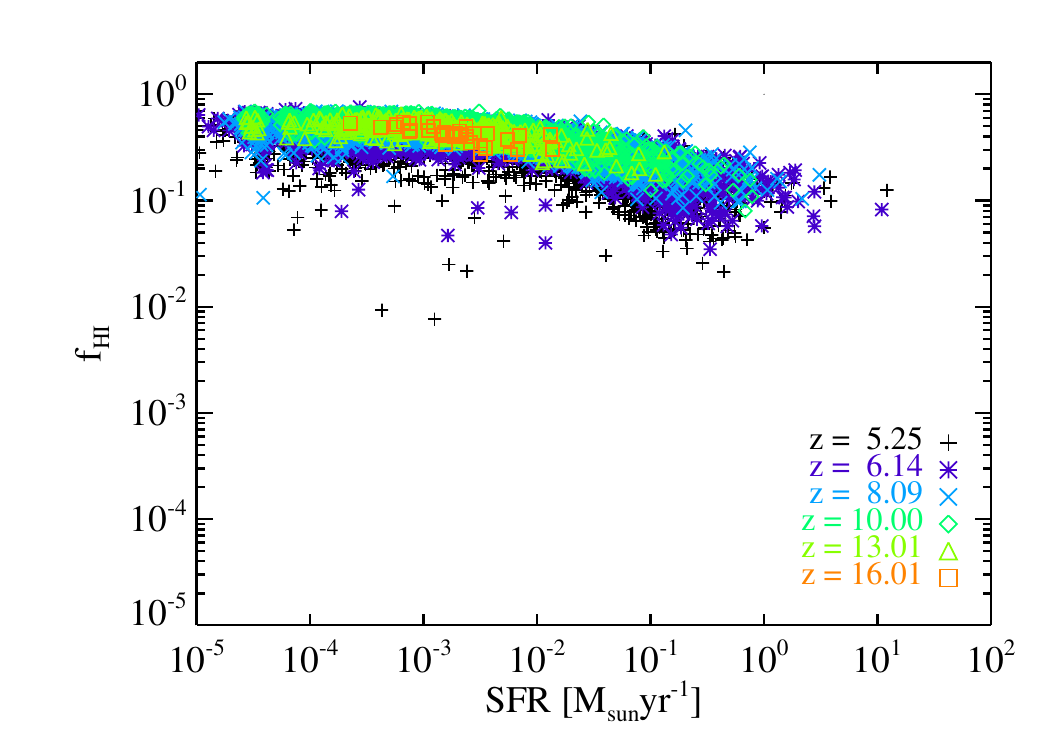}
 \hspace{-0.7cm}
 \includegraphics[width=0.25\textwidth] {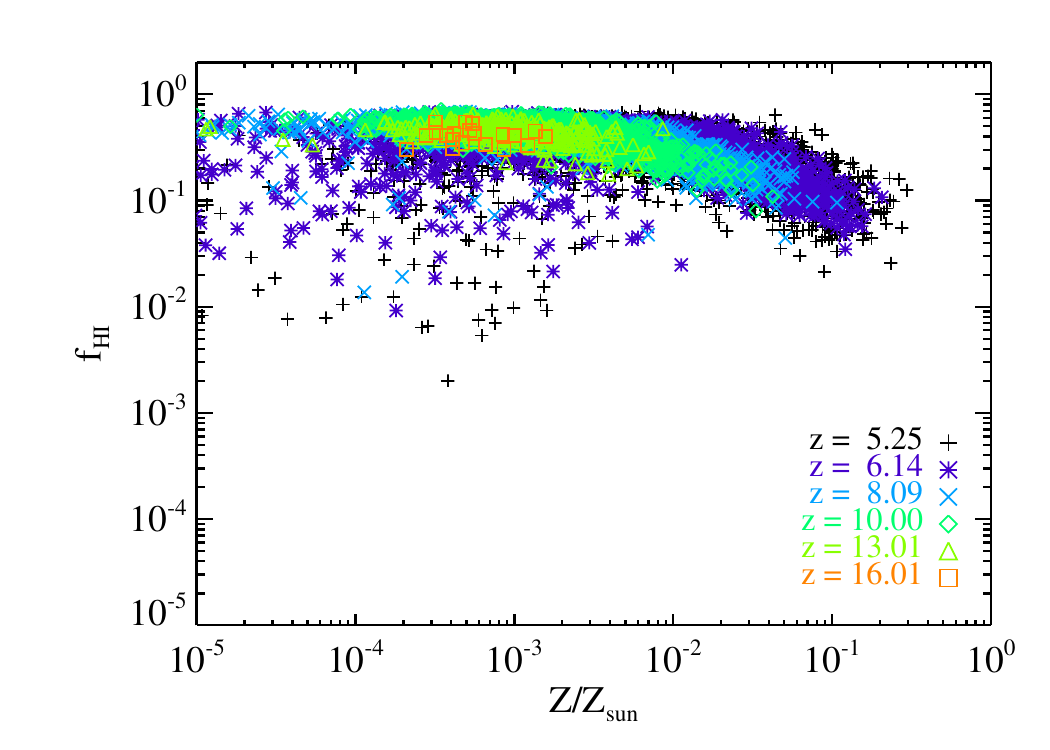}
\\
 \vspace{-0.3cm} 
 \hspace{-0.5cm}
 \includegraphics[width=0.25\textwidth] {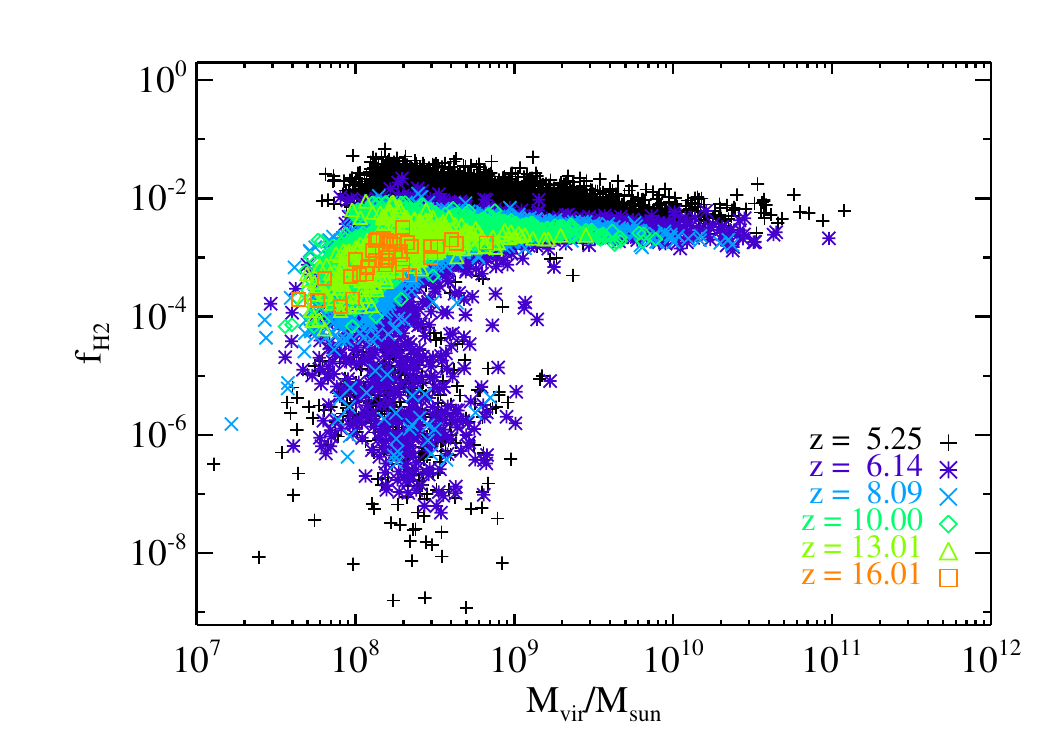}
 \hspace{-0.7cm}
 \includegraphics[width=0.25\textwidth] {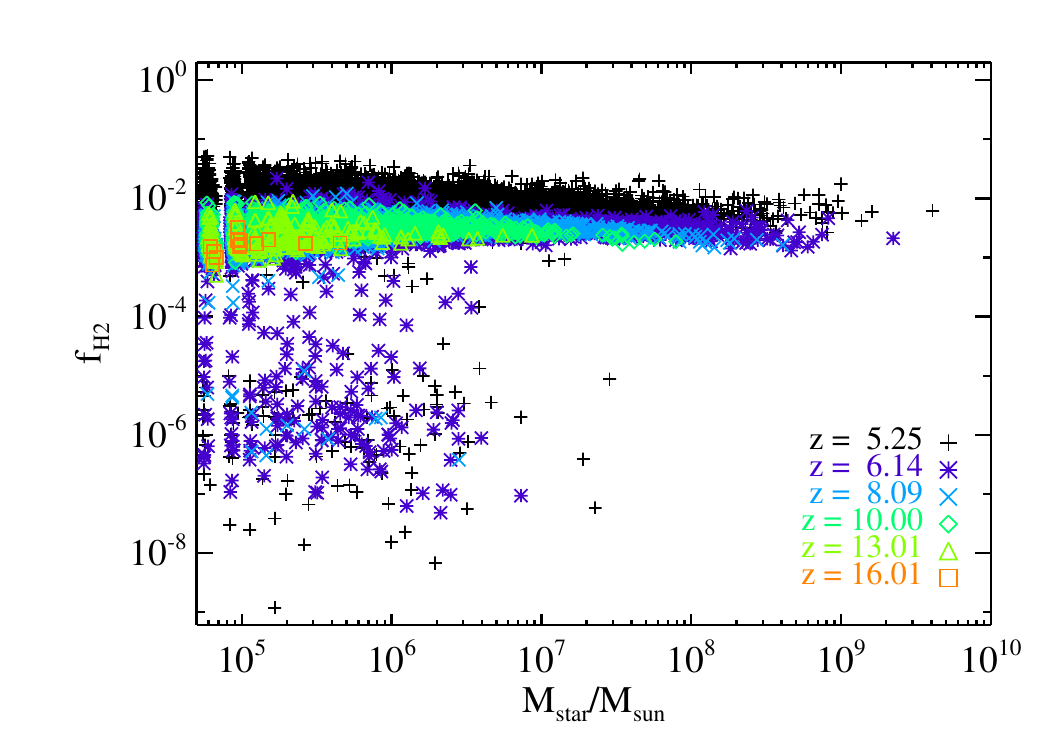}
 \hspace{-0.7cm}
 \includegraphics[width=0.25\textwidth] {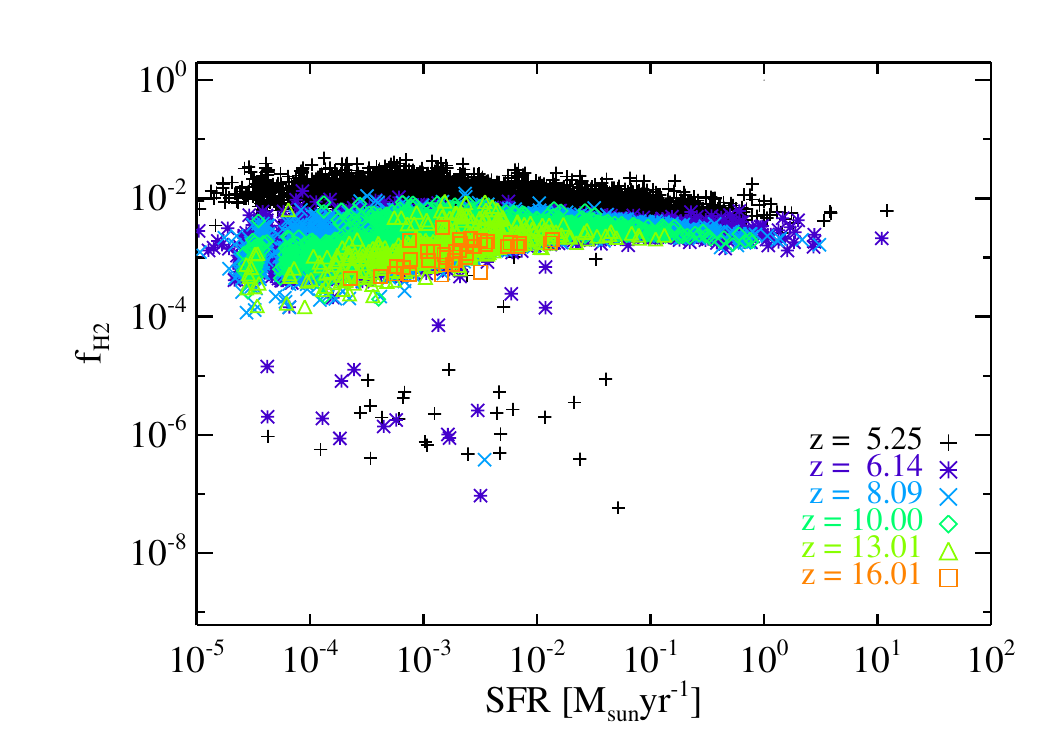}
 \hspace{-0.7cm}
 \includegraphics[width=0.25\textwidth] {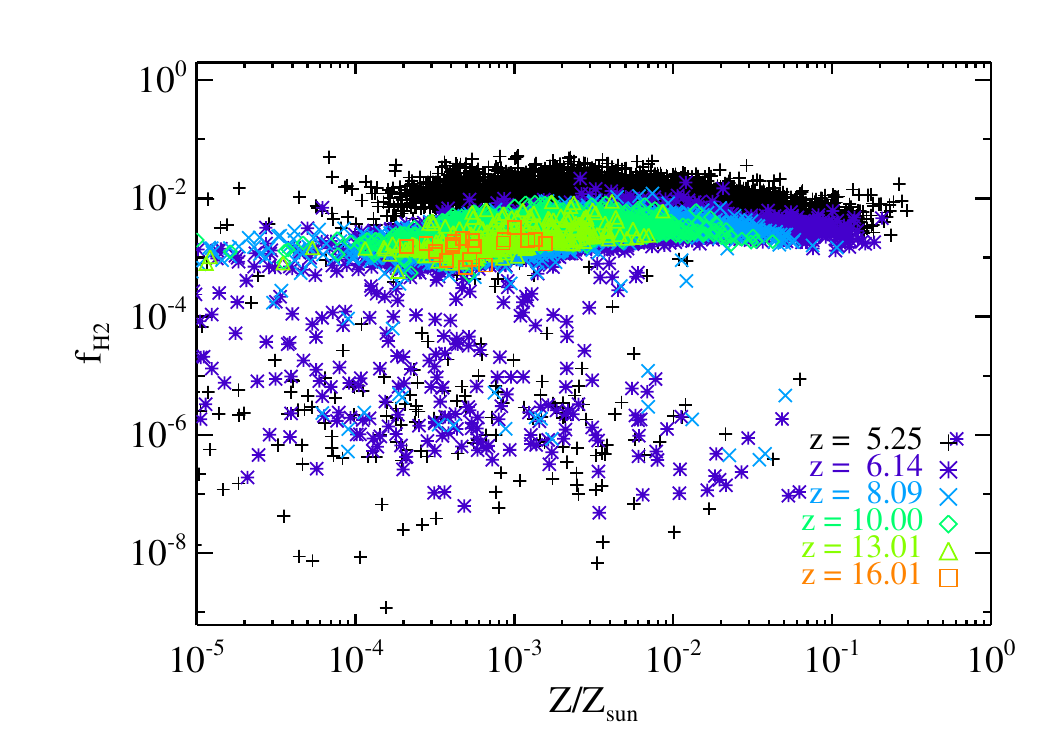}
 \caption{\small HI (top) and H$_2$ (bottom) mass fractions versus virial mass, stellar mass, SFR and gas metallicity, respectively from left to right, at $z = $5.25, 6.14, 8.09, 10.00, 13.01, and 16.01. }
 \label{fig:phasefraction2}
\end{figure*}
\begin{figure*}
 \centering
 \hspace{-0.5cm}
  \includegraphics[width=0.25\textwidth] {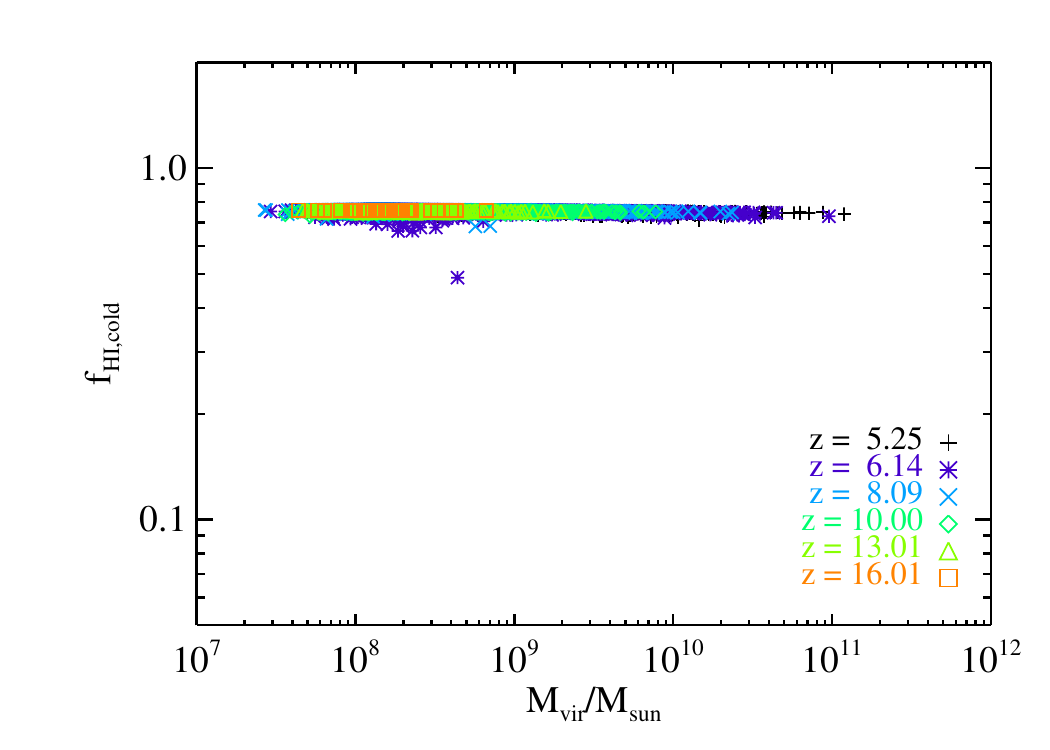}
 \hspace{-0.7cm}
 \includegraphics[width=0.25\textwidth] {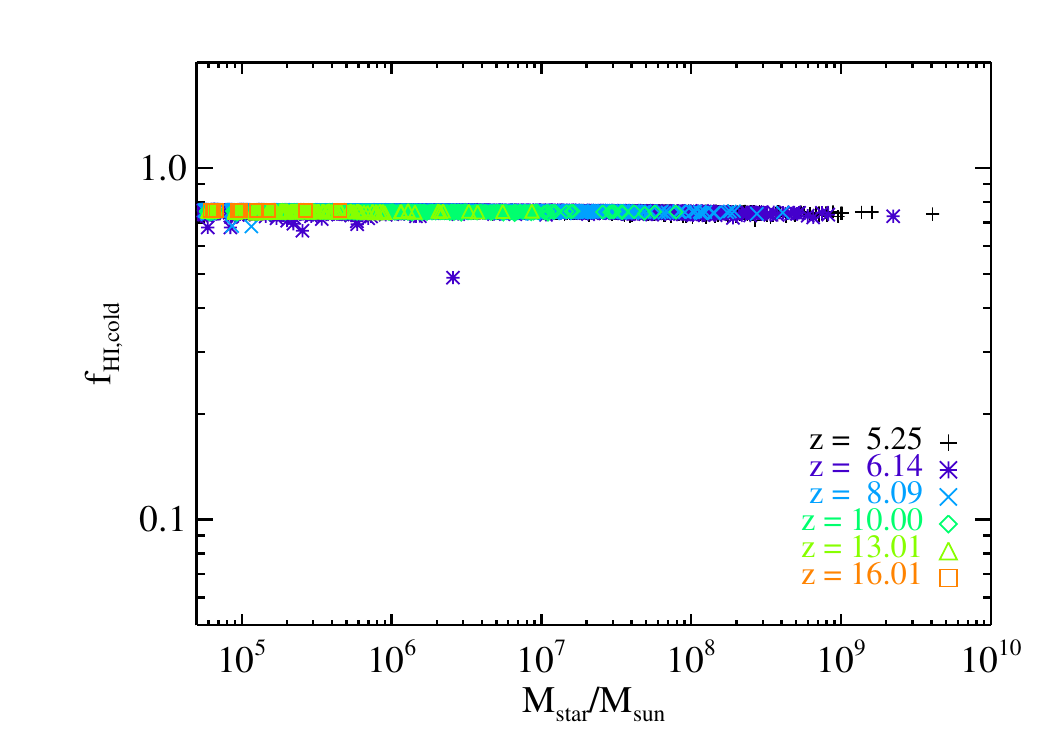}
 \hspace{-0.7cm}
 \includegraphics[width=0.25\textwidth] {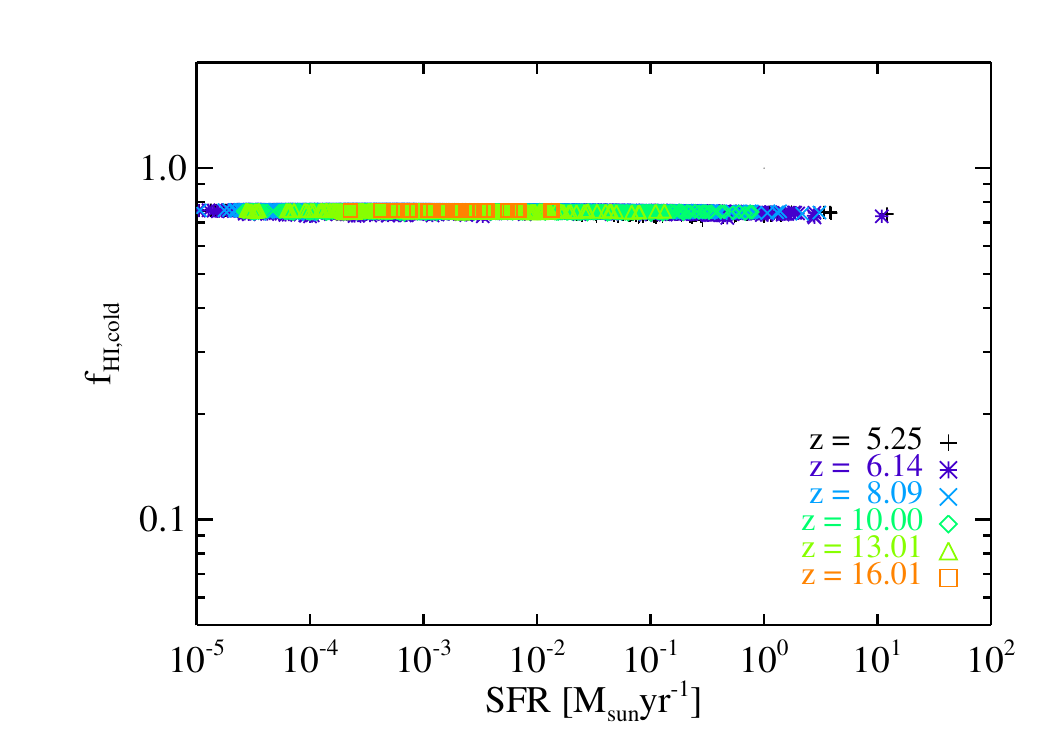}
 \hspace{-0.7cm}
 \includegraphics[width=0.25\textwidth] {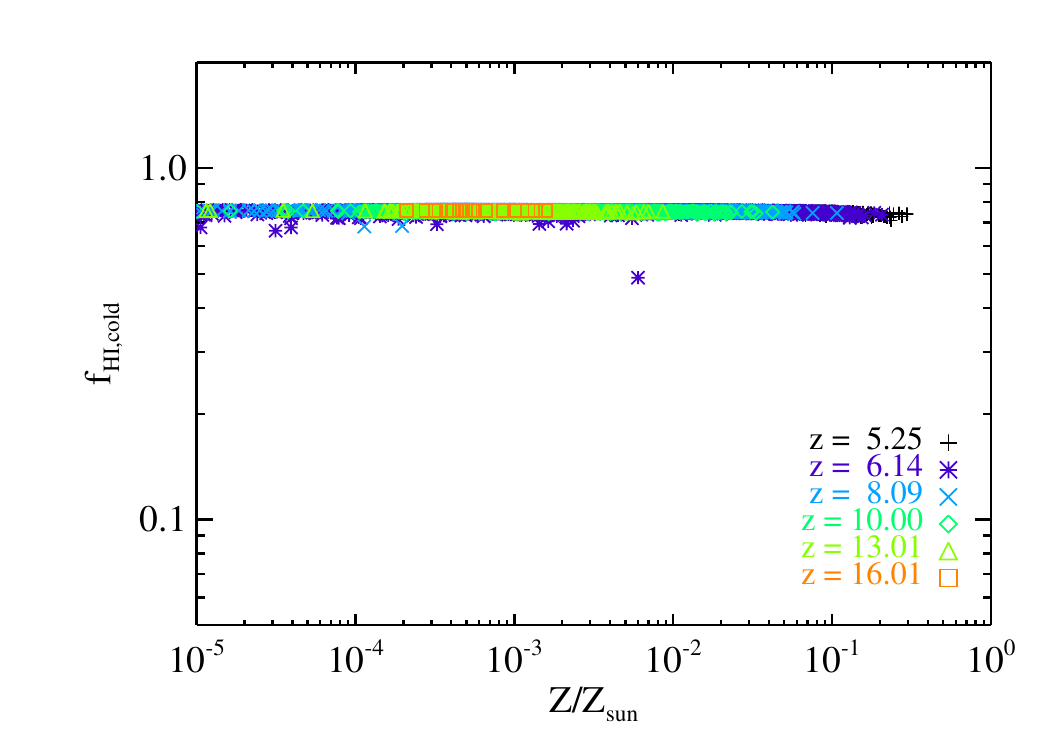} 
\\
 \vspace{-0.3cm} 
 \hspace{-0.5cm}
 \includegraphics[width=0.25\textwidth] {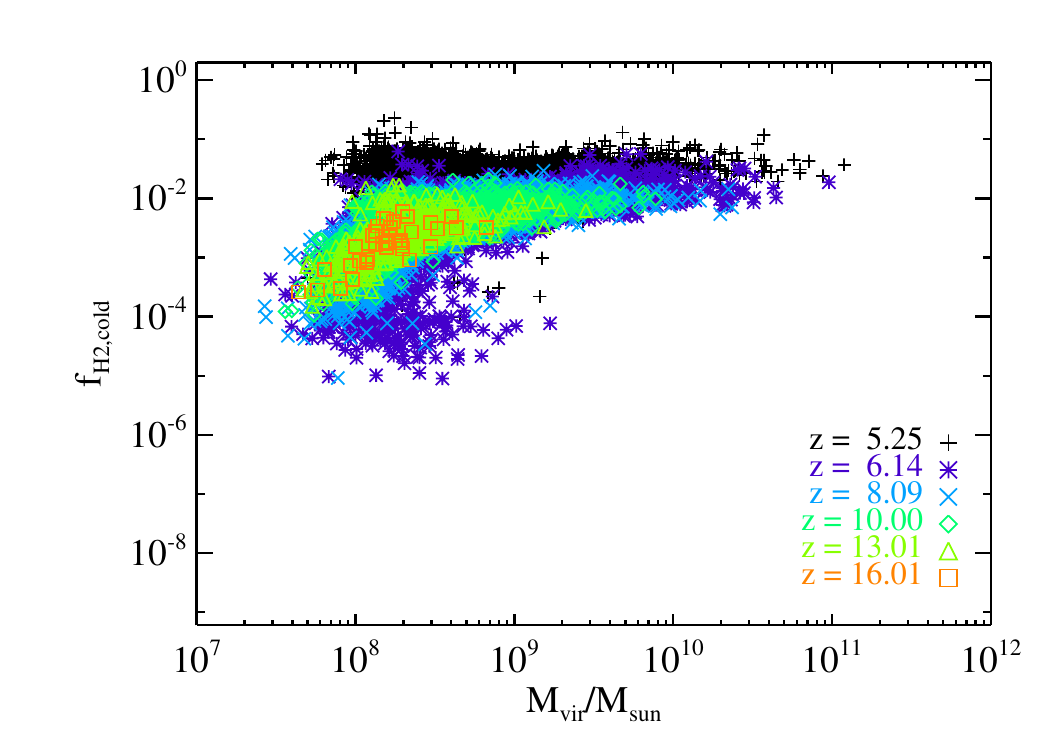}
 \hspace{-0.7cm}
 \includegraphics[width=0.25\textwidth] {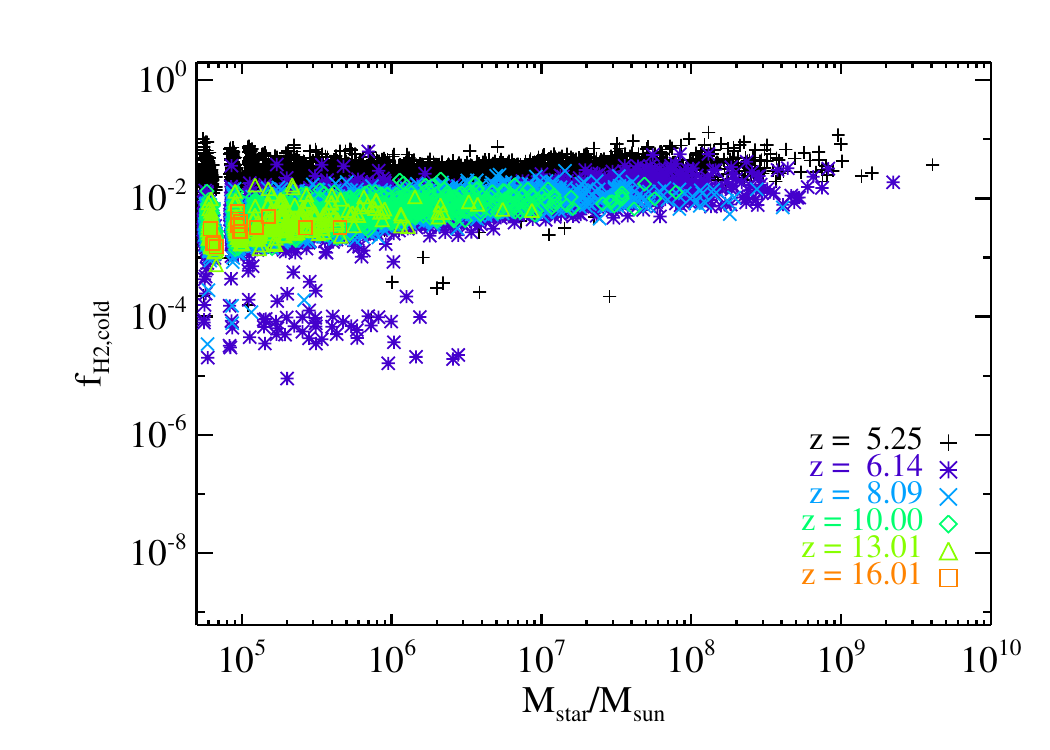}
 \hspace{-0.7cm}
 \includegraphics[width=0.25\textwidth] {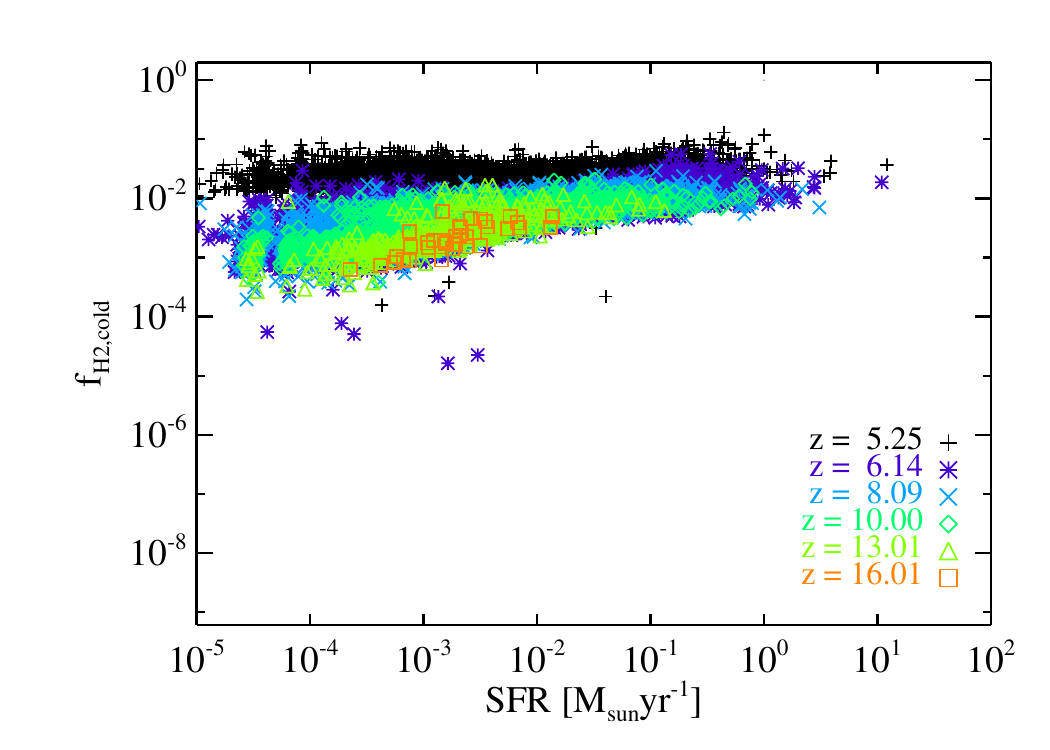}
 \hspace{-0.7cm}
 \includegraphics[width=0.25\textwidth] {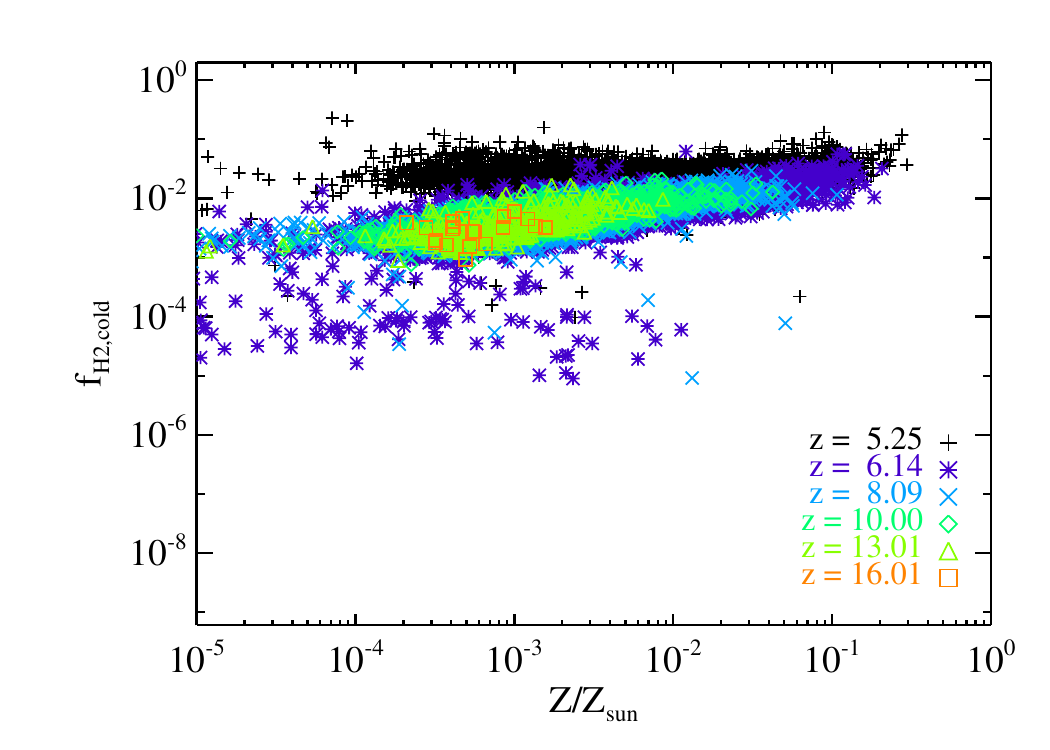} 
 \caption{\small HI (top) and H$_2$ (bottom) cold-mass fractions versus virial mass, stellar mass, SFR and gas metallicity, respectively from left to right, at $z = $5.25, 6.14, 8.09, 10.00, 13.01, and 16.01. }
 \label{fig:phasefraction3}
\end{figure*}

In the left panel of Fig.~\ref{fig:omegas2ratios} the ratios between the bound-object mass density parameters and the cosmic mass density parameters are shown for each baryon phase. Similarly, results for the IGM-to-cosmic ratios are shown in the right panel.
At early times, most of the cosmic gas is in the cold (HI) phase, but, due to the rarity of primordial objects, cold bound gas accounts for only $ \Omega_{\rm bound, cold} / \Omega_{\rm cold} \lesssim 10^{-4} $ ($z\simeq 18$).
The amount of warm cosmic gas is tiny in comparison to cold cosmic gas (Fig.~\ref{fig:omegas}) and is typically bound ($ \Omega_{\rm bound, warm} / \Omega_{\rm warm} > 50 \% $ at $z\simeq 18$).
While cosmic evolution proceeds, more and more gas experiences cooling, star formation and feedback effects in growing galaxies.
These processes boost $ \Omega_{\rm bound, H_2} /\Omega_{\rm H_2} $ up to $\sim$0.8, lower $ \Omega_{\rm IGM, H_2} /\Omega_{\rm H_2} $ down to  $\sim$0.2 and lead to $  \Omega_{\rm bound, hot} / \Omega_{\rm hot} 
\sim  0.1$--$1 \%$ at $z\simeq 6$--10.
\\
In Fig.~\ref{fig:phaseevolution}, we display the mean redshift evolution of the bound-object mass fractions for the cold, warm, hot, stellar, HI and H$_2$ baryon phase, computed as
$ < f_{\rm phase} >  = M_{\rm phase} /  M_{\rm gas} = \Omega_{\rm bound, phase} /   \left( \Omega_{\rm bound, cold} + \Omega_{\rm bound, warm} + \Omega_{\rm bound, hot} \right) $,
where, $M_{\rm phase} $ and $M_{\rm gas} $ are the total bound masses in the considered phase and in the gas, respectively.
Clearly, the cold and warm phases dominate the halo mass content at $z \gtrsim 6$ and $z\lesssim 6$, respectively, and the cold and warm fractions lie around 0.3 and 0.7 at $z\simeq 5$.
Hot gas is always subdominant, with average fractions that increase from $ \sim 10^{-6}$ at $z\simeq 10$ to about $5\times 10^{-3}$ at $z\simeq 5$, while star fractions follow star formation activity increasing from $\sim$$10^{-4}$ at $z\simeq 19$ to 0.1 at $z\simeq 5$.
HI and H$_2$ phases feature trends that are almost flat with average H$_2$-to-HI ratios $\sim$0.001-0.05.
\\
Cold, warm and hot simulated gas fractions at different $z$ as functions of virial mass, stellar mass, SFR and gas $Z$ are displayed in Fig.~\ref{fig:phasefraction}, while the trends for HI and H$_2$ gas fractions are presented in Fig.~\ref{fig:phasefraction2}.
These quantities are theoretically consistent, but they are hard to be proven observationally, since detecting all the gaseous components and estimating the total gas mass in galaxies is challenging.
So, in Fig.~\ref{fig:phasefraction3}, we normalise the HI and H$_2$ fractions by accounting for the cold component only, i.e.:
$f_{\rm HI, cold} = M_{\rm HI } / M_{\rm cold} $ 
and 
$f_{\rm H_2, cold} = M_{\rm H_2 } / M_{\rm cold} $.
These quantities are closer to observations when HI and/or H$_2$ can be considered good tracers of the gas budget \cite[e.g.][]{Bacchini2019, Bacchini2024}.
The figures show $f_{\rm HI, cold} \simeq 0.75$ for all the galaxy populations sampled at different $z$, while $f_{\rm H_2,cold} $ reaches values up to $\sim 0.2$--0.3, i.e. some dex higher than $f_{\rm H_2}$.
These results are consequences of the fact that HI is only present in the cold phase, where most H$_2$ is formed.
In general, H$_2$ molecules are susceptible to non-equilibrium, environment and feedback effects, as highlighted by the scatter in both $ f_{\rm H_2} $ (Fig.~\ref{fig:phasefraction2}) and $ f_{\rm H_2, cold} $ (Fig.~\ref{fig:phasefraction3}).
These results are consistent with Figs.~\ref{fig:phasemassrelations} and \ref{fig:phasemassrelations2}. %in the main text.

\section{Depletion times}  \label{appendixDepletionTimes}
%*****************************************************************************
%
\begin{figure*}
 \centering
 \includegraphics[width=0.33\textwidth] {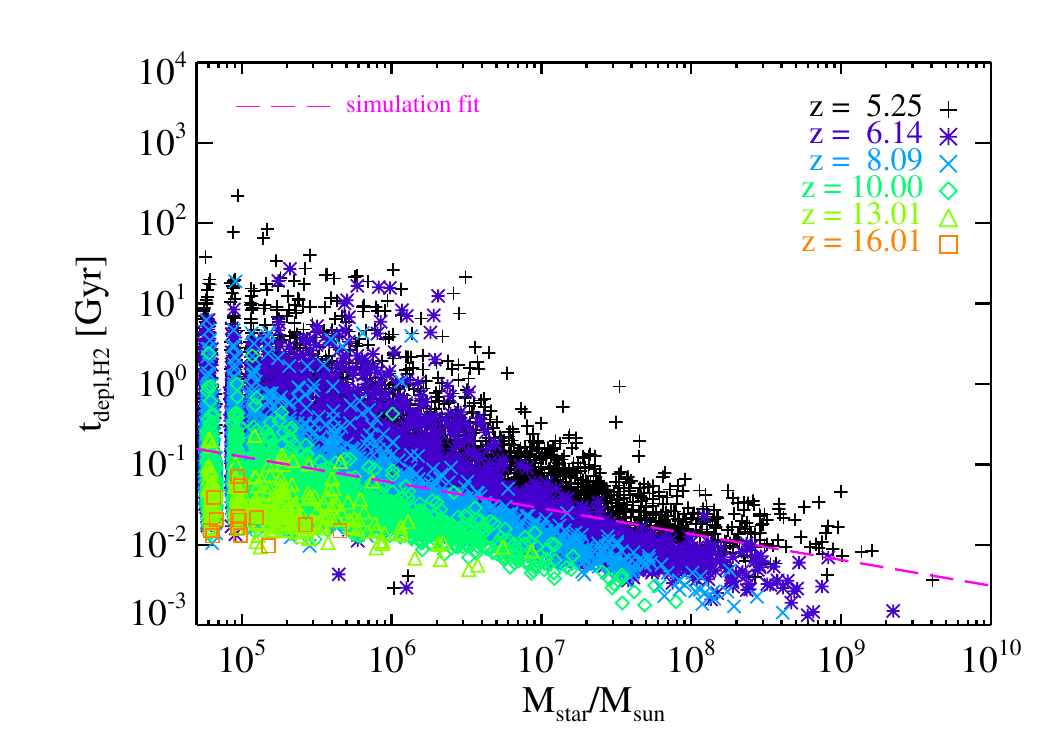}
 \includegraphics[width=0.33\textwidth] {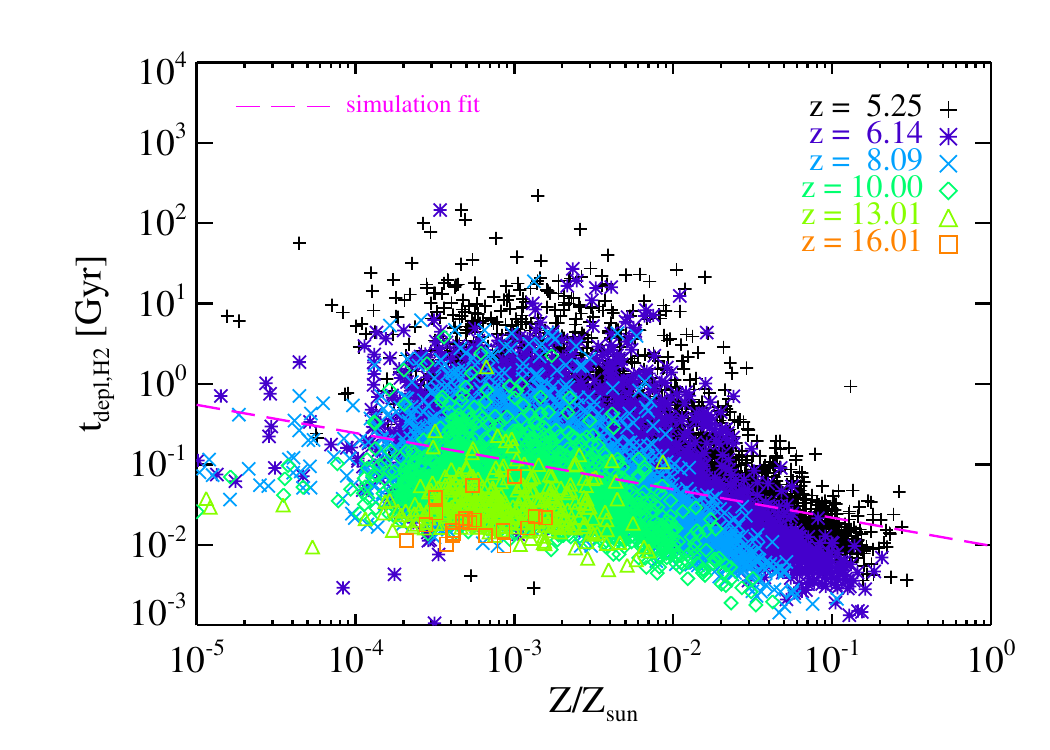}
 \includegraphics[width=0.33\textwidth] {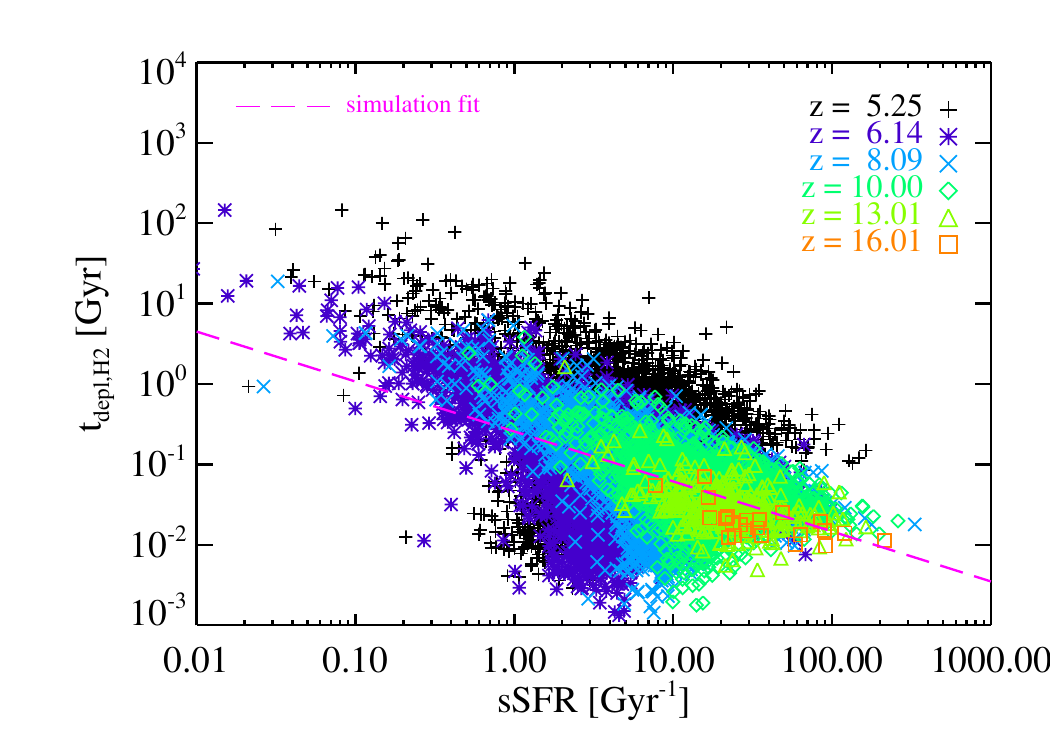}
\\
 \vspace{-0.3cm} 
 \includegraphics[width=0.33\textwidth] {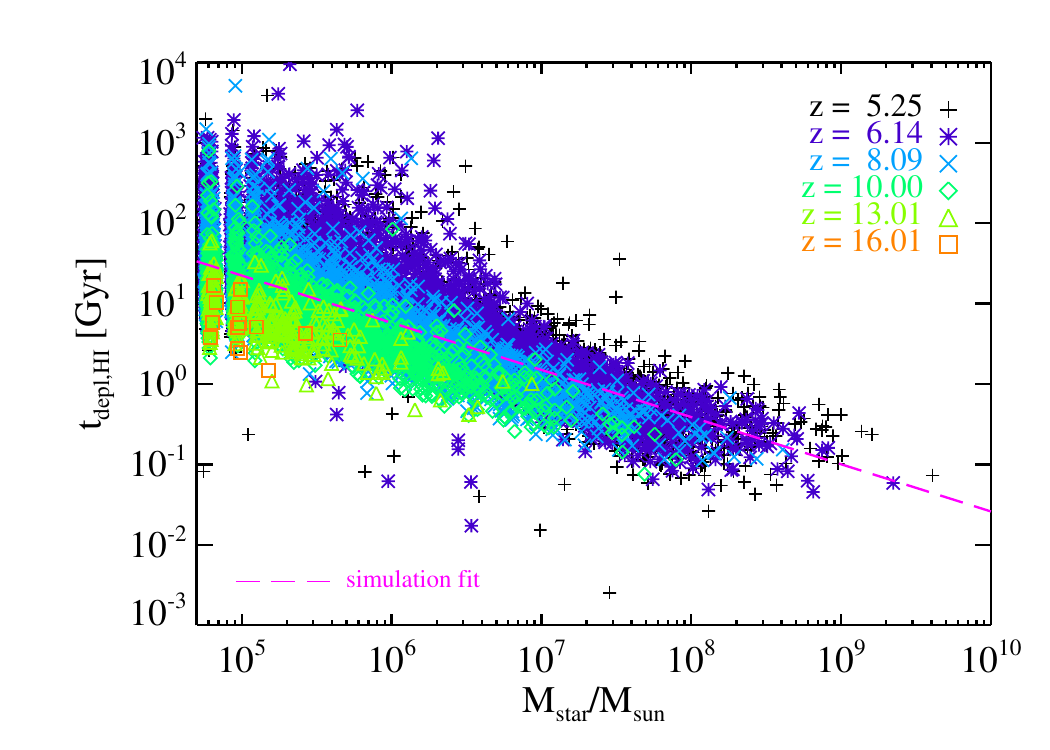}
 \includegraphics[width=0.33\textwidth] {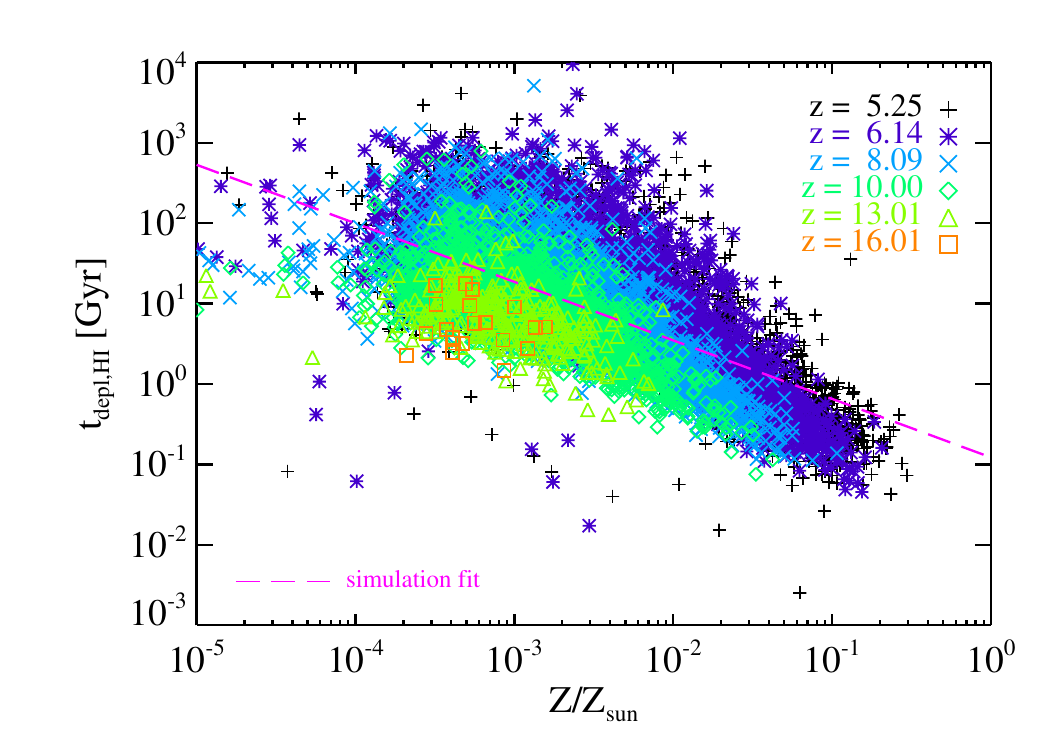}
 \includegraphics[width=0.33\textwidth] {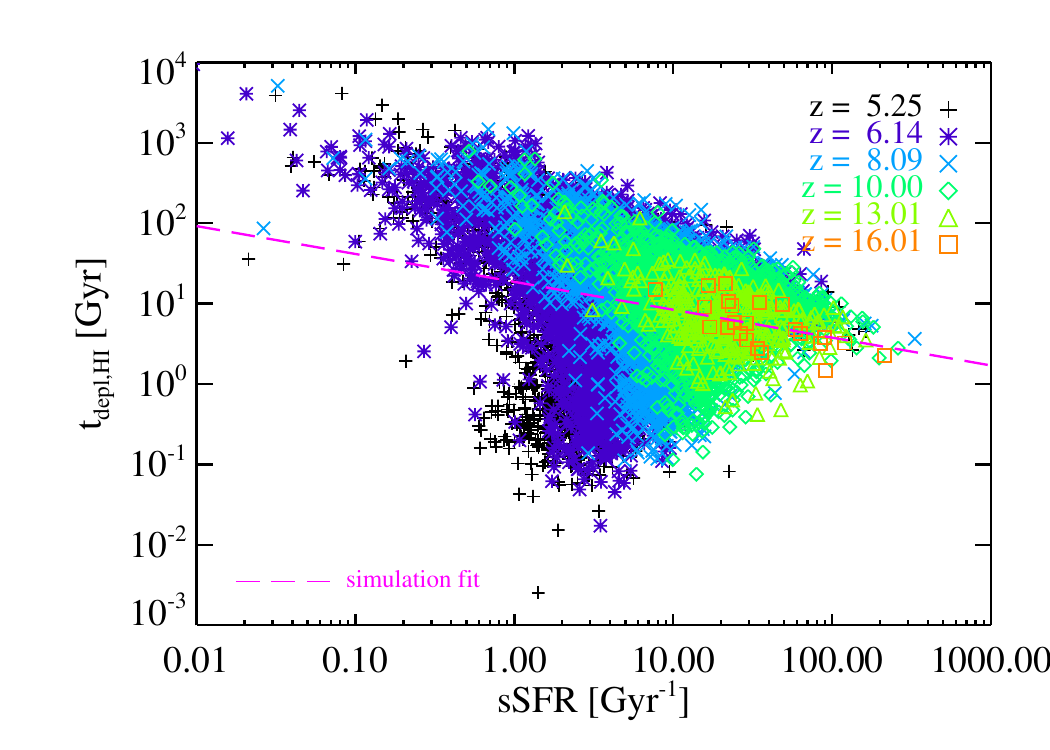}
 \caption{\small H$_2$ (top) and HI (bottom) depletion times versus stellar mass (left), gas metallicity (centre) and sSFR (right) at $z = $5.25, 6.14, 8.09, 10.00, 13.01, and 16.01. Simulation fits (dashed magenta lines) are in Appendix~\ref{appendixFits}. }
 \label{fig:tdepl2}
\end{figure*}
HI and H$_2$ gas depletion times as function of stellar mass $M_{\rm star}$, gas metallicity $Z$ and sSFR can be found in Fig.~\ref{fig:tdepl2}.
Data points show rather broad relations due to the role of feedback and environment in shaping  gas properties at all times.
The values of $t_{\rm depl, HI} $ and $t_{\rm depl, H_2} $ decrease with stellar mass and metallicity and minimum values are reached in star-forming structures with $M_{\rm star} \sim 10^8$--$10^9\,\rm M_\odot$ and $Z > 0.01\,Z_\odot$.
The similarity of these trends with the ones with SFR in Fig.~\ref{fig:tdepl} is simply understood in terms of the main-sequence and mass-metallicity relations.

\section{Fitting functions}  \label{appendixFits}

%*****************************************************************************

We fit simulated results in the base-10 logarithmic form
\begin{align}
\nonumber
{ \rm Log}\,y  &= a + b \, {\rm Log}\,x
\end{align}
where $x$ and $y$ are dimensionless, non-negative variables and $a$ and $b$ are fit parameters.\\
We consider data for masses, SFR and metallicities both at individual redshifts and in the whole $z>5$ range (all $z$).
Masses and metallicities are expressed in solar units, SFR in $\rm M_\odot yr^{-1}$, depletion times in Gyr and sSFR in Gyr$^{-1}$.
\\
In Tab.~\ref{tab:MphaseMstarSFR} we report  $a$ and $b$ parameters for the main sequence and the relations between $M_{\rm cold}$, $M_{\rm warm}$, $M_{\rm hot}$ vs $M_{\rm star}$ or SFR: \\
$ {\rm Log} ({\rm sSFR/Gyr^{-1}}) 	= a + b \, {\rm Log} (M_{\rm star} / {\rm M_\odot})$ \\
$ {\rm Log} (M_{\rm cold} / {\rm M_\odot}) 	= a + b \, {\rm Log} (M_{\rm star} / {\rm M_\odot})$ \\
$ {\rm Log} (M_{\rm warm} / {\rm M_\odot}) 	= a + b \, {\rm Log} (M_{\rm star} / {\rm M_\odot})$ \\
$ {\rm Log} (M_{\rm hot} / {\rm M_\odot}) 	= a + b \, {\rm Log} (M_{\rm star} / {\rm M_\odot})$ \\
$ {\rm Log} (M_{\rm cold} / {\rm M_\odot}) 	= a + b \, {\rm Log} [{\rm SFR} / ({\rm M_\odot\,yr^{-1}})] $ \\
$ {\rm Log} (M_{\rm warm} / {\rm M_\odot}) 	= a + b \, {\rm Log} [{\rm SFR} / ({\rm M_\odot\,yr^{-1}})] $ \\
$ {\rm Log} (M_{\rm hot} / {\rm M_\odot}) 	= a + b \, {\rm Log} [{\rm SFR} / ({\rm M_\odot\,yr^{-1}})] $.\\
Parameter determinations for the $M_{\rm hot}$ relations are missing at $z \gtrsim 10$ due to lack of statistics.
To remove resolution artefacts, we fit the data for objects with stellar masses and hot-gas masses above $ 2 \times 10^6 \,\rm M_\odot$ and SFR$ > 3 \times 10^{-2} \,\rm M_\odot\, yr^{-1} $.
\\
In Tab.~\ref{tab:mHIH2MvirMstarSFR} we quote $a$ and $b$ parameters for the relations between $M_{\rm HI}$ or M$_{\rm H_2}$ and $M_{\rm vir}$ or $M_{\rm star}$ and for the relations between $M_{\rm HI}$ or M$_{\rm H_2}$ and SFR: \\
$ {\rm Log} (M_{\rm HI} / {\rm M_\odot}) 	= a + b \, {\rm Log} (M_{\rm vir} / {\rm M_\odot})$\\
$ {\rm Log} (M_{\rm H_2} / {\rm M_\odot}) 	= a + b \, {\rm Log} (M_{\rm vir} / {\rm M_\odot})$\\
$ {\rm Log} (M_{\rm HI} / {\rm M_\odot}) 	= a + b \, {\rm Log} (M_{\rm star} / {\rm M_\odot})$\\
$ {\rm Log} (M_{\rm H_2} / {\rm M_\odot}) 	= a + b \, {\rm Log} (M_{\rm star} / {\rm M_\odot})$\\
$ {\rm Log} (M_{\rm HI} / {\rm M_\odot}) 	= a + b \, {\rm Log} [{\rm SFR} / ({\rm M_\odot\,yr^{-1}})] $ \\
$ {\rm Log} (M_{\rm H_2} / {\rm M_\odot}) 	= a + b \, {\rm Log} [{\rm SFR} / ({\rm M_\odot\,yr^{-1}})] $.\\
Results for HI and H$_2$ gas-to-star fractions $\mu_{\rm HI} = M_{\rm HI} / M_{\rm star} $ and $\mu_{\rm H_2} = M_{\rm H_2} / M_{\rm star} $ as functions of $M_{\rm star}$ are also fitted as:\\
$ {\rm Log} \mu_{\rm HI} = a + b \, {\rm Log} (M_{\rm star}/{\rm M_\odot})$ \\
$ {\rm Log} \mu_{\rm H_2} = a + b \, {\rm Log} (M_{\rm star}/{\rm M_\odot})$.\\
Corresponding $F_{\rm gas}$ values are obtained via $F_{\rm gas} = \mu_{\rm H_2} / (\mu_{\rm H_2}+1)$.
\\
Fit parameters for HI and H$_2$ depletion times, $ t_{\rm depl, HI} $ and $ t_{\rm depl, H_2} $, as functions of $M_{\rm star}$, SFR, gas $Z$ and sSFR are quoted in Tab.~\ref{tab:depletion} for the expressions below:\\
 $ {\rm Log} (t_{\rm depl, HI} / {\rm Gyr}) 	= a + b \, {\rm Log} (M_{\rm star} / {\rm M_\odot}) $\\
 $ {\rm Log} (t_{\rm depl, H_2} / {\rm Gyr}) 	= a + b \, {\rm Log} (M_{\rm star} / {\rm M_\odot}) $\\
 $ {\rm Log} (t_{\rm depl, HI} / {\rm Gyr}) 	= a + b \, {\rm Log} [{\rm SFR} / ({\rm M_\odot\,yr^{-1}})] $\\
 $ {\rm Log} (t_{\rm depl, H_2} / {\rm Gyr}) 	= a + b \, {\rm Log} [{\rm SFR} / ({\rm M_\odot\,yr^{-1})}] $\\
 $ {\rm Log} (t_{\rm depl, HI} / {\rm Gyr}) 	= a + b \, {\rm Log} (Z / {\rm Z_\odot}) $\\
 $ {\rm Log} (t_{\rm depl, H_2} / {\rm Gyr}) 	= a + b \, {\rm Log} (Z / {\rm Z_\odot}) $\\ 
 $ {\rm Log} (t_{\rm depl, HI} / {\rm Gyr}) 	= a + b \, {\rm Log} ({\rm sSFR} / {\rm Gyr^{-1}}) $\\
 $ {\rm Log} (t_{\rm depl, H_2} / {\rm Gyr}) 	= a + b \, {\rm Log} ({\rm sSFR} / {\rm Gyr^{-1}}) $.\\

\noindent
Redshift evolution of the median $ t_{\rm depl, HI} $ and $ t_{\rm depl, H_2} $ is:\\
$ {\rm Log} (t_{\rm depl,HI} /{\rm Gyr}) = 1.345 - 0.170 \, {\rm Log} (1 + z)$\\
$ {\rm Log} (t_{\rm depl,H_2} /{\rm Gyr}) = 0.943 - 2.045 \, {\rm Log} (1 + z)$.\\
Their behaviour is bracketed by the average trends\\
$ {\rm Log} (t_{\rm depl,HI} /{\rm Gyr}) = -1.177 + 1.578 \, {\rm Log} (1 + z)$\\
$ {\rm Log} (t_{\rm depl,H_2} /{\rm Gyr}) = -1.499 - 0.236 \, {\rm Log} (1 + z)$\\
and the mean trends \\
$ {\rm Log} (t_{\rm depl,HI} /{\rm Gyr}) = 2.857- 1.301 \, {\rm Log} (1 + z)$\\
$ {\rm Log} (t_{\rm depl,H_2} /{\rm Gyr}) = 2.273 - 3.065 \, {\rm Log} (1 + z)$.\\
The evolution of the average $\mu_{\rm HI}$ and $\mu_{\rm H_2}$ is given by:\\
$ {\rm Log}\,\mu_{\rm HI} = -3.379 + 4.813 \, {\rm Log} (1 + z)$ \\
$ {\rm Log}\,\mu_{\rm H_2} = -3.700 + 2.300 \, {\rm Log} (1 + z)$\\
and $ F_{ \rm gas} = \mu_{\rm H_2} / (\mu_{\rm H_2} + 1) $.\\

\noindent
We fit the recovery timescale for 15-$200\,\rm M_\odot$ stars as:\\
${\rm Log} (t_{\rm rec}/{\rm Myr}) = 3.295 - 2.349 \, {\rm Log} (t_{\rm age}/{\rm Myr})$, equivalent to\\
${\rm Log} (t_{\rm rec}/{\rm Myr}) = -0.424 + 1.307 \, {\rm Log} (M_{\rm star}/{\rm M_\odot})$.
\\

\noindent
The fitting function for the stellar return fraction $R$ (eq.~\ref{eq:fit}) is:
\begin{align}
\nonumber
R (t_{\rm age}) &=  0.0885 \, {\rm Log} \left( \frac{t_{\rm age}}{ 3.853\,\rm Myr} \right)
\end{align}
for stellar ages $t_{\rm age} > 3.853 \,\rm Myr$ and $R = 0$ otherwise. \\

\clearpage
\newpage

\begin{table*} %  fits
\tiny
\centering
\caption{\small Fit parameters for the main sequence (sSFR vs. $M_{\rm star}$) and the relations between $M_{\rm cold}$ or $M_{\rm warm}$ or $M_{\rm hot}$ and $M_{\rm star}$ or SFR. Masses are in solar units, SFR in solar masses per year and sSFR in Gyr$^{-1}$.}
\label{tab:fits}
\label{tab:MphaseMstarSFR}
\begin{tabular}{c | c c | c c | c c | c c | c c | c c | c c}
$z$	 & \multicolumn{2}{c|}{sSFR-$M_{\rm star}$}     & 	\multicolumn{2}{c|}{$M_{\rm cold}$-$M_{\rm star}$}   	& 	\multicolumn{2}{c|}{$M_{\rm warm}$-$M_{\rm star}$} & \multicolumn{2}{c|}{$M_{\rm hot}$-$M_{\rm star}$} &  \multicolumn{2}{c|}{$M_{\rm cold}$-SFR} &  \multicolumn{2}{c|}{$M_{\rm warm}$-SFR} &  \multicolumn{2}{c}{$M_{\rm hot}$-SFR} \\
   	& 	$a$ 			& 	$b$ 			& 	$a$ 			& 	$b$ 			& 		$a$		& 	$b$			& 	$a$ & 	$b$ & $a$ 	& 	$b$ & $a$ 	& 	$b$ & $a$ & $b$ \\
\hline
     all   &	1.871	&	-0.187 & 6.199 & 0.237 	& 4.615 & 0.495 	& 1.602 & 0.645 	& 8.156 & 0.239 	& 8.715 & 0.492 	& 7.139 & 0.625\\
 16.01 &	3.370 	&	-0.376 & 5.275 & 0.407 	& 4.891 & 0.469 	& - & - 			& 7.942 & 0.271 	& 8.741 & 0.601 	& - & - \\
 13.01 &	2.212	&	-0.175 & 6.029 & 0.262 	& 4.512 & 0.526 	& - & - 			& 8.021 & 0.257 	& 8.440 & 0.480 	& - & - \\
 10.00 &	1.787	&	-0.115 & 6.153 & 0.254 	& 4.179 & 0.581 	& - & - 			& 8.111 & 0.242 	& 8.598 & 0.531 	& - & - \\
   8.09 &	1.464	&	-0.091 & 6.031 & 0.274 	& 4.499 & 0.529 	& -2.716 & 1.090 	& 8.178 & 0.252 	& 8.615 & 0.474 	& 6.369 & 0.484\\
   6.14 &	1.185	&	-0.093 & 6.109 & 0.254 	& 4.646 & 0.491 	& -0.366 & 0.852 	& 8.193 & 0.230 	& 8.746 & 0.463 	& 6.853 & 0.849\\
   5.25 &	1.531	&	-0.158 & 6.185 & 0.228 	& 4.380 & 0.517 	&  1.033 & 0.723 	& 8.130 & 0.234 	& 8.818 & 0.487 	& 7.398 & 0.905\\
\hline
\end{tabular}
\end{table*}
\begin{table*}  %  fits
\tiny
\centering
\caption{\small Fit parameters for $M_{\rm HI}$ or M$_{\rm H_2}$ and $M_{\rm vir}$, $M_{\rm star}$ and SFR relations, as well as for gas-to-star fractions $ \mu_{\rm HI}$ and $ \mu_{\rm H_2}$ versus $M_{\rm star}$ with masses in solar units and SFR in solar masses per year.}
\label{tab:mHIH2MvirMstarSFR}
\begin{tabular}{c | c c | c c | c c | c c | c c | c c| c c | c c}
$z$	& 	\multicolumn{2}{c|}{$M_{\rm HI}$-$M_{\rm vir}$}   	& 	\multicolumn{2}{c|}{$M_{\rm H_2}$-$M_{\rm vir}$} & \multicolumn{2}{c|}{$M_{\rm HI}$-$M_{\rm star}$} &  \multicolumn{2}{c|}{$M_{\rm H_2}$-$M_{\rm star}$} &  \multicolumn{2}{c|}{$M_{\rm HI}$-SFR} &  \multicolumn{2}{c|}{$M_{\rm H_2}$-SFR} &	\multicolumn{2}{c|}{$\mu_{\rm HI}$-$M_{\rm star}$} &	\multicolumn{2}{c}{$\mu_{\rm H_2}$-$M_{\rm star}$} \\
   	& 	$a$ 			& 	$b$ 			& 		$a$		& 	$b$			& 	$a$ & 	$b$ & $a$ 	& 	$b$ & $a$ 	& 	$b$ & $a$ & $b$ & $a$ & $b$  && \\
\hline
  all 	   &	1.710  &	0.659 &	-2.577 & 	0.926 & 	6.084 & 	0.236 &	2.156 & 	0.505 & 8.031 & 0.238 &	6.496 &	0.484 & 6.084 & -0.764 & 2.443 & -0.489\\
 16.01 &	-0.029 &	0.862 &	-4.530 & 	1.115 & 	5.160 & 	0.406 &	1.240 & 	0.708 & 7.820 & 0.271 &	6.625 &	0.773 & 5.160 & -0.594 & 1.240 & -0.293\\
 13.01 &	0.303 &	0.828 &	-3.041 & 	0.956 & 	5.916 & 	0.261 &	2.038 & 	0.561 & 7.896 & 0.256 &	6.517 &	0.647 & 5.916 & -0.739 & 2.038 & -0.439\\
 10.00 &	1.126 &	0.737 &	-3.619 & 	1.030 & 	6.035 & 	0.253 &	2.455 & 	0.501 & 7.986 & 0.241 &	6.392 &	0.519 & 6.035 & -0.747 & 2.455 & -0.499\\
  8.09  &	0.911 &	0.759 &	-3.546 & 	1.024 & 	5.916 & 	0.273 &	2.143 & 	0.549 & 8.054 & 0.251 &	6.478 &	0.526 & 5.916 & -0.727 & 2.157 & -0.451\\
  6.14  &	1.300 &	0.707 &	-2.881 & 	0.945 & 	5.993 & 	0.253 &	1.862 & 	0.534 & 8.068 & 0.229 &	6.451 &	0.476 & 5.993 & -0.747 & 2.249 & -0.481\\
  5.25  &	2.106 &	0.603 &	 0.923 & 	0.577 & 	6.073 & 	0.226 &	1.053 & 	0.652 & 8.004 & 0.233 &	6.506 &	0.298 & 6.073 & -0.774 & 3.899 & -0.671\\
\hline
\end{tabular}
\end{table*}
 \begin{table*} %  fits
 \tiny
 \centering
 \caption{\small Fit parameters for HI and H$_2$ depletion times $t_{\rm depl, HI} $ and $t_{\rm depl, H_2}$ in Gyr versus $M_{\rm star}$, SFR, gas $Z$ and sSFR with masses in solar units, SFR in solar masses per year, $Z$ in units of $Z_\odot$ and sSFR in Gyr$^{-1}$. }
 \label{tab:depletion}
 \begin{tabular}{c | c c | c c | c c | c c | c c| c c | c c| c c}
 $z$	& \multicolumn{2}{c|}{$t_{\rm depl, HI}$-$M_{\rm star}$} & \multicolumn{2}{c|}{$t_{\rm depl, H_2}$-$M_{\rm star}$} & \multicolumn{2}{c|}{$t_{\rm depl, HI}$-SFR} & \multicolumn{2}{c|}{$t_{\rm depl, H_2}$-SFR} & \multicolumn{2}{c|}{$t_{\rm depl, HI}$-$Z$} & \multicolumn{2}{c|}{$t_{\rm depl, H_2}$-$Z$} & \multicolumn{2}{c|}{$t_{\rm depl, HI}$-sSFR}& \multicolumn{2}{c}{$t_{\rm depl, H_2}$-sSFR}\\
  	& $a$  & $b$ &	 $a$	& $b$ & $a$ & 	$b$ & $a$ & $b$ & $a$ & $b$	& $a$ & $b$ & $a$ & $b$ & $a$ & $b$\\
\hline
  all 	   	& 4.283 & -0.587 &  0.704 & -0.321  & -0.969 &	-0.761 &	-2.443 &	-0.508 & -0.913 & -0.728 & -2.013 & -0.350 & 1.270 &-0.347& -0.591 & -0.621\\
 16.01 	& 1.790 & -0.218 & -2.131 &  0.083  & -1.180 &	-0.729 &	-2.375 & 	-0.227 &  0.388 & -0.100 & -1.168 &  0.168  & 1.679 &-0.605& -1.071 & -0.423\\
 13.01 	& 3.704 & -0.564 & -0.174 & -0.264  & -1.104 &	-0.744 &	-2.483 & 	-0.353 & -0.386 & -0.393 & -1.724 & -0.070 & 1.466 &-0.468& -0.946 & -0.426\\
 10.00 	& 4.251 & -0.632 &  0.672 & -0.385  & -1.014 &	-0.759 &	-2.608 & 	-0.481 & -1.146 & -0.714 & -2.444 & -0.379 & 1.506 &-0.473& -0.734 & -0.510\\
   8.09  	& 4.472 & -0.638 &  0.731 & -0.366  & -0.946 &	-0.749 &	-2.519 & 	-0.473 & -1.195 & -0.799 & -2.405 & -0.420 & 1.600 &-0.586& -0.603 & -0.656\\
  6.14  	& 4.962 & -0.676 &  1.318 & -0.425  & -0.932 &	-0.771 &	-2.516 & 	-0.527 & -1.038 & -0.840 & -2.375 & -0.497 & 1.444 &-0.575& -0.681 & -0.688\\
  5.25  	& 4.617 & -0.626 &  2.749 & -0.565  & -0.996 &	-0.767 &	-2.350 & 	-0.706 & -1.071 & -0.840 & -2.357 & -0.746 & 0.955 &-0.209& -0.522 & -0.244\\
\hline
\end{tabular}
\end{table*}
%

% =============================================================================

\end{document}